\documentclass[fleqn,usenatbib]{mnras}
\usepackage{newtxtext,newtxmath}

\usepackage[T1]{fontenc}
\usepackage{ae,aecompl}

\usepackage{graphicx}	
\usepackage{amsmath}	
\usepackage{amssymb}	
\usepackage{nccmath}
\usepackage{rotating}
\usepackage{tabularx}
\usepackage{longtable}
\usepackage{lastpage}
\usepackage{caption}
\usepackage{multicol}

\def\arcsec{$\,^{\prime\prime}$~}

\def\Chandra{${\it Chandra}$}

\def\ergss{erg s$^{-1}$}

\def\simge{\mathrel{%
   \rlap{\raise 0.511ex \hbox{$>$}}{\lower 0.511ex \hbox{$\sim$}}}}
\def\simle{\mathrel{
   \rlap{\raise 0.511ex \hbox{$<$}}{\lower 0.511ex \hbox{$\sim$}}}}

\newcommand{\Msun}{\ifmmode {M_{\odot}}\else${M_{\odot}}$\fi}
\newcommand{\Lsun}{\ifmmode {L_{\odot}}\else${L_{\odot}}$\fi}
\newcommand{\Rsun}{\ifmmode {R_{\odot}}\else${R_{\odot}}$\fi}

\newcommand{\lsim }{{\lower0.8ex\hbox{$\buildrel <\over\sim$}}}
\newcommand{\gsim }{{\lower0.8ex\hbox{$\buildrel >\over\sim$}}}
\newcolumntype{Y}{>{\centering\arraybackslash}X}
\newcolumntype{b}{>{\hsize=.75\hsize}X}
\newcolumntype{B}{>{\hsize=1.2\hsize}X}
\newcolumntype{m}{>{\centering\arraybackslash\hsize=.6\hsize}X}
\newcolumntype{s}{>{\centering\arraybackslash\hsize=.5\hsize}X}
\newcolumntype{S}{>{\centering\arraybackslash\hsize=.35\hsize}X}

\title[X-ray Emissivity of Stellar Populations]{The X-ray Emissivity of Low-Density Stellar Populations}

\author[C.~O. Heinke et al.]{C.~O. Heinke$^{1}$\thanks{E-mail: heinke@ualberta.ca}, M.~G. Ivanov$^{1}$,   E.~W. Koch$^{1}$, R. Andrews$^{1}$, L. Chomiuk$^{2}$, \newauthor H.~N. Cohn$^{3}$, S. Crothers$^{1}$,  T. de Boer$^{4}$,  N.~Ivanova$^{1}$,  A.~K.~H. Kong$^{5}$, \newauthor N.~W.~C.~Leigh$^{1,6,7}$,  P.~M. Lugger$^{3}$,  L.~Nelson$^{8}$,  C.~J. Parr$^{1}$, E.~W. Rosolowsky$^{1}$, \newauthor A.~J. Ruiter$^{9}$,  C.~L. Sarazin$^{10}$,  A.~W. Shaw$^{1,11}$, G.~R. Sivakoff$^{1}$, M. van den Berg$^{12}$
\\
$^{1}${Dept.\ of Physics, University of Alberta, CCIS 4-183, Edmonton, AB T6G 2E1, Canada; heinke@ualberta.ca}\\
$^{2}${Dept.\ of Physics \& Astronomy, Michigan State University, East Lansing, MI, USA}\\
$^{3}${Dept.\ of Astronomy, Indiana University, 727 E. Third Street, Bloomington, IN 47405, USA}\\
$^{4}${Dept.\ of Physics, University of Surrey, Guildford GU2 7XH, UK}\\
$^5${Institute of Astronomy, National Tsing Hua University, Hsinchu 30013, Taiwan}\\
$^{6}${Dept. of Astrophysics, American Museum of Natural History, Central Park West and 79th Street, New York, NY 10024-5192, USA,} \\
$^7${Departamento de Astronom\'ia, Facultad de Ciencias F\'isicas y Matem\'aticas,
Universidad de Concepci\'on, Concepci\'on, Chile}\\
$^8${Dept. of Physics \& Astronomy, Bishop's University, 2600 rue College, Sherbrooke, Qu\'{e}bec J1M 1Z7, Canada}\\
$^9${ARC Future Fellow, School of Physical, Environmental and Mathematical Sciences, University of New South Wales, Australian Defence Force Academy, Canberra, ACT 2600, Australia}\\
$^{10}${Dept. of Astronomy, University of Virginia, PO Box 400325, Charlottesville, VA, 22904, USA}\\
$^{11}${Department of Physics, University of Nevada, Reno, NV 89557, USA}\\
$^{12}${Harvard-Smithsonian Center for Astrophysics, 60 Garden Street, Cambridge, MA 02138, USA}\\
}

\date{Accepted XXX. Received YYY; in original form ZZZ}

\pubyear{2020}

\begin{document}
\label{firstpage}
\pagerange{\pageref{firstpage}--\pageref{lastpage}}
\maketitle

\begin{abstract}
The dynamical production of low-mass X-ray binaries and brighter cataclysmic variables (CVs) in dense globular clusters is well-established.  We investigate how the X-ray emissivity of fainter X-ray binaries (principally CVs and coronally active binaries) varies  between different environments. 
We compile  calculations (largely from the literature) of the X-ray emissivity of old stellar populations, including open and globular clusters and several galaxies.
We investigate three literature claims of unusual X-ray sources in low-density stellar populations. We show that a suggested quiescent neutron star in the open cluster NGC 6819 is a foreground M dwarf. We show that the suggested diffuse X-ray emission from an old nova shell in the globular cluster NGC 6366 is actually a background galaxy cluster. And we show that a suggested population of quiescent X-ray binaries in the Sculptor Dwarf Galaxy is mostly (perhaps entirely) background galaxies. 
We find that above densities of $10^4$ \Msun/pc$^3$, the X-ray emissivity of  globular clusters increases, due to dynamical production of X-ray emitting systems. Below this density, globular clusters have lower X-ray emissivity than the other populations, and we do not see a strong dependence of X-ray emissivity due to density effects.  We find  significant correlations between X-ray emissivity and binary fraction,  metallicity, and density. Sampling these fits via bootstrap techniques gives less significant correlations, but confirms the effect of metallicity on low-density populations, and that of density on the full globular cluster sample.
\end{abstract}




\section{Introduction}\label{s:intro}

The effects of dynamical processes in dense environments on X-ray emitting binaries were recognized early in X-ray astronomy \citep{Katz75,Clark75}. It is now well understood that the high densities in the core of globular clusters permit various interactions that place single neutron stars (and black holes) into close binaries, elevating the number of low-mass X-ray binaries (LMXBs) per unit mass in globular clusters to $\sim$100 times the Galactic field rate \citep{Fabian75,Hills76,Verbunt87b,Hut91}.  Studies of the distribution of bright ($L_X>10^{36}$ ergs/s) LMXBs in Galactic and extragalactic globular clusters vs. the characteristics of those clusters essentially confirm the theoretically expected dependence upon cluster dynamical properties \citep{Verbunt87, Jordan04, Sivakoff07}, while also identifying a metallicity dependence \citep{Kundu03, Sarazin03}. Observations of Galactic globular clusters identify roughly ten times as many quiescent LMXB candidates (soft, blackbody-like X-ray sources with $L_X\sim10^{32-33}$ ergs/s) as bright LMXBs \citep{Pooley03,Heinke03d}. The quiescent LMXBs are thought to be transient LMXBs observed between outbursts \citep[e.g.][]{Verbunt84,Campana98a}.  Their numbers in globular clusters clearly show the predicted dependence on cluster properties \citep{Pooley06,Heinke06b}. Radio millisecond pulsars (MSPs), the progeny of LMXBs, are also distributed among clusters roughly as expected by the theory of collision rates \citep{Johnston92,Ransom05,Bahramian13,Lorimer15}, though their X-ray luminosities are rarely high enough to dominate the cluster $L_X$ (e.g. \citealt{Bogdanov06}; but cf. \citealt{Becker03}).

With the capabilities of {\it Hubble} and \Chandra\ to identify low-luminosity X-ray sources in dense clusters, it is possible to study the relation between dynamical processes and fainter X-ray emitting binaries. Observations of the numbers of faint ($L_X<10^{33.5}$ ergs/s) X-ray sources vs. globular cluster properties indicate that the brightest such sources, largely cataclysmic variables (CVs) containing white dwarfs (WDs), are more frequent in denser clusters \citep{Johnston96,Pooley03,Heinke03d,Bahramian13}, though the dependence on cluster density may be less steep than for LMXBs \citep{Pooley06,Heinke06b}. Theoretical work \citep{Davies97, Ivanova06, Shara06} indicates that dynamical effects both destroy primordial CVs and produce new, dynamically formed CVs, with some work suggesting that CVs are rarer in medium-density globular clusters than in the field due to binary destruction \citep{Shara06}. This last prediction has received  support from a deep study of $\omega$ Centauri's X-ray sources \citep{Haggard09}, which limited the number of CVs in the core of $\omega$ Centauri to $<1/2$ the number expected from field CV density measurements \citep{Patterson98,Schwope02,Grindlay05,Pretorius07}.

Coronally active binaries (ABs) are (typically) tidally-locked partially convective stars, with rotation rates (and thus magnetic fields, and X-ray activity) enhanced over other stars of similar ages.  They make up the majority of X-ray sources by number in old stellar populations, though their X-ray luminosities are low (typically $L_X<10^{31.5}$ ergs/s, \citealt{Dempsey93a,Dempsey97}). They have been detected in many globular clusters, and have generally been thought to be primordial in origin and their numbers to scale simply with cluster mass \citep{Grindlay01a,Verbunt02,Bassa04,Kong06}. Studies of nearby, moderate-density clusters including M4 and M71 indicate that they have more ABs than expected for their masses, compared to denser globular clusters \citep{Bassa04,Elsner08, Huang10}.  Several studies have shown that X-rays from lower-density globular clusters are likely dominated by ABs of primordial origin, rather than by dynamically formed binaries \citep{Kong06,Verbunt08,Bassa08,Lu11,Cheng18}.

Substantial work has been done to identify the nature of the X-ray sources in open clusters, particularly M67 \citep{Belloni93,Belloni98,vandenBerg02,vandenBerg04}, identifying many ABs in close binaries (1-10 days), one CV, and several long-period (30-$10^4$ day) binaries whose X-rays cannot be satisfactorily explained.  
\citet{Verbunt99,Verbunt01} noted from ROSAT observations that globular clusters contain fewer ABs per unit mass than open clusters such as M67; in fact, the total X-ray luminosity per unit mass was lower in most globular clusters without bright LMXBs than in M67.  Verbunt suggested two possible explanations:  that binaries are destroyed efficiently in globular clusters, and/or that M67 has lost a very large fraction of its mass (and of its non-binary content), thus concentrating X-ray luminous ABs.  A key point of this paper is to address this question observationally, by considering the X-ray luminosity per unit mass in a range of environments of different stellar densities.  We take advantage of a series of studies establishing the X-ray emissivity of a variety of old stellar populations \citep{Sazonov06,Revnivtsev07,Revnivtsev08,Warwick14,Ge15}.  

The question of binary destruction in dense environments vs. variations in the initial binary fraction in different environments has been a topic of substantial work. Binaries with binding energy less than the typical thermal energy of stellar systems (``soft'' binaries; in massive globular clusters, typically these have separations larger than an AU) are easily destroyed early in clusters' lives \citep{Heggie75,Kroupa01,Parker09}. Observationally, dense clusters are found to have much lower photometric binary fractions than low-density globular clusters \citep{Bellazzini02,Clark04,Zhao05,Sollima07,Milone08,Davis08,Sommariva09,Milone12}, and open clusters are found to have even higher binary fractions \citep{Sollima10}. Studies have disagreed over whether the fraction of binaries in cluster cores diminishes over time in dense clusters or grows with time \citep{Ivanova05b,Hurley07}, with newer studies giving more complex answers, depending on how soft binaries are treated and whether binaries containing two compact objects are included \citep{Sollima08,Fregeau09}. 

In this paper, we combine information from  the literature and new analyses to compare the X-ray emissivity (X-ray luminosity per unit mass) from a variety of stellar populations, and investigate possible trends with density, metallicity, binary fraction, and age. In section 2, we compute X-ray luminosities per unit mass for various stellar environments.
Then in section 3, we investigate in detail three particular claims of intriguing objects in relatively low-density stellar systems, where the nature of the object(s) is crucial for our results. 
In section 4, we identify a density regime where dynamical production of X-ray binaries dominates X-ray emissivity, and then investigate the effects of metallicity, binary fraction, and age within the lower-density regime, where dynamical production of X-ray binaries is not dominant. In section 5, we compare our results to some recent literature on this topic, and conclude in section 6.

\section{Calculations of X-ray Emissivities}\label{s:calculations}

Here we compute X-ray luminosities per unit mass for a variety of stellar environments. 
We calculate all X-ray luminosities per unit mass in the 0.5-2.0 keV band, and exclude luminous LMXBs (with $L_X$(0.3-8 keV)$>10^{36}$ ergs/s).  Where possible we work with values from the literature. We also obtain (typically from the literature) a variety of other quantities that may be relevant to understanding the X-ray emissivity of each population; the metallicity, age, binary fraction, and stellar density of each population. 
Binary fractions are defined as $f_b$ = $N_b$/($N_s$+$N_b$), where $N_s$ is the number of single stars, and $N_b$ is the number of binary systems (we include higher-order multiples within binaries).

We consider the relevant density in each case to be the highest density which typical binaries will have experienced for a significant fraction (e.g. $>1$ Gyr) of  their lives.  In standard King-model globular clusters, this will be the central (core) density, as typical clusters have relaxed, allowing binaries to segregate into the cluster core.  The density experienced by binaries in core-collapsed globular clusters is likely to have a complex history.  For these clusters, we also use the standard central density; this choice does not have a large effect on our results, as such clusters are few in number and are the highest-density systems.  
Below we give some details about the values we used for some of these stellar populations.

\subsection{Galaxies}

We use estimates of the X-ray emissivity of local space (near the Sun), the bulge of M31, the giant elliptical NGC 3379, and the dwarf ellipticals M32, NGC 147, NGC 185, and NGC 205.
Galaxies typically have relaxation times longer than the Hubble time, and thus the density we consider is the density of the region where the binaries are found, and presumably were formed.  
We compute the mass density of the analyzed parts of M32
using its de Vaucouleurs profile \citep{Mazure02}.  
It is important to recognize that most binaries have been processed in the  slightly higher density environments of the young clusters in which most stars are born \citep[e.g.][]{Parker09}, but we use their current densities for this exercise. The parameters we use in our calculations, and their references, are summarized in Table~\ref{table*:nonglobs}. Galaxies often have complex star formation histories; we plot average ages of the bulk of the star formation for our galaxies, using the references in Table~\ref{table*:nonglobs}.

We take measurements of the 0.5-2.0 keV X-ray luminosity per unit mass for old stars (ABs and CVs, with LMXBs and young stellar objects excluded) in our local Galactic disc by averaging two independent measurements; \citet{Warwick14}, who use the XMM-Newton slew survey, and \citet{Sazonov06}, who use the ROSAT All-Sky Survey and RXTE Slew Survey (updated by \citealt{Revnivtsev07}).  Both estimates partly rely on 2-10 keV measurements, and thus there is some uncertainty in the extrapolation to the 0.5-2.0 keV band. We calculate the 0.5-2.0 keV X-ray emissivity, using Warwick's 2-10 keV estimates and Warwick's assumed spectral shapes, as $1.2\pm0.3\times10^{28}$ ergs/s/\Msun.
\citet{Revnivtsev07} excluded young stars from Sazonov's work to find $9\pm3\times10^{27}$ ergs/s/\Msun.  Thus we choose $1.05\pm0.3\times10^{28}$ ergs/s/\Msun\ for old stars in the local Galactic disc.  For the Galactic field, the binary fraction (including multiples) is taken from \citet{Raghavan10}. 


For NGC 3379, \citet{Revnivtsev08} find $8.2\pm2.6\times10^{27}$ ergs/s/\Msun.  Since then, \citet{Trinchieri08} have identified a portion of this emission as being produced by hot gas.  However, the inferred diffuse gas component only produces $L_X$(0.5-2.0)$=4\times10^{37}$ ergs/s, or 13\% of the total diffuse emission in the 0.5-2.0 keV band.  The diffuse X-ray emission from NGC 3379 also includes unresolved LMXBs with $L_X>10^{36}$ ergs/s, as the deep \Chandra\ observations do not resolve all sources at this $L_X$.  To remove these, we extrapolate the $dN/d{\rm ln}L$ of NGC 3379 in \citet{Kim09}'s Fig. 6, and find a total $L_X$(0.3-8 keV)$=2.1^{+0.8}_{-0.6}\times10^{38}$ ergs/s from unresolved LMXBs. The errors are derived from assuming either that all LMXBs below $6\times10^{36}$ ergs/s, or all LMXBs below $3\times10^{36}$ ergs/s, are unresolved.  Assuming a power-law index of 1.7 \citep{Brassington10}, this gives $L_X$(0.5-2.0)$=7.5^{+2.9}_{-2.1}\times10^{37}$ ergs/s, or a further reduction of 24\% compared to Revnivtsev's result, giving $5.2\pm1.6\times10^{27}$ ergs/s/\Msun.

We omit the Sculptor dwarf spheroid from our calculations, as discussed in \S \ref{s:Sculptor}, as we are not certain that any of the X-ray sources identified by \citet{Maccarone05} are actually associated with this galaxy. Studies of other dwarf galaxies (e.g. Draco, \citealt{Sonbas16}) have also failed to confidently identify X-ray sources (except for one symbiotic star, \citealt{Saeedi18}). We consider the state of the literature on X-ray binaries in dwarf spheroid galaxies to be inadequate to constrain their X-ray emissivity yet.

\subsection{Open clusters}

We consider five old ($>2$ Gyrs) open clusters, for which deep X-ray observations have been published, and for which estimates of mass and cluster structure exist.  Generally we use a 2 keV thermal plasma (MEKAL in XSPEC; \citealt{Liedahl95}) to convert reported X-ray luminosities to 0.5-2 keV, as typically assumed for conversions in the open cluster studies mentioned below.  We calculate $L_X$ both from the reported point sources of members, and by extracting a spectrum from within the full half-mass radius. We use these two measurements to identify the plausible range of $L_X$, with the identified members providing a lower limit.  We estimate central densities for the open clusters using a simple Plummer model, $\rho(r_0)=\frac{3 M_{Pl}}{4 \pi R_{Pl}^3}$, where $M_{Pl}$ is the total mass and $R_{Pl}$ is the half-mass radius \citep[e.g.][]{Chumak10}.

ROSAT observations of M67 and NGC 188 revealed numerous X-ray sources \citep{Belloni98}. A \Chandra\ observation of M67 \citep{vandenBerg04} allowed clear identification of counterparts (within the \Chandra\ field), which  combined with proper motion information identified which X-ray sources are members; summing the 0.5-2 keV luminosities of the members gives a total of  $3.6\times10^{31}$ ergs/s.
 Alternatively, we extract a spectrum from the 10.5' half-mass radius \citep{Fan96} of M67 from XMM-Newton ObsID 0109461001 (using MOS2 and pn data) and fit it simultaneously with a local background spectrum, finding $L_X=5.9\pm0.5\times10^{31}$ erg/s. 
M67's binary fraction and total mass are taken from \citet{Geller15}.   We estimate a central density of 32 \Msun/pc$^3$ using 2100 \Msun and a half-mass radius of 2.5 pc (10.5' at 820 pc).

A full catalogue of X-ray sources in NGC 188 observed by \Chandra\ and XMM-Newton is reported by \citet{Vats18} (\citealt{Gondoin05} gives a partial analysis), giving an estimated 0.5-2 keV $L_X$ for members of $2.45\pm0.55\times10^{31}$ erg/s.  We also extract a spectrum from the half-mass region using the \Chandra\ observation described in \citet{Vats18}. We subtract stowed background data covering the same detector region from the half-mass spectrum, and also from a nearby background spectrum, and simultaneously fit the half-mass and background spectra to determine the emission from the half-mass region. We adjust the exposure to match the spectra of the source and stowed spectra in the 9-12 keV energy range,  scale the sky background by the BACKSCAL values, and fit the additional source component with an APEC spectrum. This gives us $L_X$(0.5-2)$=3.3^{+2.6}_{-2.9}\times10^{31}$ erg/s. (The high uncertainty is largely due to the small area imaged by Chandra outside the half-mass radius.)

Mass estimates for NGC 188 range from 1500$\pm400$ \Msun\ \citep{Chumak10} to 2850$\pm$120 \Msun\ \citep{Geller13a}; we use the estimate of 2300$\pm$460 from \citet{Geller08} as a median that agrees with both.
We can estimate a core density of 9$\pm$2 \Msun/pc$^3$ from the Plummer model of \citet{Chumak10},  using their half-mass radius of 4 pc and our mass estimate above.   
 \citet{Geller12} estimated NGC 188's binary fraction, covering about 13 core radii from the cluster centre.

NGC 6791 has been studied in detail using \Chandra\ by \citet{vandenBerg13}, who identify three to four CVs, and twenty likely or candidate active binaries (including "red stragglers" or "sub-subgiants"). From \citet{vandenBerg13}'s proper motion and X-ray membership identifications, we estimate the 0.5-2 keV luminosity of NGC 6791 is $(1.04\pm0.11)\times10^{32}$ ergs/s. 
 Fitting the \Chandra\ spectrum extracted from the half-mass radius (using the double subtraction procedure as above), we measure $L_X$(0.5-2)=$1.12\pm0.25\times10^{32}$ erg/s, which is nicely consistent. 
\citet{Platais11} estimate NGC 6791's mass to be no lower than 5000 \Msun\ (\citealt{vandenBerg13} propose 6000$\pm$1000 \Msun).  Using a Plummer model with half-mass radius 5.1 pc \citep{Platais11}, 
gives a central mass density of 11 \Msun/pc$^3$.  
\citet{Bedin08} estimate NGC 6791's binary fraction (they quote it for the core, but say it is very similar throughout the rest of the cluster.)

For NGC 6819, \citet{Gosnell12} report an XMM-Newton X-ray study.
The certain cluster members among Gosnell et al's sources are X5, X6, and X9, using the proper motion membership information of \citet{Platais13}, while the membership of the sources X2 and X4 remains uncertain. Given this membership information, we estimate the 0.5-2 keV luminosity as $(4.8\pm2.4)\times10^{31}$ ergs/s. Note that in section 2.1 we argue that source X1 is not a cluster member. 
 Extracting spectra from the half-mass region of the XMM-Newton MOS data gives $L_X$(0.5-2)$=6.4\pm1.1\times10^{31}$ erg/s.
The binary frequency of NGC 6819 was found to be 22$\pm3$\% by \citet{Milliman14}.  From its mass of 2600 \Msun \citep{Kalirai01} and half-mass radius of 2.25 pc \citep[e.g.][]{Gosnell12}, we estimate a central density of 54 \Msun/pc$^3$.

Using \Chandra, \citet{Vats17} found a total of 151 X-ray sources in Collinder 261, to a limiting luminosity of $4\times10^{29}$ ergs/s (0.3-7 keV). Thirty-three of these sources are active binaries and ten to eleven are CVs. From \citet{Vats17}'s X-ray membership identification, the total X-ray luminosity of point source members of Cr 261 can be estimated as $(8.6\pm1.0)\times10^{31}$ ergs/s (0.5-2 keV).
 The \Chandra\ spectrum from the half-mass radius of Collinder 261, using the double subtraction method, gives a total $L_X$(0.5-2)$=1.1\pm0.4\times10^{32}$ erg/s, in agreement. 
For Cr 261, we use an estimated age of $6.5\pm0.5$ Gyr and a distance of $2.45\pm0.25$ kpc from \citet{Gozzoli96}. Using a King profile, \citet{Vats17} report a mass of 5800-7200 \Msun.   We estimate the central mass density at 15 \Msun/pc$^3$.

\subsection{Globular Clusters}

For the cluster parameters distance,  metallicity,  central density and reddening, we generally use the Harris catalog\footnote{http://physwww.physics.mcmaster.ca/~harris/mwgc.dat} of globular cluster parameters \citep[][ 2010 edition]{Harris96}, as the distances computed there have the advantage of uniformity, for comparisons between clusters. 
For mass estimates, 
 we use the calculations of \citet{BaumgardtHilker18}, which are in general agreement with the mass estimates of \citet{Watkins15}.\footnote{https://people.smp.uq.edu.au/HolgerBaumgardt/globular/parameter.html}

The M/$L_V$ ratios of clusters differ from each other, showing a standard deviation of $\sim$20\% among the M/$L_V$ ratios in \citet{Watkins15}.
Thus, we add an error of 20\% in quadrature to the errors on $L_X$/M on each cluster. 
Following the lead of the literature \citep[e.g.][]{vandenBerg13}, we include only X-ray sources within the half-mass radius of each globular cluster, and thus divide the inferred masses by 2.  \footnote{An argument can be made against doing so, as binaries tend to sink into the core during clusters' evolution; not dividing all cluster masses by 2 would only strengthen our key results.}   

We use two different sets of globular cluster age estimates, both using the HST-ACS survey to compare the relative positions of the horizontal branch and main-sequence turnoff; \citet{Marin-Franch09}, and \citet{VanDenBerg13A}.
  We assign GC binary fractions to be those measured by \citet{Milone12} in the annulus between the core and the half-mass radius (wherever possible), as the most representative of the cluster as a whole.

We list in Table~\ref{table*:globs}  globular clusters for which detailed (typically \Chandra) X-ray studies have been published identifying the X-ray source content, excluding bright LMXBs.  Since this group suffers from  observational selection effects (known sources were more likely to be observed), we also focus on a set composed of all globular clusters within 6 kpc, and analyze archival X-ray observations where necessary to produce constraints on all of these clusters.  We give details on each globular cluster's particular properties (including which X-ray sources we regard as secure members) in Appendix A. Here we describe our general procedures.

\citet{Trager93} separate globular clusters by photometric data quality. We exclude those with the poorest data quality, Terzan 1, Terzan 11/12, and NGC 6540/Djorg 3, along with recently discovered highly obscured clusters such as GLIMPSE-C01 \citep{Kobulnicky05}, as for these clusters we cannot reliably determine basic globular cluster parameters such as distance, total stellar mass, and central density. We do include Terzan 5, since its parameters have now been well-determined through infrared and radio pulsar timing studies \citep{Lanzoni10,Prager17}. 

 Estimating the X-ray emission from actual cluster members is complicated. The most difficult questions are the X-ray emission from non-members, and unresolved X-ray emission.
Due to the relatively flat luminosity functions of rich globular clusters \citep{Pooley02b}, most of the X-ray flux comes from the most luminous sources.  \Chandra\ and XMM exposure times for these clusters vary significantly, but the effect on the total flux normalization of unresolved sources is typically small for richer clusters.  Sources below the detection limits of M28 and M80 ($L_X=4\times10^{30}$  and $L_X=6\times10^{30}$ ergs/s)  were constrained to produce less than 15\% and 5\% of the 0.5-2 keV X-ray flux in these clusters, respectively \citep{Becker03,Heinke03c}. \citet{Eger10} and \citet{Wu14} identify  likely diffuse emission from the outskirts of Terzan 5 and 47 Tuc, but the flux is $<$10\% of the total from these clusters, and is not attributed to the kind of point sources we study, so we do not include it.
\citet{Hui09} do not find evidence for  diffuse X-ray emission in \Chandra\ observations of the globular clusters M5, M13, M3, M71, M53, and M4.  On the other hand, \citet{Bassa04} find roughly equal $L_X$ from M4's core below $L_X$(0.5-2.5)$=6\times10^{30}$ ergs/s, vs. sources from $6\times10^{30}$ up to $=6\times10^{31}$ erg/s, indicating that unresolved emission can produce a significant contribution in the analysis of X-ray faint clusters.  

 \begin{figure*}
\includegraphics[width=0.82\linewidth]{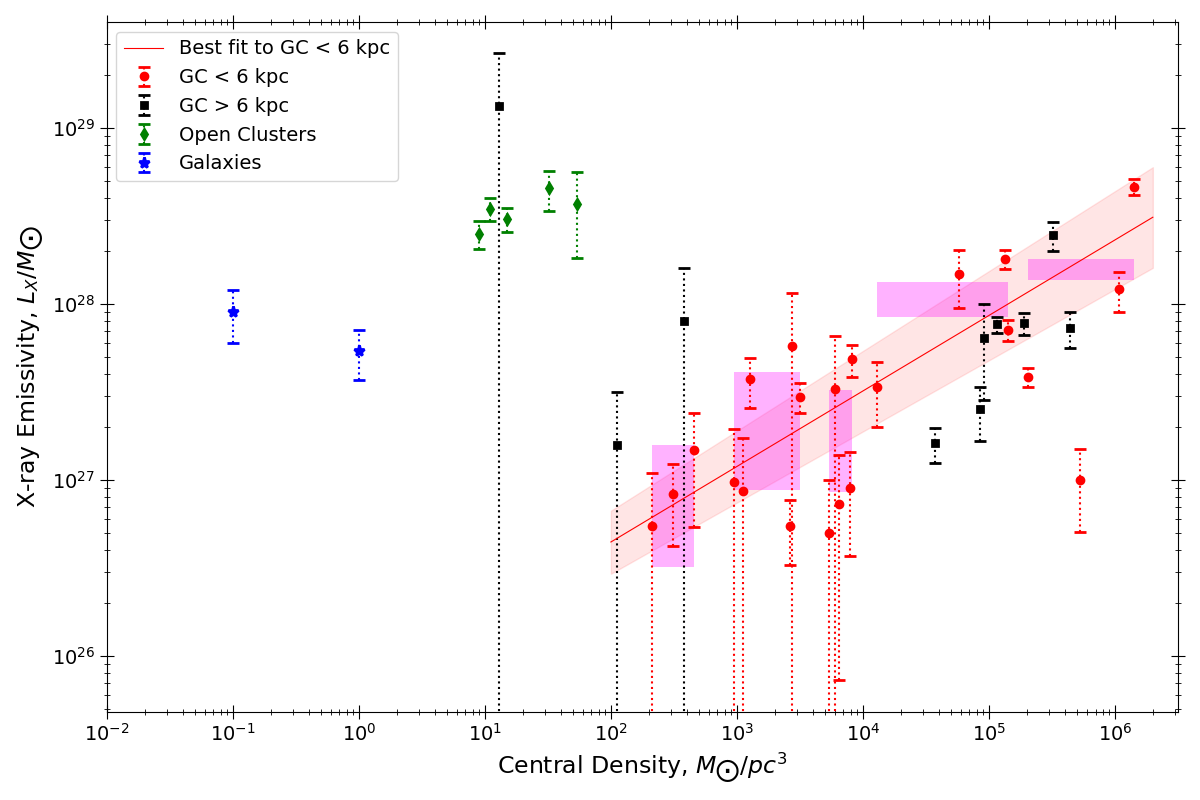}
\caption[]{ \label{fig:RoLx}
X-ray luminosity (0.5--2 keV) per unit mass for a variety of old systems at various densities, excluding bright low-mass X-ray binaries.  Filled circles represent individual populations (red for the GCs within 6 kpc, black for farther GCs), with dotted lines indicating errorbars when the membership of X-ray sources is unclear, or X-ray observations provide only an upper limit on the X-ray source content.  Magenta rectangles average the GC $L_X/M$ ratios for each of five density ranges for the complete sample of GCs within 6 kpc, including the uncertainties of each cluster.  See text for details.
} 
\end{figure*}

 Unresolved, diffuse emission can be difficult to clearly associate with a cluster, while point sources can be linked to particular stars and their membership in the cluster identified. We illustrate the issue with diffuse sources using the case of NGC 6366 (\S 3.2), for which the X-ray emission is dominated by a background galaxy cluster. 
We approach this problem by calculating X-ray emission from each cluster by both methods--identifying emission from individual sources (and measuring their membership), and using the X-ray emission enclosed within the entire half-mass radius.  In most cases, we use estimates for X-ray emission within the half-mass radius from \citet{Cheng18}, and convert them to 0.5-2 keV using either the power-law photon index best fit of \citet{Cheng18}, or a photon index of 2. For clusters not studied by \citet{Cheng18}, we perform this analysis ourselves. When the two estimates (from adding up point sources, or extracting the full flux within the half mass radius) disagree, we encompass both estimates within our error bars. When we have only upper limits, we impose a minimum log$L_X/M$ of 25.5 (enabling symmetric errors in the log, necessary for our Monte Carlo fitting). 

 We identify the maximum $L_X$ for each cluster from all sources within the half-mass radius (column 9 of Table~\ref{table*:globs}). We also compute estimates of the resolved extragalactic X-ray background flux for the half-mass radius of each globular cluster.  For typical \Chandra\ observations, the resolved soft-band source flux  varies from $1.4\times10^{-15}$ ergs cm$^{-2}$ s$^{-1}$ arcmin$^{-2}$ for a 20-ks exposure, which reaches \citep[e.g. ][]{Lu09} a limiting 0.5-2 keV flux of $10^{-15}$ ergs cm$^{-2}$ s$^{-1}$ \citep{Mushotzky00,Giacconi01}, to $1.8\times10^{-15}$ ergs cm$^{-2}$ s$^{-1}$ arcmin$^{-2}$ for a 1-2 Ms exposure resolving sources  to $5.5\times10^{-17}$ ergs cm$^{-2}$ s$^{-1}$ \citep[][Table 5]{Hickox06}.

 We compute an expected background flux using $1.7\times10^{-15}$ ergs cm$^{-2}$ s$^{-1}$ arcmin$^{-2}$ for each cluster, reported in column 10 of Table~\ref{table*:globs}. The actual background flux is stochastic, depending on what bright AGN happen to lie in this direction.
For globular clusters which lie on the Galactic plane, the foreground X-ray source flux can often be higher than that from background sources, and is often not easy to calculate. 
 If the X-ray sources within the half-mass radius are clearly highly concentrated (e.g. most of the flux is within the core), and the maximal $L_X$ is much larger ($>$3 times) than the expected background flux, then we subtract twice the expected background flux to estimate the minimum $L_X$, in column 11 (used for NGC 104, NGC 6397, NGC 6544, NGC 6626, NGC 6656, NGC 6752, Terzan 5, NGC 2808, NGC 6093, NGC 6341, NGC 6388, NGC 6440, NGC 6715, NGC 7099).
If the X-ray source distribution is not so clearly concentrated, and/or the expected background is of the same order as the observed $L_X$, then we make reference to identifications (typically optical) of secure cluster members in the literature to determine a minimal X-ray luminosity produced by secure cluster members (column 11 of Table~\ref{table*:globs}). We identify which members we take as secure in \S~2.3.1.

For some nearby clusters,  some or all archival \Chandra\ observations had not been published  (or had only been partly published) when we undertook our analysis; for these clusters ( NGC 3201, NGC 4372, NGC 6254/M10, NGC 6304, NGC 6352, NGC 6553, M28/NGC 6626, NGC 6544, Palomar 6, Palomar 10) we analyzed the archival data within the half-mass radius.  In these cases, we convert the measured cluster E(B-V) to $N_H$ (using  \citealt{Guver09}), and use a power-law of photon index 2 to convert the observed 0.5-2 keV counts to flux with the CXC's PIMMS utility\footnote{http://asc.harvard.edu/toolkit/pimms.jsp}, accounting for the cycle the data were taken.    
 No \Chandra\ or XMM-Newton data of IC 1276 has been taken yet; we calculate limits from analyzing an archival Swift observation.  
The point of analyzing all clusters within 6 kpc is to have a (nearly) complete, distance-limited sample, which we use to investigate the behaviour of $L_X/M$ with central density.
Using the observed X-ray fluxes, and minimal X-ray fluxes from known members, for each cluster, we calculate the minimum and maximum $L_X$ values for each cluster in this distance-limited sample. We group these clusters into five ranges in cluster central density (with 4-5 clusters in each group), and calculate the maximal possible ranges in $L_X/M$ for each group (taking the maximum $L_X$ for each cluster, and the minimum $L_X$, for the extrema), and plot these ranges as magenta rectangles in Fig.\ref{fig:RoLx}. 
A significant trend of increasing $L_X/M$ among globular clusters can be seen at high densities ($>10^4$ \Msun/pc$^3$).

\section{Three Intriguing Claims}\label{s:claims}

The key to our analysis is the careful collation of data on a variety of stellar populations.  
Recent work has added significantly to our understanding of X-ray emissivity in old open clusters \citep{Gosnell12,vandenBerg13,Vats17,Vats18} and dwarf ellipticals \citep{Ge15}.
To understand the nature of X-ray sources projected onto low-density populations, we feel it is crucial to undertake careful investigation of each source.  
In section 5 below, we discuss how a different curation of data have led \citet{Cheng18} to significantly different conclusions than ours.
In this section, we investigate three intriguing claims in the literature, about X-ray sources in 
(the direction of) 
the low-density populations of the open cluster NGC 6819, the sparse globular cluster NGC 6366, and the Sculptor dwarf galaxy.

\subsection{A Neutron Star in the Open Cluster NGC 6819?}\label{s:6819}

The identification of a candidate quiescent neutron star LMXB in the low-density, old, open cluster NGC 6819 by \citet{Gosnell12} was quite unexpected.  X-ray binaries may be formed either primordially or dynamically, but neither method seems plausible here. NGC 6819 is such a low-mass cluster (2600 \Msun, \citealt{Kalirai01}) that the chance of finding a primordial NS binary is extremely small.  We can roughly estimate this chance, from the following numbers; $\sim$200 quiescent LMXBs in  Galactic globular clusters  \citep{Pooley03,Heinke03d,Heinke05b}, vs. $\sim$20 known transient or persistent NS LMXBs in globular clusters \citep[e.g.][]{Bahramian14}, gives $\sim$10 times more quiescent than active NS LMXBs in globular clusters.  If this can be extrapolated to the Galaxy, then the $\sim$200 known transient or persistent LMXBs \citep[e.g.][]{Liu07} imply $\sim$2000 quiescent LMXBs in the Galaxy. \citet{Kiel06} indeed estimate $1900$ LMXBs in the Galaxy, with other population syntheses predicting $10^3$ to $10^5$ LMXBs \citep{Pfahl03,Jonker11}. Values of $10^3$--$10^4$ LMXBs were favoured by \citet{Britt13} from analysis of the Chandra Galactic Bulge Survey.  For a Galactic stellar mass of $\sim5\times10^{10}$ \Msun\ \citep{Cox00}, the full plausible range ($10^3$ to $10^5$ Galactic LMXBs) predicts a probability of $5\times10^{-5}$ to $5\times10^{-3}$ of finding a primordial quiescent NS LMXB in a cluster of 2600 \Msun, consistent (on the upper end) with Gosnell et al's  estimate, made via a very different method. We speculate that, if this open cluster were initially more massive, the probability of hosting a NS LMXB would increase accordingly.  However, the hosting probability should also be reduced by the retention fraction of NSs, which should be quite small for a low-mass cluster \citep[e.g.][]{Pfahl02b}.

The low core density of NGC 6819 suggests that strong dynamical effects are unlikely.  We use the corrected stellar number counts, down to $V$=21.5, of \citet{Kalirai01} to estimate that 20\% of its stars are projected within the core. We then estimate the core density of NGC 6819 using this estimate, Kalirai et al's estimated total cluster mass of 2600 \Msun, NGC 6819's core radius $r_c$ of 1.64 pc (adjusting Kalirai's 1.75 pc for the closer 2.34 kpc distance of \citealt{Basu11}, as used by \citealt{Gosnell12}), giving a rough stellar density $\rho$ of 28 \Msun/pc$^3$ (assuming an average stellar mass of 1 \Msun).  Using the simplified prescription for stellar interaction of $\Gamma$=$\rho^{1.5}/r_c^2$ \citep{Verbunt87}, we find that NGC 6819 has a stellar interaction rate only $2\times10^{-7}$ as large as that of 47 Tuc.  (Using a Plummer model as in \S~2, we get a slightly larger density of 54 \Msun/pc$^3$, and rate $5\times10^{-7}$ that of 47 Tuc.) As 47 Tuc has 5 known quiescent NS LMXBs \citep{Heinke05b}, this would suggest a probability of $\sim10^{-6}$ of a dynamically formed NS LMXB existing in NGC 6819. As for the primordial origin pathway, the history of the open cluster could affect this rate (e.g. if it were much denser at early times), but N-body simulations of open clusters \citep[e.g.][]{Hurley05} do not suggest that densities comparable to globular clusters are reached. Within this paradigm, it seems that new dynamical pathways to forming LMXBs would be needed \citep[e.g. interactions with triple stars,][]{Leigh13a}.

In contrast to Gosnell et al's conclusions, therefore, we find that any NS LMXB in NGC 6819 would be more likely to be primordial in origin, though the probability of finding one would be quite small.

This motivates us to check whether the identification of X1 as a quiescent NS LMXB is correct. We first review the arguments of \citet{Gosnell12}. 
\citet{Gosnell12} analyzed an XMM-Newton observation of the moderately old (2-2.4 Gyr, \citealt{Basu11}) open cluster NGC 6819.  Their analysis of the X-ray spectra and multiwavelength counterparts of the brightest X-ray source projected within the cluster half-mass radius, X1, indicated that it was a quiescent LMXB containing a neutron star.  The principal arguments in favor of this classification were: (1) the X-ray spectrum was inconsistent with single-temperature plasma spectra typical of coronally active binaries or cataclysmic variables, but consistent with a quiescent LMXB spectrum; (2) the X-ray position was compatible with a very blue ultraviolet source (in XMM-Newton Optical Monitor UVW1 and UVM2 filters), argued to be associated; (3) the ultraviolet position was incompatible with the position of a nearby bright ($V$=16.4) optical source, identified as a foreground star.

Next, we consider the X-ray spectral information. 
Figure 9 of \citet{Gosnell12} does not appear to show a good fit of the NS atmosphere model to the spectrum from XMM-Newton's pn camera. 
We downloaded the relevant XMM-Newton observation (ObsID 0553510201) and, using SAS 13.5.0, extracted spectra of X1 from the pn, MOS1, and MOS2 cameras, from 15$"$ radius regions, and nearby background spectra.
We constructed response matrices and effective area functions, and grouped each spectrum to 15 counts bin$^{-1}$.  We ignored data below 0.2 keV, and used the {\it tbabs} interstellar absorption model, with abundances from \citet{Wilms01}.

\begin{figure}
\includegraphics[angle=270,scale=0.32]{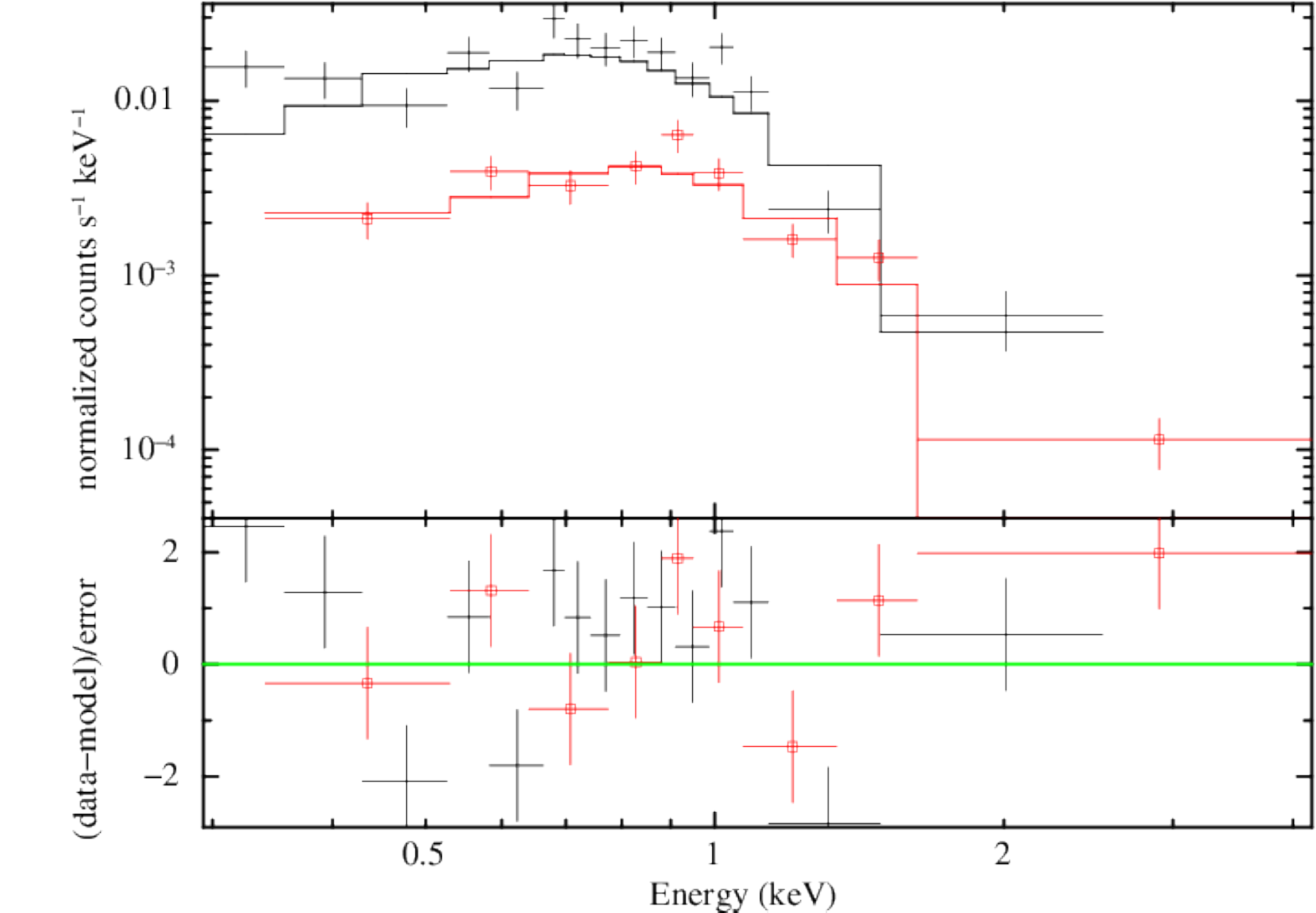}
\caption[]{ \label{fig:6819ns}
Spectral fit of NGC 6819 X1 to a NS atmosphere, with $N_H$ free. Combined MOS data and model in red (with squares), pn data and model in black (data rebinned for plotting purposes). Note the correlated residuals (lower panel), especially to the pn data, below 2 keV.
} 
\end{figure}

\begin{figure}
\includegraphics[angle=270,scale=.32]{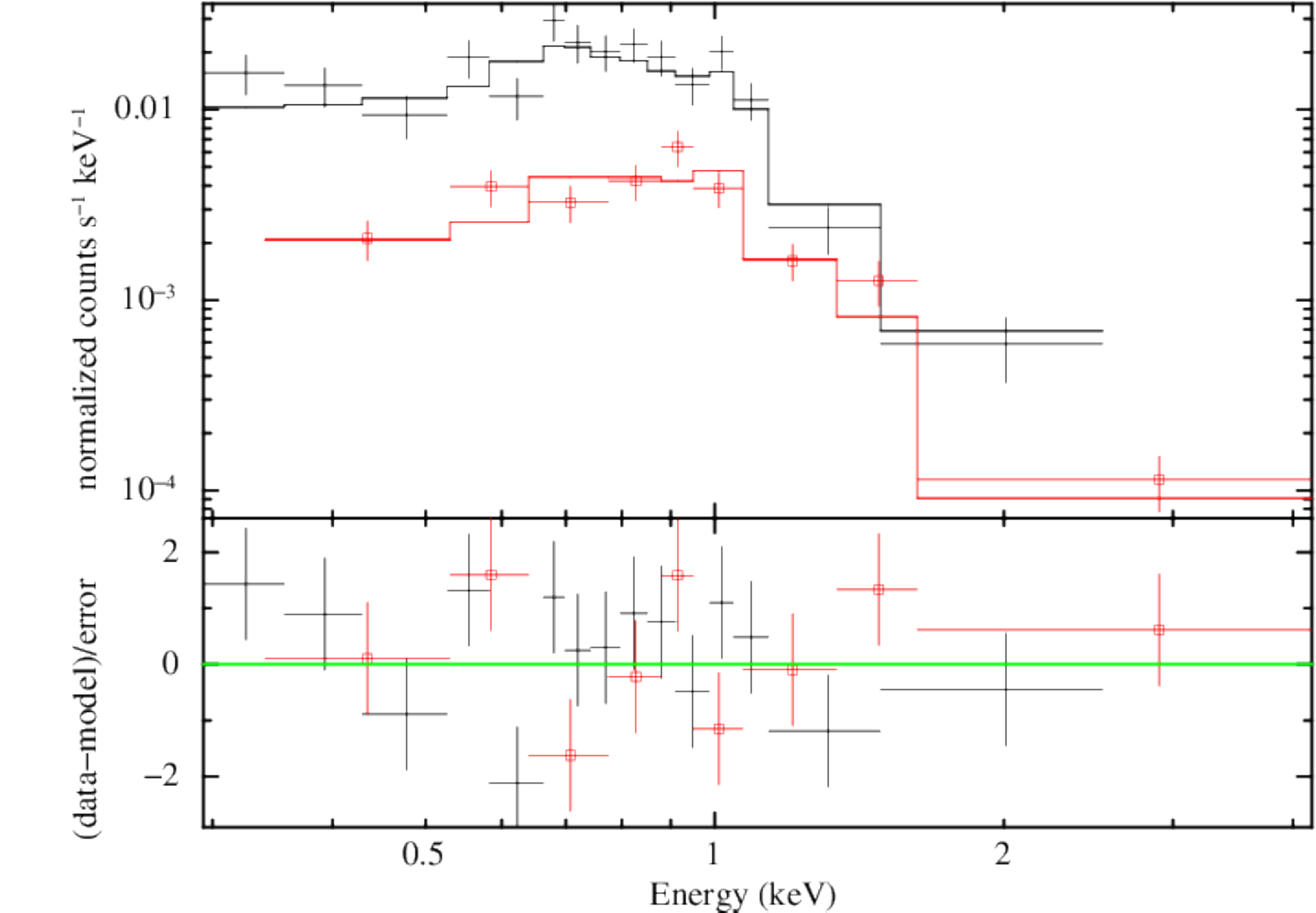}
\caption[]{ \label{fig:6819twoT}
Spectral fit of NGC 6819 X1 to a two-temperature thermal plasma ({\it mekal} models in XSPEC, $kT$=0.34$^{+0.07}_{-0.06}$ and 1.2$\pm0.2$ keV), with $N_H$ free (it tends to zero). Combined MOS data and model in red (with squares), pn data and model in black (data rebinned for plotting purposes). Note the  reduction in residuals compared to Fig.~\ref{fig:6819ns}.
} 
\end{figure}

We begin by trying to fit (using XSPEC\footnote{http://xspec.gsfc.nasa.gov}) an absorbed NS atmosphere ({\it nsatmos}, of \citealt{Heinke06a}), using a 10 km radius, 1.4 \Msun\ mass, 2.34 kpc distance, and normalization=1 (indicating emission from the full surface). 
We find that, in accord with the appearance of Gosnell et al's Figure 9, a NS atmosphere is an exceptionally bad fit to the XMM-Newton spectra. If $N_H$ is fixed to the cluster value of 8.3$\times10^{20}$ cm$^{-2}$, the reduced $\chi^2$ is 4.3; when allowed to be free, $N_H$ rises to 2.7$\times10^{21}$ cm$^{-2}$, but the reduced $\chi^2$ remains 2.05, with 42 degrees of freedom and a null hypothesis probability of $7\times10^{-5}$. In Figure \ref{fig:6819ns} we show this latter spectral fit; note the large residuals near 0.5 keV (as mentioned by \citealt{Gosnell12}) and generally below 2 keV. 

We tried alternative fits using one or two {\it mekal} \citep{Liedahl95} thermal plasma models, as appropriate for coronally active binaries.  As noted by \citet{Gosnell12}, a single {\it mekal} or APEC model gives a poor fit (reduced $\chi^2$ of 2.4).  However, coronally active binaries are typically fit with thermal plasma of at least two temperatures.  \citet{Dempsey97} fit ROSAT 0.2-2.5 keV spectra of 35 BY Dra dwarf binaries using two-temperature thermal plasmas.  The range of best-fit temperatures for the two components was 0.13 to 0.31 keV for the low-temperature component, and 1.1 to 2.8 keV for the high-temperature component.  We found a reasonable ($\chi^2$=55.95 for 39 degrees of freedom, null hypothesis probability of 3.8\%) fit to the XMM-Newton spectra with a two-temperature model (Figure \ref{fig:6819twoT}), using $N_H$=0$^{+2.9}_0 \times10^{20}$ cm$^{-2}$, $kT_1$=$0.34^{+.07}_{-.06}$ keV, $kT_2$=$1.2^{+0.2}_{-0.2}$ keV.  

The two-temperature thermal plasma fit is much better than the fit to a neutron star atmosphere model.  The temperatures of the two components are quite compatible with the ROSAT results for BY Dra systems. 
The emission measures for the two components of X1 in our fit are equal, which is also consistent with the ROSAT results on BY Dra systems. The preferred $N_H$ is significantly lower than the cluster value, indicating a foreground source for this interpretation.  Thus we conclude that the X-ray spectrum favours a BY Dra binary interpretation, rather than a quiescent NS LMXB.

Next, we consider the arguments relating to potential optical and UV counterparts.
\citet{Gosnell12} argued against an association between X1 and the nearest optical source, a star far redder than the cluster main sequence that is 1.56'' from the XMM-Newton position for X1 (which has a listed positional error of 1.07''). This star, known as UCAC4 652-075493 or WOCS 65005 \citep{Hole09}, is clearly not a member of the cluster \citep{Platais13}. 
This star (WOCS 65005) has optical and infrared colors, derived from 2MASS \citep{Skrutskie06} and USNO B magnitudes, suggestive of an M0-M1 dwarf.  
Gaia's Data Release 2 \citep{GaiaDR2} \footnote{https://www.cosmos.esa.int/web/gaia/dr2} identifies a source with Gaia magnitude $G$=15.2, at J2000 position $\alpha$=19:41:14.14, $\delta$=+40:14:06.58 (consistent with UCAC4 652-075493's J2000 position of $\alpha$=19:41:14.13, $\delta$=+40:14:06.74), at a distance of 155 pc. The observed $V$ magnitude of 16.4 \citet{Gosnell12} then gives an absolute $V$ magnitude of 10.45. Consulting the stellar tables of \citet{PecautMamajek13,Pecaut12} \footnote{http://www.pas.rochester.edu/$\sim$emamajek/\break EEM\_dwarf\_UBVIJHK\_colors\_Teff.txt}, we see that this absolute magnitude matches an M2V star, which predicts $B-V$=1.5, $J-K_S$=0.834; the measured values are $B-V$=1.5 \citep{Gosnell12} and $J-K_S$=0.841 \citep[from the PPMXL catalog, and thus 2MASS,][]{Roeser10,Skrutskie06}, in nice agreement with the M2V classification.

Instead, Gosnell et al.\ preferred to associate X1 with a relatively bright UV source detected in the XMM-Newton optical monitor UVW1 and UVM2 filters, located 1.45$"$ from the X-ray position of X1. They do not associate the UV and optical sources with each other, because the UV and optical source positions disagree by 0.34$"$,  a 2$\sigma$ separation from the quoted errors on each.  However, the optical imaging used was performed in 1999 \citep{Kalirai01}, while the XMM-Newton UV imaging was taken in 2008.  The proper motion for this star (UCAC4 652-075493) is known to be 38.3 mas/yr in declination \citep{Roeser10}, or 32.3 mas/yr in declination \citep{GaiaDR2}, which gives a shift of 0.29-0.34$"$ in declination over the 9-year interval. Thus, the optical and UV positions appear consistent with each other.  A full astrometric analysis of the UV, optical, and X-ray observations (including more recent Chandra imaging) is outside the scope of the current paper.

If X1 is located at 155 pc (the distance of the M2V dwarf), its inferred X-ray luminosity  is $4.4\times10^{29}$ erg/s, quite consistent with the range of BY Dra binaries in \citet{Dempsey97}. X1's spectrum is much cooler than the active binaries which are members of NGC 6819 identified by \citet{Gosnell12,Platais13}.  We attribute this difference in temperature to the tight correlation seen between coronal X-ray luminosity and coronal X-ray temperature \citep{Vaiana83,Gudel04}. The low $N_H$ from our 2-temperature model fit also agrees with the nearby Gaia distance.

To determine whether the UV emission observed by \citet{Gosnell12} is unusual, we consulted the analysis of GALEX and ROSAT data on active M stars by \citet{Stelzer13}.  The near-UV GALEX filter covers the wavelength range of 1771 to 2831 \AA, comparable to the XMM OM filters (UVM2: 2000-2600 \AA, UVW1: 2500-3500 \AA). The UVW1 flux density from this star is $4.9\times10^{-17}$ ergs cm$^{-2}$ s$^{-1}$ \AA$^{-1}$, translating to $5.1\times10^{-14}$ ergs cm$^{-2}$ s$^{-1}$ for the GALEX NUV band, or a GALEX NUV luminosity of $1.5\times10^{29}$ ergs s$^{-1}$.  This luminosity is normal for young active M stars, which typically have soft X-ray  luminosities a factor of a few higher than the near-UV, as does our object.  Finally, the unusually blue UVM2-UVW1 color of the X1 counterpart is not unusual for chromospheric, rather than photospheric, emission \citep{Stelzer13}.  

We conclude that all the properties of NGC 6819 X1 are perfectly consistent with a foreground M dwarf binary, and thus treat X1 as a nonmember.

\subsection{An Old Nova in the Globular Cluster NGC 6366?}\label{s:6366}

\citet{Bassa08} reported X-ray ({\it Chandra X-ray Observatory}, observed 2002 July 5) and optical ({\it Very Large Telescope; VLT} and {\it Hubble Space Telescope; HST}) observations of the low-density globular cluster NGC 6366. The brightest X-ray source located within the half-mass radius of NGC 6366, CX1, was unusual, containing both a point source (CX1a; $L_X$(0.5-6 keV)$=1.3\times10^{31}$ erg/s) and an extended source (CX1b, $L_X$(0.5-6 keV)$=1.4\times10^{32}$ erg/s), where the quoted $L_X$ values assume NGC 6366's distance of 3.5 kpc \citep[][2010 update]{Harris96}. Both have relatively soft spectra (power-law with photon index 3.5$\pm$0.6, and bremsstrahlung with kT=2 keV, respectively; \citealt{Bassa08}.) \citet{Bassa08} identified an optical counterpart for CX1a, a $V$=17.75 star slightly to the red of NGC 6366's giant branch in $B-R$ and $V-I$ color-magnitude diagrams, and with no detectable H$\alpha$ excess. No diffuse optical counterpart to CX1b was seen, in continuum or H$\alpha$ light. \citet{Bassa08} discussed whether CX1b could be a planetary nebula, supernova remnant, group or cluster of galaxies, or a nova remnant, and argued in favor of a nova remnant. However, this would be a very unusual nova remnant, as CX1b has a much harder X-ray spectrum than the (presumably younger) nova remnant around GK Per \citep{Balman05}, and the lack of H$\alpha$ emission would also be unusual. Their arguments against a background galaxy cluster were that CX1b's flux and diameter would imply a distance of more than a Gpc (which they thought unlikely), and that their {\it VLT}/FORS2 images showed no obvious galaxies. We investigate the nature of CX1b and CX1a in turn.

\subsubsection{Diffuse X-ray Source CX1b: A Background Galaxy Cluster}

\begin{figure}
\includegraphics[angle=0,scale=.45]{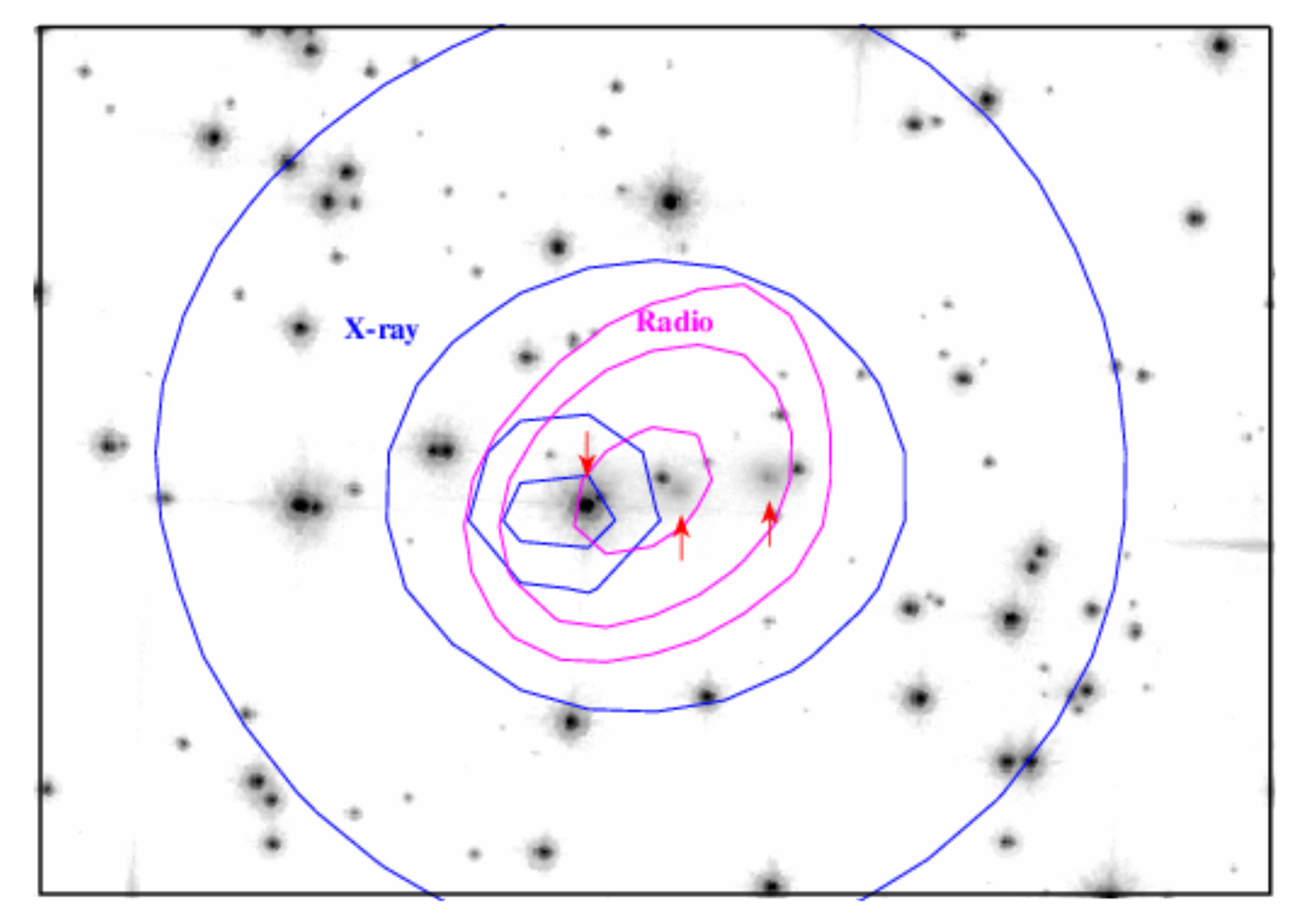}
\caption[]{ \label{fig:6366}
{\it Hubble Space Telescope} ACS F814W image of part of NGC 6366, with VLA radio (magenta) and Chandra X-ray (blue) contours overlaid. North is up, east to the left, and the black box is 26.0" by 18.8". The point source CX1a is identified by the smallest blue contours, and the bright star within it (indicated by a downward-pointing arrow) is its previously suggested optical counterpart (an alternative counterpart lies just to its NW). The diffuse X-rays of CX1b (two outer blue contours), and the radio emission, are consistent with being centred on the eastern (left) galaxy, of the two galaxies visible within the radio contours (indicated by upward-pointing arrows). This suggests a galaxy cluster as their origin.  
} 
\end{figure}

Inspection of the 10\arcsec by 10\arcsec {\it HST} finding charts in \citet{Bassa08} reveals the presence of two faint ($V>21$), extended galaxies, 2\arcsec and 4\arcsec to the west of CX1a (see Figure \ref{fig:6366}).

One of these galaxies (the western one) is detected in the ACS Cluster Survey catalog \citep{Anderson08a}, with magnitude $I_{814W}$=21.5, which for an assumed absolute magnitude of $\sim$-21 (typical of bright cluster galaxies), and extinction of $A_I\sim$1.04, implies a luminosity distance of 1.9 Gpc, or a redshift of z=0.35. The other (eastern) galaxy is consistent with Bassa's reported center of CX1b (see Figure \ref{fig:6366}).

The inferred X-ray luminosity of CX1b at this distance would be $L_X$(0.5-6 keV)$=4\times10^{43}$ erg/s, or $\sim2\times10^{43}$ erg/s in 2-10 keV.  This suggests X-ray emission from a (relatively poor) cluster of galaxies, which fits perfectly with its measured $\sim$2 keV X-ray spectrum (see, e.g., \citealt{David93}, their Fig. 5), similar to the cluster of galaxies that \citet{Yuasa09} identified in the background of the globular cluster 47 Tuc.

We analyzed an archival VLA observation of NGC 6366 obtained under observation ID AG627 (PI: B. Gaensler), observed on 2002 Sept. 4. The data were taken at 1.4 GHz in B configuration, with 2 IFs each of 12.5 MHz bandwidth, sampled by 16 channels. A total on-source time of 3.9 hours was acquired. We edited, calibrated, and imaged the data in AIPS using standard routines, and an imaging robust value of 0 yields an image rms of 50 microJy beam$^{-1}$ and a synthesized beam of $5.5^{\prime\prime} \times 4.7^{\prime\prime}$.

One source is clearly detected in this image, at 503$\pm$49 $\mu$Jy, centered at J2000 RA=17:27:42.74, Dec=-05:05:04.8.  There is no evidence for extension in this image, but the location coincides with that of the eastern galaxy, located at the center of CX1b. 

Figure~\ref{fig:6366} shows the radio (magenta) and X-ray (blue) contours, overlaid on an archival HST ACS F814W ($I$) 570-s image, showing that the diffuse X-ray and radio contours are centered on the eastern of the two galaxies. We used the {\tt csmooth} command in CIAO to adaptively smooth \citep{Ebeling06} the Chandra image.

Comparing the ratio of radio and X-ray fluxes (log $L_R/L_X=-13.3$, assuming the radio flux is attributed to CX1b) to the tabulations for different source classes in \citet{Maccarone12}, we see that clusters of galaxies typically have log $L_R/L_X$=-13 to -14, nicely matching CX1b, while CX1b would be inconsistent with typical values for coronally active stars (-15.5), or cataclysmic variables ($<$-15). Thus, the X-ray, radio, and optical data of CX1b are all consistent with a background cluster of galaxies, projected onto NGC 6366.

\subsubsection{X-ray Point Source CX1a: A Star in the Cluster?}

The nature of CX1a remains unclear; presumably it is either an AGN associated with the galaxy cluster, or a star in NGC 6366. 

\begin{figure}
\includegraphics[angle=0,scale=.4]{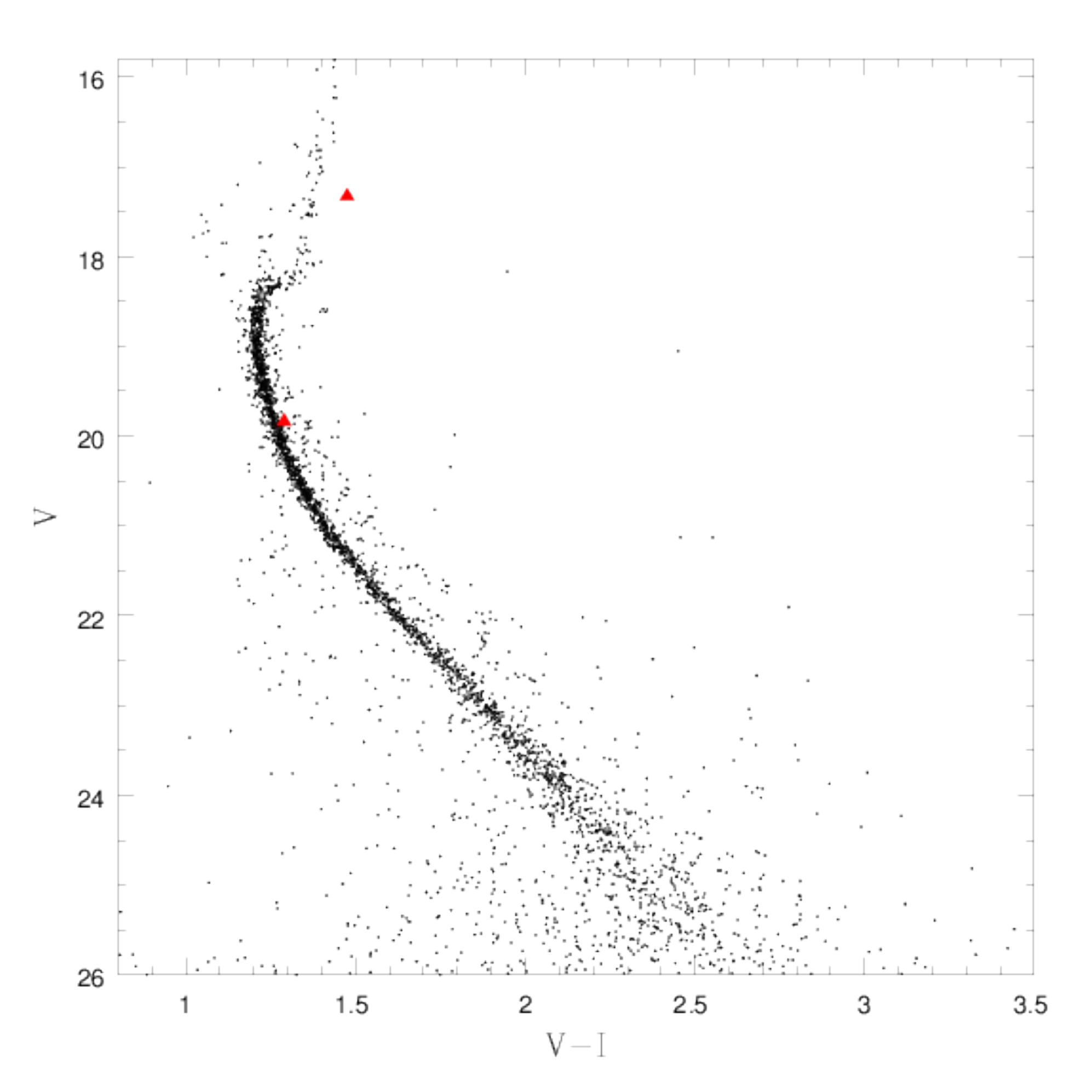}
\caption[]{ \label{fig:6366_cmd}
Color-magnitude diagram from {\it Hubble Space Telescope} ACS F606W and F814W photometry of NGC 6366, from \citet{Anderson08a}. Two red triangles indicate the potential counterparts for CX1a. 
} 
\end{figure}

Our inspection of the HST F606W and F814W imaging of NGC 6366 \citep{Anderson08a} reveals two stars in the error circle; the bright ($V$=17.33) star identified by Bassa et al., and a fainter star 0.3\arcsec to the NW, at $V$=19.8 (see Fig.\ref{fig:6366}). Fig. \ref{fig:6366_cmd} uses the Anderson et al. photometry to show a colour-magnitude diagram illustrating the location of the two possible counterparts for CX1a. The fainter star lies on or very near the main sequence, while the brighter star lies to the red of the giant branch, a region with very few stars. The chance of such an unusual star lying within the X-ray error circle is small, while the chance of a normal main-sequence star in the error circle is larger, so we agree with Bassa et al. that the brighter star is the more likely counterpart. However, it is not clear from the CMD position of the bright candidate counterpart that this star is actually a member of the cluster.

To determine its nature, we obtained a longslit optical spectrum of the brighter star in the X-ray error circle (the fainter star will contribute only $\sim$10\% to the light), using the {\it Gemini} Multi-Object Spectrograph (GMOS) on {\it Gemini} South, on 2012 Sept 5 and 10. We took six 900-second exposures with the B600 grating, using a slit width of 2\arcsec (since we used the poor weather queue) and a central wavelength of 520 nm for the first three, 525 nm for the rest (to ensure coverage across detector gaps). The first spectroscopic exposure was taken in bad weather, and the telescope tracking failed; the final three frames have noticeably less noise than the first three. 

We reduced the data using the {\sc IRAF} Gemini package. Two flat field frames were combined with {\tt GSFLAT} for each central wavelength. The arc frames were combined for each central wavelength, then bias subtracted, overscan trimmed, and wavelength calibrated using {\tt GSWAVELENGTH}. The wavelength calibration was refined to yield rms error values of $\sim$2 pixels. The science images were bias-subtracted, flat-fielded, overscan trimmed, and wavelength calibrated with the arc frames. A cosmic ray rejection algorithm was applied to each frame during the reduction process. A median frame of the 3 science frames for each central wavelength was created with {\tt GEMCOMBINE}, taking positional offsets into account, to get the best trace for source extraction. Sky-subtraction was performed on the combined science frames. We extracted the spectrum of  the brighter of the two candidate counterparts to CX1a  interactively using {\tt GSEXTRACT} in both images, cross-referencing the slit position with the slit image stamp of NGC 6366 taken at the time of the observations. Additionally, we extracted spectra of the five brightest stars in the slit as reference spectra.

Observations of the standard star (the white dwarf EG21) were reduced in the same manner, except that cosmic ray rejection on the single frame was performed with the L.A. Cosmic algorithm \citep{vanDokkum01}. The sensitivity function was created using {\tt GSSTANDARD}. The spectra for each central wavelength were flux calibrated using the sensitivity function, and combined using {\tt SCOMBINE}. The spectra centered at 525 nm proved to have a higher flux than those at 520 nm, and on inspection showed sharper features. Therefore, we focus on the spectra using the 525 nm central wavelength.  

Finally, telluric features were removed by hand using the deblending feature in {\tt SPLOT}, and the spectra were corrected for velocity dispersion using {\tt DISPCOR}.

\begin{figure}
\centering
\includegraphics[angle=0,width=0.48\textwidth]{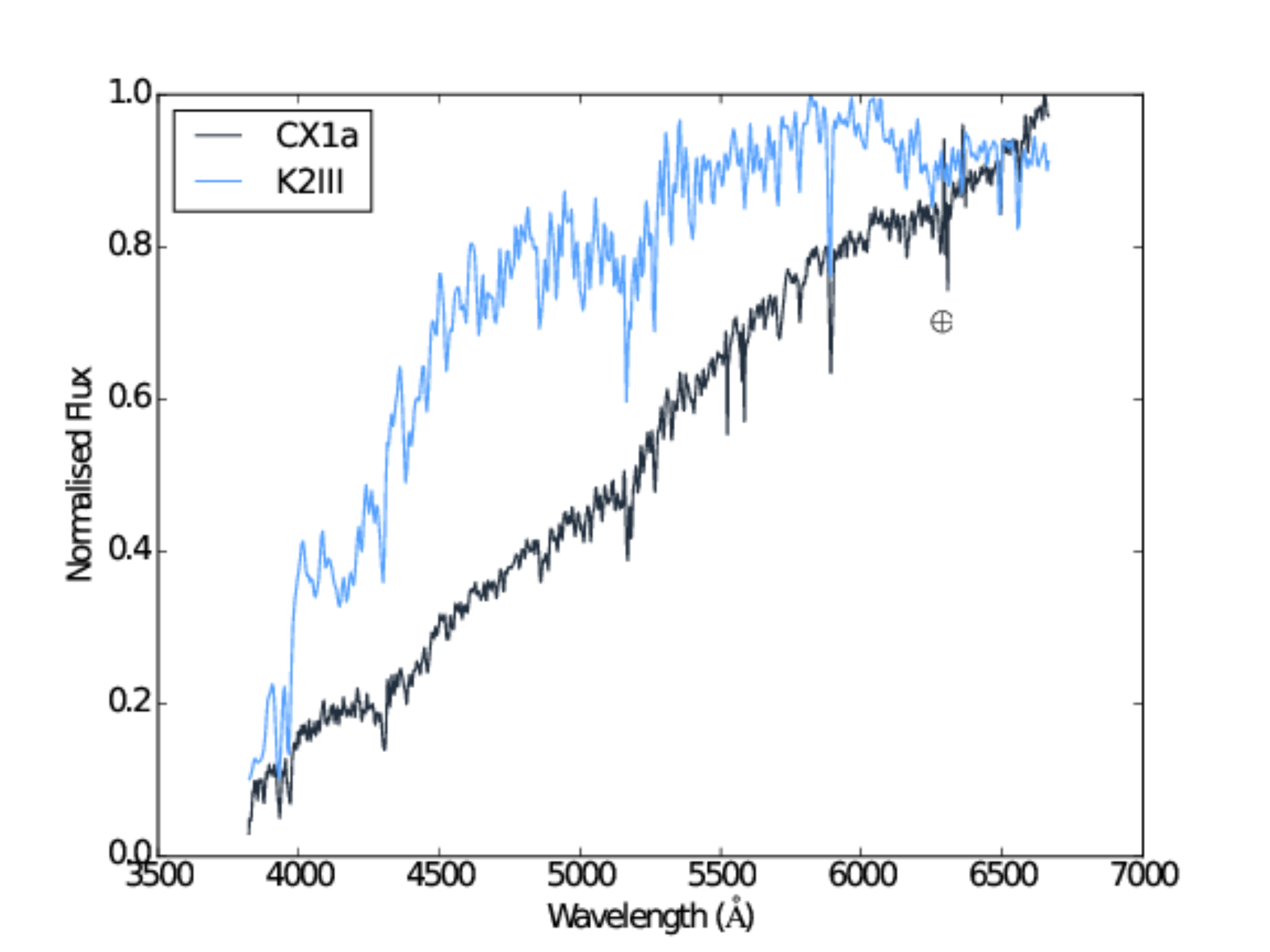}
\caption[]{ \label{fig:6366a_K2}
Comparison of our Gemini/GMOS spectrum of CX1a (black) with the normalized library spectrum of a K2 III star from \citet{Silva92} (blue). The location of telluric lines are highlighted by the $\oplus$ symbol.
} 
\end{figure}

\begin{figure}
\centering
\includegraphics[angle=0,width=0.48\textwidth]{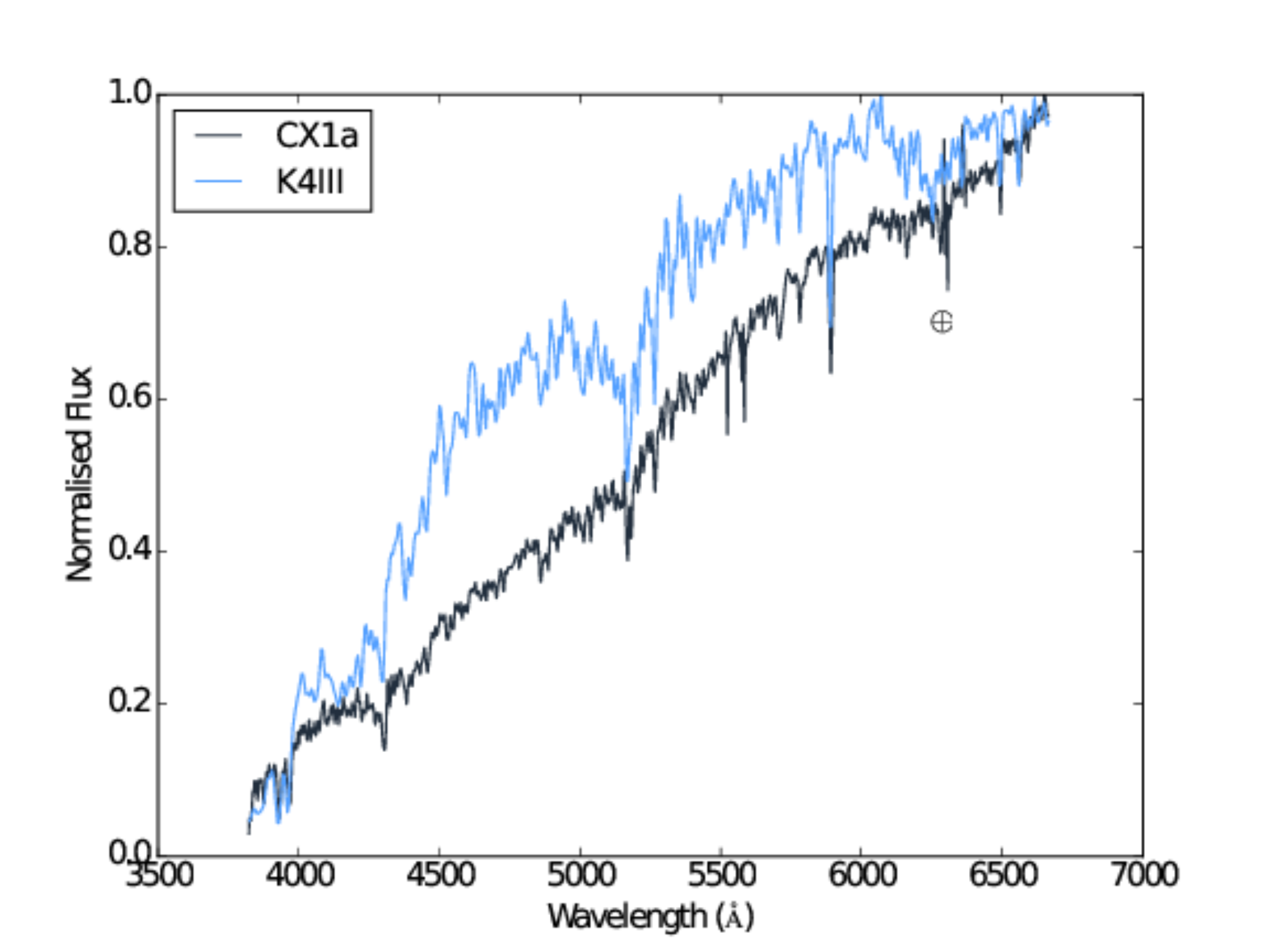}
\caption[]{ \label{fig:6366a_K4}
 Comparison of our Gemini/GMOS spectrum of CX1a (black) with the normalized library spectrum of a K4 III star from \citet{Silva92} (blue). The location of telluric lines are highlighted by the $\oplus$ symbol.
} 
\end{figure}

We compared the final spectrum of the CX1a optical counterpart candidate to digital stellar spectra from \citet{Silva92}. We do see various spectral lines, which resemble those from a K2-K4 star (Figs. \ref{fig:6366a_K2}, \ref{fig:6366a_K4}).  In particular, the relative depths of the Ca II K \& H, G band, Mg I, and Na I lines are comparable to mid-K stars. 

However, the spectral features are much weaker than in the library spectra. This is especially true of the Balmer drop at 4000\AA, and the H$\alpha$ and H$\beta$ absorption lines. This suggests that the (K-class) stellar spectrum may be combined with another, brighter, featureless spectrum, though this hypothesized spectrum would have to be quite red to avoid making the full spectrum bluer.

\begin{figure}
    \centering
    \includegraphics[width=0.45\textwidth]{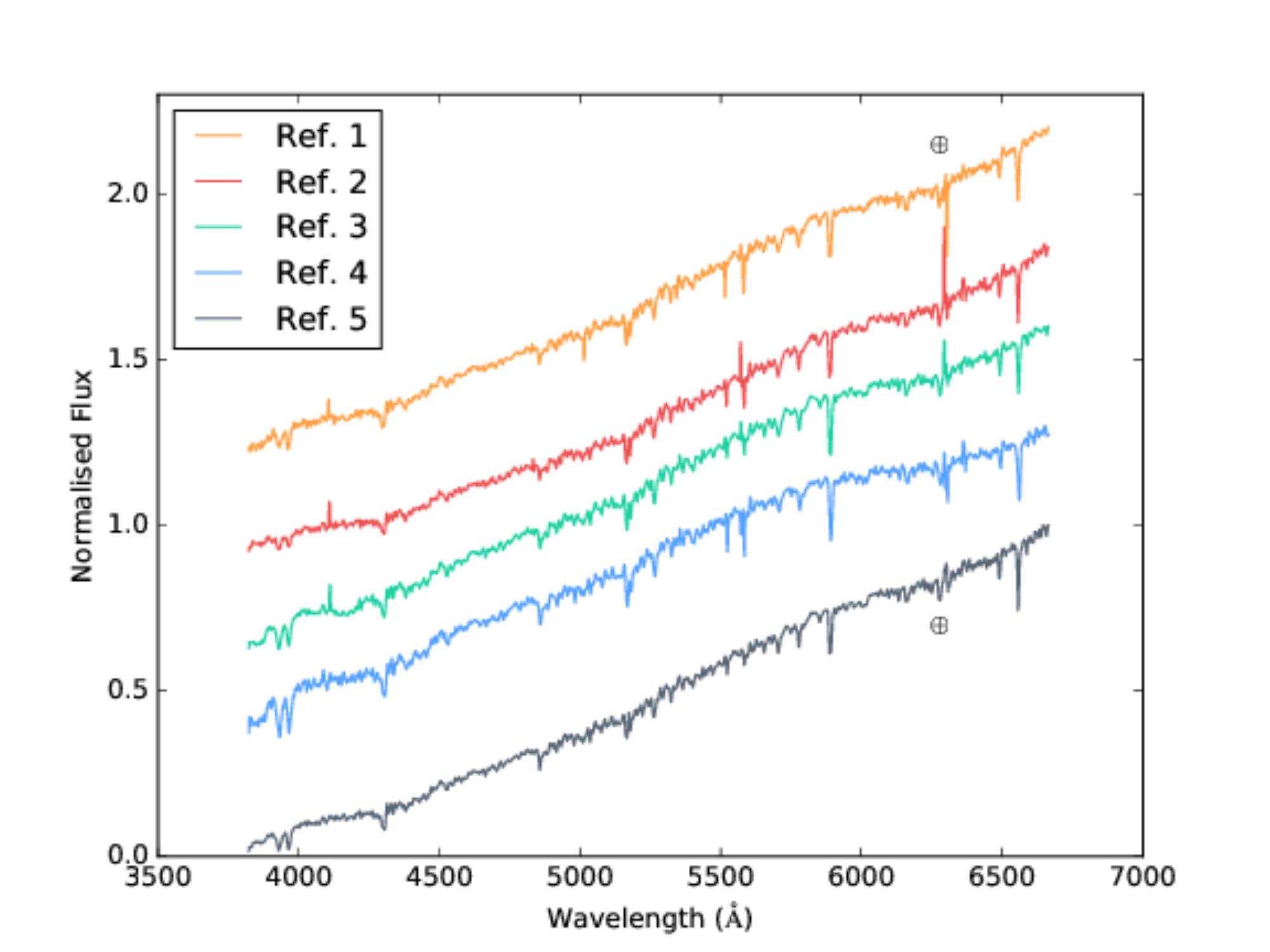}
    \caption{Gemini/GMOS spectra of the five brightest stars in the 2\arcsec slit used to observe CX1a (order as in the legend). The location of telluric lines are highlighted by the $\oplus$ symbol. Apparent emission lines in the spectra are a result of imperfect modeling and subtraction of sky emission lines by the GMOS reduction tools.}
    \label{fig:ref_stars}
\end{figure}

A plausible alternative is that the lack of metals in NGC 6366 reduces the strength of the metal lines, compared to the library spectra. Fig. \ref{fig:ref_stars} shows the spectra of the five brightest stars also in the slit, all of which exhibit weaker metal features than the library spectra, similar to CX1a.

Another method to test the nature of NGC 6366 CX1a is to compare the relative radial velocity of the optical counterpart with that of other bright objects within the slit. Given NGC 6366's radial velocity with respect to the Earth of -122 km/s \citep{Rutledge97}, we should be able to discriminate cluster members from foreground stars, and thus determine if the candidate counterpart to CX1 is a foreground interloper star. Given the density of stars in the cluster, it is likely that most or all bright objects in the slit will be cluster members.

Spectra of the 5 brightest objects within the slit were extracted to serve as reference stars.
The spectra were cross-correlated with templates using the IRAF task {\tt FXCOR}. This returned the relative radial velocities uncorrected for dispersion due to position off the centre of the slit. 

The 2" slit width was larger than the seeing of our 525 nm data, introducing  shifts in the radial velocities of the reference stars. 
We attempted to correct for the off-centre positions of the reference objects using the slit image to centroid the location of each star.  However, this correction was not fully satisfactory, as relative shifts among the reference stars of $\sim$200 km/s remained.

We therefore attempted an alternate wavelength calibration, using telluric and/or interstellar lines in the raw spectra to directly wavelength-calibrate the spectra (using "raw" spectra before telluric-line removal). We were able to identify an O$_2$ telluric feature near 6282 \AA\ in each spectrum, near a bright sky line whose subtraction leaves prominent residuals. This feature also coincides with a diffuse interstellar band, though for our purposes we do not need to discriminate between the two. We fit the absorption feature (carefully avoiding the residuals from the sky line subtraction) with IRAF's {\tt SPLOT} command using a Voigt profile, and estimated uncertainties by selecting slightly larger or smaller regions around the line to include in the fit. We then subtracted the radial velocities of the telluric features from the measured radial velocities of the stellar spectra themselves, giving the values in Table 1. We have not attempted to place these spectra on an absolute basis.

\begin{table}
\centering
\begin{tabular}{l|c}
Star & Relative RV (km s$^{-1}$) \\
\hline
 CX1a    &  0$\pm$19 \\
 Ref. 1  &  -16$\pm$19 \\
 Ref. 2  &  3$\pm$19 \\
 Ref. 3  &  -17$\pm$24 \\
 Ref. 4  &  -55$\pm$39 \\
 Ref. 5  &   5$\pm$24 \\
\end{tabular}
\caption{Relative radial velocities of reference stars, vs. the candidate optical counterpart of CX1a in NGC 6366.}
\end{table}

As is clear from the table, the radial velocity of CX1a's candidate optical counterpart is consistent with the radial velocity of the five reference stars, within the errors of the calibration. Since the typical errors ($\sim20$ km s$^{-1}$) are much smaller than the radial velocity of the cluster (-122 km s$^{-1}$), we conclude that this star is certainly a member of NGC 6366. 

 We also obtained (lower quality) spectra on 2012 June 01 using the 1.6-m telescope at Observatoire Astronomique du Mont-M\'{e}gantic (OMM). Three 1800-second exposures were taken through 1.55 to 1.65 air masses during the middle of the night. The long-slit spectrometer consisted of an STA0520 (blue) CCD and a 1200-line/mm grating (no filters), yielding a resolving power at 5000\AA\ of about 5500. Cosmic ray rejection was handled manually and the wavelength calibration was carried out with a CuAr lamp.  The fit was better than 0.1\AA\ (rms) and the calibration was also checked against the strong airglow line [OI] at 5577\AA.   The two best spectra were background subtracted, corrected for the quantum efficiency of the detector and the reflectance of the grating, corrected for Rayleigh scattering in the atmosphere, and co-added.  Despite a lower S/N ratio than that for the GMOS spectrum shown in Figure \ref{fig:6366a_K2}, the spectrum is qualitatively  similar to a K5-type star.  In particular, there are many shallow absorption lines associated with the Fraunhofer group including lines of Mg I, Fe I, and Ca I.  Using a Voigt profile for absorption and by correlating several of the cleanest (i.e., minimal line overlap) Fe I, Mg I, and H$\beta$ lines in the range of 4300 to 5300\AA, we were able to estimate the radial velocity of CX1a as $\simeq -90 \pm 15$ km/s.  This result helps strengthen the argument that CX1a is a member of NGC 6366.  The possible deviation from the cluster average of -122 km/s, if real, is well within the expected range for binary orbital motion. 

 Finally, we were able to identify this star in Gaia Data Release 2 \citep{GaiaDR2}, where its proper motion is consistent with that of NGC 6366 (\citealt{GaiaGC};the Gaia parallax measurement is not significant yet.)  Thus, we confirm that all three components of the candidate optical counterpart's velocity are consistent with the motion of NGC 6366.

This exercise does not prove that this star is the true optical counterpart of CX1a, as it remains possible that this star is simply an accidental interloper in the error circle, with the true counterpart being fainter. However, this star is in an unusual position on its CMD, redward of the giant branch (see Fig. \ref{fig:6366_cmd}). Stars in this position ("red stragglers") are frequently associated with coronally active X-ray sources in open and globular clusters \citep{Belloni98,Kaluzny03c,Cool13,Geller17}. Therefore, we think it very likely that CX1a is associated with this star, a cluster member. The nature of its X-ray emission is not certain, but its X-ray luminosity ($L_X=1.3\times10^{31}$ erg s$^{-1}$) and X-ray/optical flux ratio \citep{Bassa08} are similar to those of other coronally active stars, including some in similar CMD positions \citep{Geller17}.

\subsection{X-ray Binaries in the Sculptor Dwarf Galaxy?}\label{s:Sculptor}

The Sculptor dwarf galaxy  is old \citep[$>$10 Gyrs,][]{Monkiewicz99,Mapelli09} and metal-poor \citep{Mateo98}, with a total mass of possibly  $1.4\times10^7$ \Msun\ \citep{Queloz95}, of which roughly $2\times10^6$ \Msun\ is stars. A satellite of the Milky Way, it is 86$\pm5$ kpc away \citep{Pietrzynski08}. 

\citet{Maccarone05} used 126 ks of Chandra time (in 21 exposures) to identify X-ray sources towards the Sculptor dwarf galaxy. They identified five moderately faint ($L_X\sim10^{33}-10^{35}$ ergs s$^{-1}$) X-ray sources, and three slightly fainter sources, with optical sources, which they argue are on Sculptor's giant and horizontal branches.  
No bright LMXBs are known in this dwarf galaxy.  

Maccarone et al. argued that these systems were quiescent LMXBs, although they are somewhat brighter than typical quiescent LMXBs in our Galaxy. They then inferred that the number of quiescent X-ray binaries per unit mass in Sculptor is $>$1/10 that of the dense globular cluster NGC 6440 \citep{Pooley02b,Heinke03d}. Globular clusters in general contain $\sim$100 times more bright LMXBs per unit mass than the Galaxy as a whole \citep{Clark75}, and NGC 6440 is an unusually dense and LMXB-rich globular cluster \citep{Pooley02b}, so this might suggest that Sculptor has $\gg 10$ times more LMXBs per unit mass than the Milky Way.  
Maccarone et al. suggest an alternative explanation, that LMXBs in the Milky Way (and Sculptor) have lower duty cycles (there are more quiescent LMXBs per observed bright LMXB) than LMXBs in globular clusters (the only population where we can measure both, and thus have a rough estimate).
In this case, quiescent LMXBs in the Milky Way could be as common (per unit mass) as in Sculptor. However, this explanation conflicts with current observational constraints on the numbers of quiescent LMXBs in the Galaxy \citep[e.g.][]{Britt14}.

In the 0.5-2 keV band, the lower limit on Sculptor's $L_X$ per unit mass would be $3.5\times10^{28}$ ergs s$^{-1}$ \Msun$^{-1}$, if the five XRBs confidently identified by \citet{Maccarone05} are accepted. This would give Sculptor the highest X-ray luminosity per unit mass of the stellar populations we have studied. This suggests investigation of whether the optical counterparts identified by Maccarone et al. are robust members of Sculptor. 
 
Maccarone et al. make the following argument: among the nine brightest X-ray sources in their Chandra observations of Sculptor, five are within 0.4" of optical sources with $V<$20.5 in the catalog of \citet{Schweitzer95}. The expected number of chance matches of optical sources in this catalog with the error circles of 9 bright X-ray sources is quite low, 0.04. Maccarone et al. argue that these five stars (listed in their Table 1) are located on the giant and horizontal branches in Schweitzer et al.'s photometry, and that they have proper motion estimates that indicate probability of at least 96\% of being members of Sculptor. They also identify a further four matches among fainter X-ray sources (their Table 2), one characterized as a background galaxy.
 
 \begin{figure}
\includegraphics[angle=0,scale=.43]{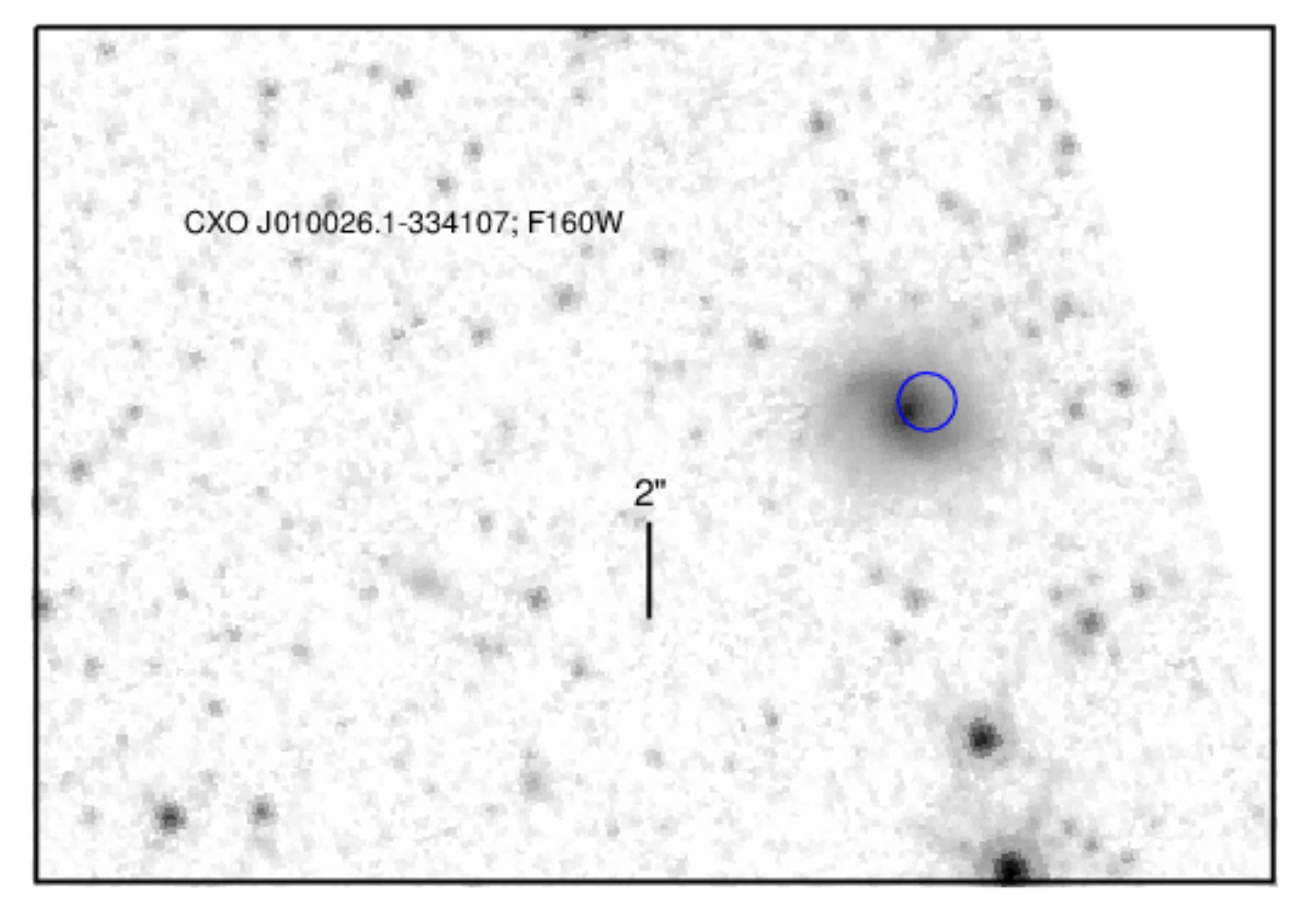}
\caption[]{ \label{fig:Sculptor_gal2}
HST/WFC3 F160W image of a portion of the Sculptor Galaxy, containing Maccarone et al.'s candidate \#2 (position indicated by a blue circle of radius 0.6\arcsec). A spiral galaxy is clearly present at the location of source \#2.
} 
\end{figure}

\begin{figure}
\includegraphics[angle=0,scale=.43]{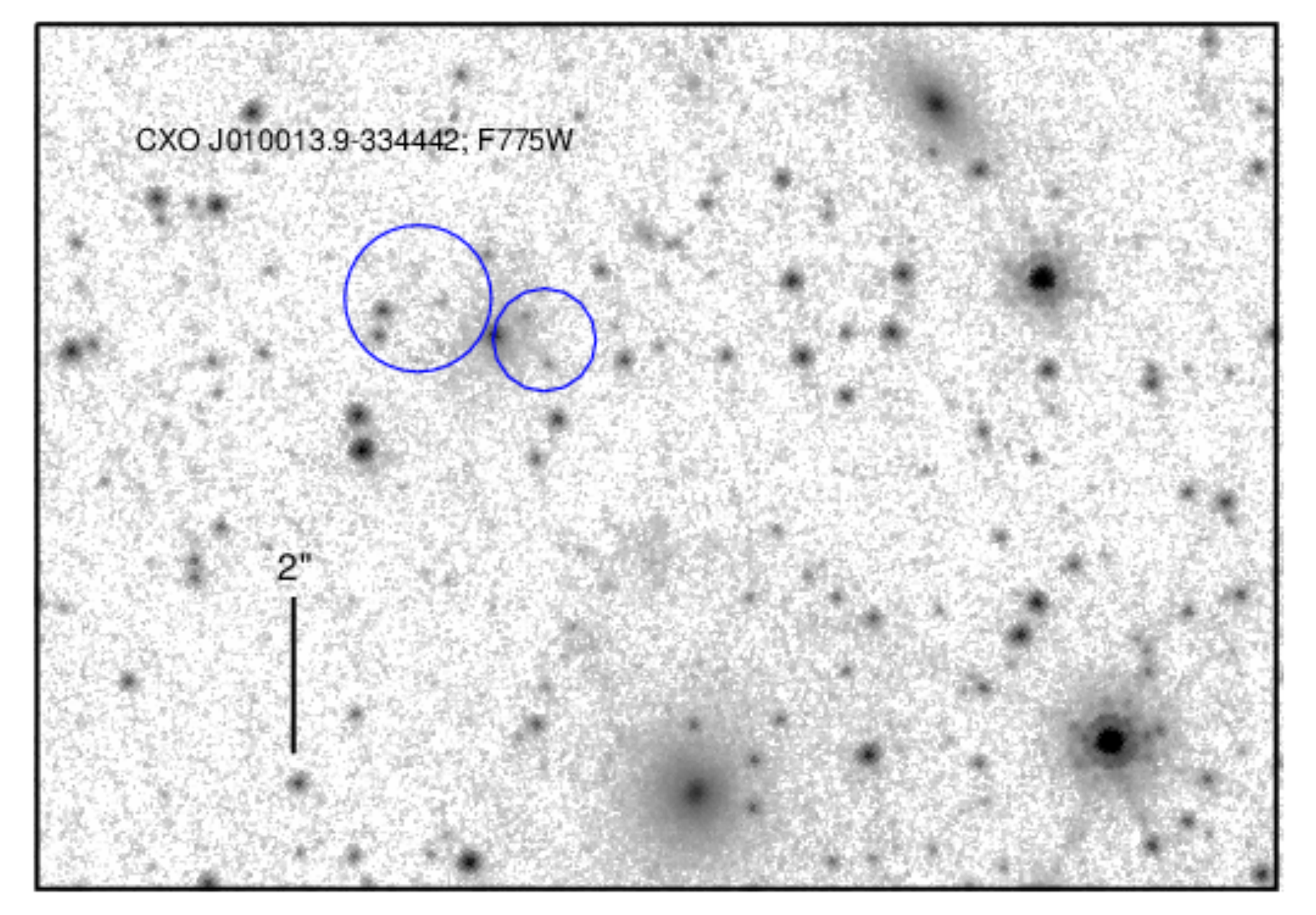}
\caption[]{ \label{fig:Sculptor_gal5}
HST/WFC3 F775W image of a portion of the Sculptor Galaxy, containing Maccarone et al.'s candidate \#5. The position of the X-ray source is indicated by two blue circles, representing two entries in the Chandra Source Catalog v.1.1. A point source surrounded by diffuse emission is clearly present at the intersection of the circles, indicating that source \#5 is almost certainly a galaxy.
} 
\end{figure}
 
However, we note that Schweitzer et al. do not have sufficient accuracy in their proper motion to measure the Sculptor dwarf's proper motion; they only distinguish between foreground stars with large proper motions, vs. objects with small proper motions--such objects could be either Sculptor members, or background objects. Consultation of Schweitzer's CMD reveals that it barely reaches 20.5, the level of the horizontal branch, and that errors are quite large there.  Thus, any object near the detection limit with colours bluer than the giant branch could be a potential member of the horizontal branch, or a background galaxy. 
 
We first searched for evidence from HST archival images, and then proceeded to investigate archival optical photometry, to assess the plausibility of these objects as cluster members. 

We located archival HST WFC3 images containing two of the suggested optical counterparts, the 2nd and 5th in Table 1 of Maccarone et al. (2005), and downloaded the drizzled, cosmic-ray-cleaned images from the STScI archive\footnote{http://archive.stsci.edu/hst/search.php}. We used the photometry of \citet{deBoer11},  aligned with 2MASS, to correct the astrometry of the HST images. For the location of the relevant X-ray sources in Sculptor, which are not provided directly by \citet{Maccarone05}, we use the Chandra Source Catalog, v. 1.1\footnote{http://cxc.cfa.harvard.edu/csc/}, to obtain positions and uncertainties. 

The location of Maccarone et al.'s X-ray source \#2 coincides with a clearly extended object in a WFC3-IR F160W image, identifying X-ray source \#2 with a galaxy showing clear spiral structure (Fig. \ref{fig:Sculptor_gal2}). \citet{Arnason19} confirms this conclusion from Gemini optical spectroscopy.

Their  source \#5 appears twice in the Chandra Source Catalog (from different observations), with slightly different positions but similar fluxes ($F_X$(0.5-7 keV)$=5\times10^{-14}$ erg s$^{-1}$ cm$^{-2}$), roughly consistent with Maccarone et al's reported flux. We plot both positions in Fig. \ref{fig:Sculptor_gal5}, on a WFC3-UVIS F775W image. The two reported positional uncertainties nearly overlap, suggesting they refer to the same source. Indeed, an optical source (with $I$=22 in the de Boer catalog) is visible at the intersection of these circles, suggesting it is the likely optical counterpart to the X-ray source. Although the optical source initially appears pointlike, close inspection reveals diffuse emission surrounding it which is not present around stars of similar brightness, indicating an AGN nature. (In addition, two other galaxies are clearly visible within 7".) Thus we conclude that sources \#2 and \#5 are background galaxies. \citet{Arnason19} identifies an ATCA radio source and a Spitzer infrared source with colours typical of AGN, which we find to be $<$1" from our suggested optical counterpart. The combined evidence strongly indicates an AGN nature for this X-ray source.


The other suggested counterparts in Maccarone et al. do not have available HST imaging. We therefore utilized deeper $B$, $V$, and $I$ photometry of the Sculptor dwarf with the MOSAIC-II camera on the CTIO 4-m, presented in \citet{deBoer11}, which has a limiting magnitude of $V$=24.8.

\begin{figure}
\includegraphics[angle=0,scale=.48]{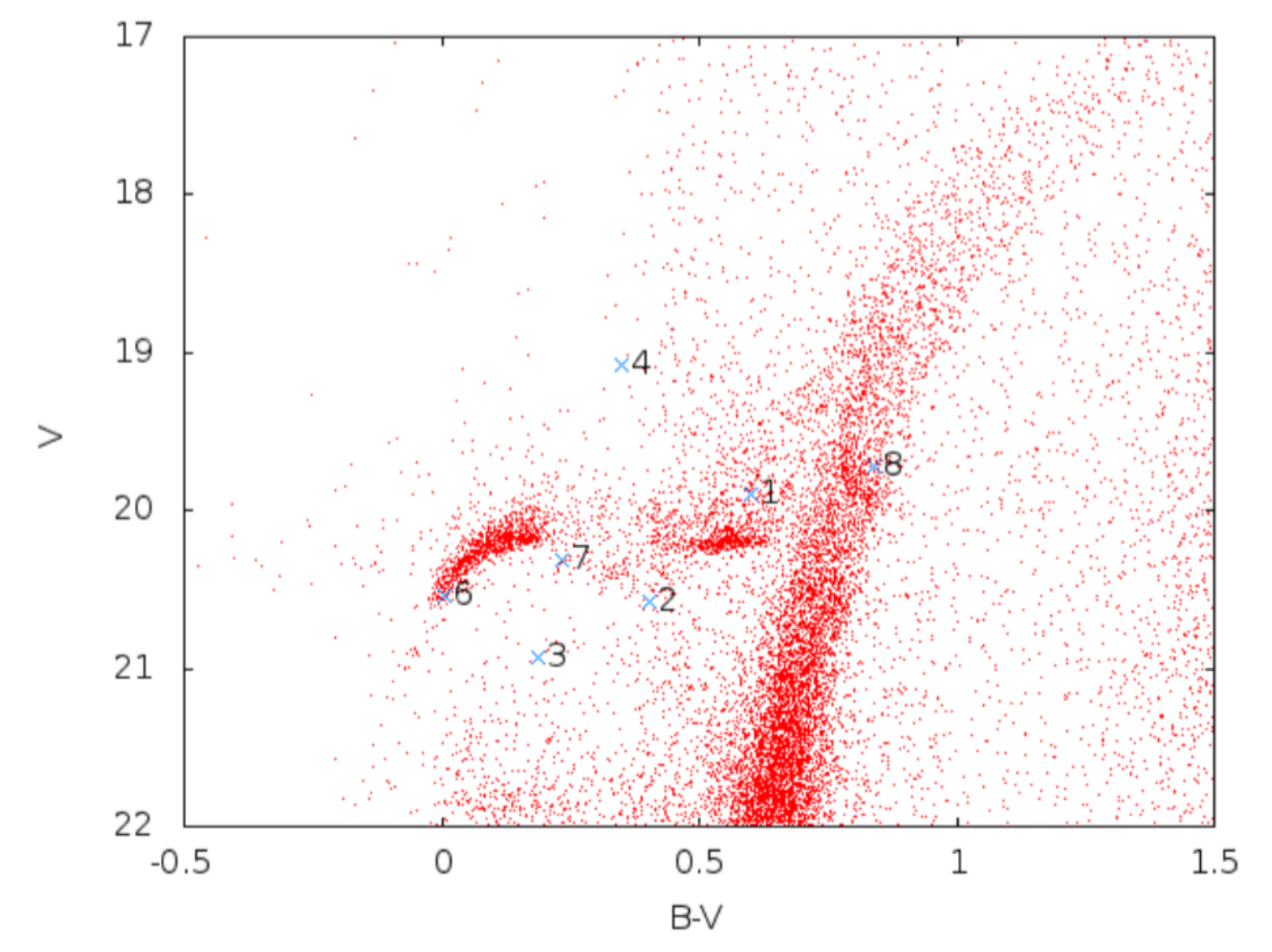}
\caption[]{ \label{fig:SculptorBV_CMD}
B-V color-magnitude diagram from the photometry of \citet{deBoer11} of the Sculptor Dwarf Galaxy, showing the positions of the optical counterparts to \citet{Maccarone05} X-ray sources.
} 
\end{figure}

\begin{figure}
\includegraphics[angle=0,scale=.48]{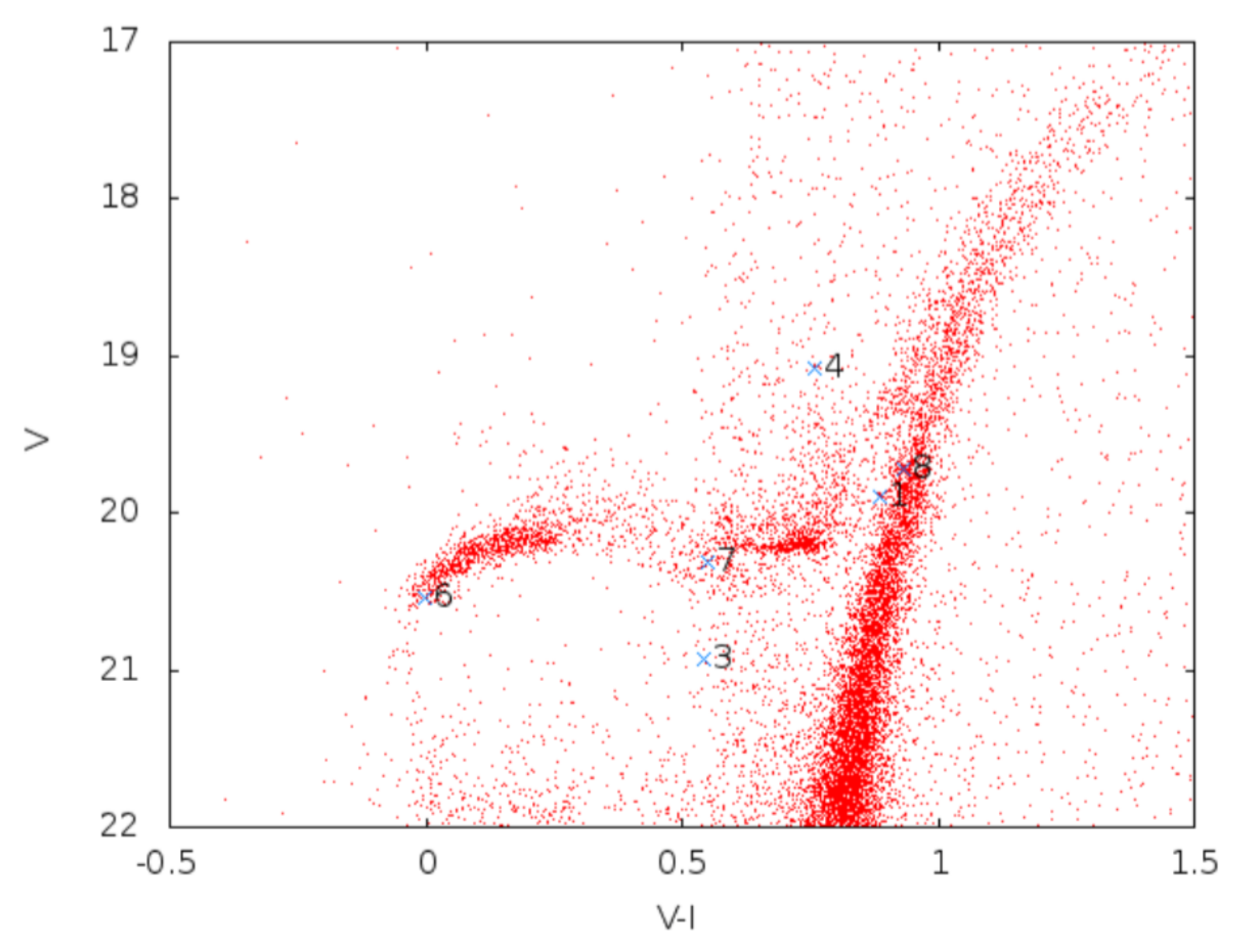}
\caption[]{ \label{fig:SculptorVI_CMD}
V-I color-magnitude diagram from the photometry of \citet{deBoer11} of the Sculptor Dwarf Galaxy, showing the positions of the optical counterparts to \citet{Maccarone05} X-ray sources.
} 
\end{figure}

A search radius of 1\arcsec revealed matches to Maccarone et al. sources \#1-8 (from their Tables 1 and 2). (Note that Maccarone's source \#9 was identified by them as a background galaxy). However, not all of these had photometry in all filters; \#2 lacks an $I$ magnitude, while \#5 lacks $V$ and $I$. We show the locations of those sources with the relevant photometry in Fig. \ref{fig:SculptorBV_CMD} and Fig. \ref{fig:SculptorVI_CMD}.

The photometry of \citet{Schweitzer95} was not accurate enough to determine if these objects actually lay upon the red giant branch and horizontal branch of the Sculptor dwarf. Using the de Boer photometry, we are able to see that the two faintest sources identified by \citet{Maccarone05}, their source \#6 and source \#8 ($L_X\sim3\times10^{32}$ and $2\times10^{33}$ erg/s, respectively), are consistent with the blue end of the horizontal branch, and with the red giant branch, respectively. The remaining objects, sources \#1, 2, 3, 4, and 7, do not appear associated with  
Sculptor's horizontal branch or red giant branch.  
We note that sources \#1, 3, 4, and 7 are substantially redder in the $V$-$I$ CMD than the $B$-$V$ CMD, when nearby features in the Sculptor dwarf CMD are used for comparison.  (Source \#2 does not have an $I$ magnitude in this photometry.)

What is the nature of these sources?  The unusual colors indicate that either these are not Sculptor members, or that their light is a combination of a (red) Sculptor member plus a hotter component.  The estimated luminosities of $6\times10^{33}$ to $6\times10^{34}$ erg/s (if members of Sculptor) of these five sources are generally too luminous to be explained as cataclysmic variables, so assuming they are Sculptor members, they would have to be low-mass X-ray binaries with neutron star or black hole accretors (in agreement with the argument of \citealt{Maccarone05}). However, the optical light of  low-mass X-ray binaries during quiescence is generally dominated by the secondary. For light from an accretion disc to dominate the light of a star at or above the horizontal branch would require a very large mass transfer rate, which would then generate an X-ray luminosity well above $10^{36}$ erg/s, which is not seen. We therefore conclude that the optical counterparts to sources \#1, 2, 3, 4, and 7 are almost certainly background galaxies (for \#2, this agrees with our conclusion from HST imaging, above).  \citet{Arnason19} reports  Gemini spectra that confirm an AGN nature for objects \#1, 2, 3, and 4, and Spitzer infrared photometry indicative of an AGN nature for \#7.

The nature of sources \#6 and 8 are less clear, as their optical counterpart colors and luminosities are consistent with members of the Sculptor dwarf. 
The potential optical counterpart to source \#6 is almost certainly a star in Sculptor, as its position on the blue horizontal branch in two filter combinations would be extremely unlikely to occur by chance for a background galaxy. \citet{Arnason19} identify the likely counterpart to \#6 in their Spitzer infrared and Gemini optical photometry, with colours typical of similar stars (including H$\alpha$ absorption typical of stars of this temperature), and discuss an optical spectrum to the possible counterpart to \#8 that shows no emission lines. 
A background galaxy nature is quite plausible for these two sources, as the Sculptor stars could be chance alignments (the stellar density in this region is such that each 1.5" error circle should have a roughly 10\% chance of capturing a spurious match). Further spectroscopy, high-resolution imaging (to rule out association with extended galaxies, as we found for \#2 and 5 above), and/or proper motion studies could clarify their nature. 

Since we have no robust identifications of X-ray sources associated with the Sculptor dwarf, we do not attempt to calculate the X-ray emissivity of the Sculptor dwarf galaxy in this work.

\section{What Influences X-ray Emissivity?}

From the data assembled in \S 2 and 3, we investigate several possible variables that may affect X-ray emissivity. 

 We explore the dependence of X-ray emissivity on four interesting variables: age, binary fraction, metallicity, and density. We utilize several methods to test the importance of each variable (and combinations of variables), including (1) least-squares regression analysis using a Markov Chain Monte Carlo (MCMC) method and the No-U-Turn Sampler (NUTS) algorithm to handle errors in both variables by sampling from the uncertainty distributions; (2) bootstrapping (random sampling with replacement); and (3) Pearson and Spearman correlation tests.

Stellar interactions are well-known to produce more X-ray binaries in higher-density globular clusters \citep[e.g.][]{Verbunt87}. We plot central density vs. X-ray emissivity in Fig.\ref{fig:RoLx}, grouping globular cluster populations at similar central densities together to improve statistics. Among globular cluster populations, we see evidence of a significant increase in X-ray emissivity with central density,  which becomes clear above $10^4$ \Msun\ pc$^{-3}$. Below $10^4$ \Msun\  pc$^{-3}$, it is unclear whether there is any trend with central density. We do see that populations outside globular clusters have substantially higher X-ray emissivity than in lower-density globular clusters, as first identified by \citet{Verbunt00}, and discussed by several other authors (see the Introduction). Fig.\ref{fig:RoLx} shows that this discrepancy occurs among populations  that in some cases have similar central densities. 

For several of the low-density systems plotted, we know the major contributor(s) to the X-ray luminosity.  In our local region of the Milky Way disk, the old stellar population's X-ray emission is dominated by ABs, though there are numerous unidentified sources in all surveys  \citep{Sazonov06,Agueros09, Warwick14}.  In M67, the total X-ray luminosity is dominated by short-period (1-10 days) ABs ($\sim$half the $L_X$), and by longer-period binary stars whose X-ray emission is not well understood \citep[most of the remainder, ][]{vandenBerg04}.  
NGC 6791 hosts four known CVs, which produce a substantial fraction of its X-ray luminosity  \citep{Kaluzny97,deMarchi07,vandenBerg13}.
In order to reduce the X-ray luminosity per unit mass by an order of magnitude in globular clusters compared to other systems, both types of X-ray emitting binaries must be reduced in frequency.  

\begin{figure}
\includegraphics[width=\linewidth]{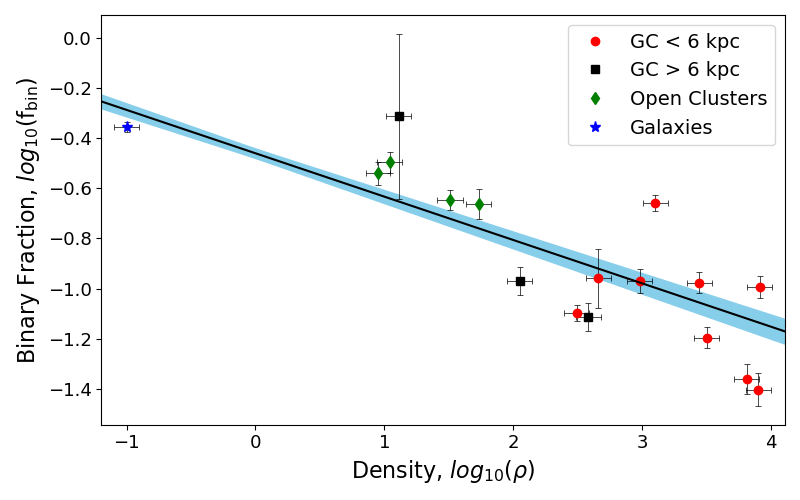}
\caption[]{ \label{fig:MC_BRo}
Central density plotted against binary fraction of clusters, for central densities less than $10^4~\mathrm{\Msun / pc^3}$.  The observational errors were assumed to be 1-$\sigma$ error bars for all the properties of the clusters, which are required to be symmetric in our code.
 NUTS was used to sample the parameter space and obtain best fit values along with its error. From that, we generate a 1-$\sigma$ MCMC error region of the fit, shaded in blue. We observe a negative correlation between binary fraction and central density.
}
\end{figure}

In this work, we concentrate on trying to understand the parameters that affect the X-ray emissivity at central densities below $10^4$ \Msun\ pc$^{-3}$, leaving the analysis of how density-dependent stellar interactions produce X-ray binaries for other works.
We consider several possible parameters; age, metallicity, binary fraction, and density of the stellar populations.
Binary fraction and density are  significantly anti-correlated (see Figure~\ref{fig:MC_BRo}  for a demonstration among our sample), as shown by \citet{Milone12}, though this may be a side effect of the even stronger correlation between binary fraction and total cluster mass.  Unfortunately, density, metallicity, and age tend to be strongly correlated (as globular clusters are dense, old and metal-poor). Our goal is to identify what parameters seem to most strongly influence X-ray emissivity, and to what degree.

\subsection{Bayesian Analysis}

We look for the strongest correlations between X-ray emissivity and combinations of the parameters of age, binary fraction, metallicity and/or density.
We first compare the parameters individually against X-ray emissivity, then try more complicated comparisons  of combinations of parameters. We use two methods of performing  linear fits: i) likelihood maximization 
using the MCMC sampling method and
ii) bootstrapping, a random sampling of the data with replacement.
We also assess correlations using two statistical measures: i) Pearson correlation coefficient (r) and ii) Spearman's rank-order coefficient ($\rho$).

MCMC is a sampling method that deals well with ``nuisance" parameters or data 
with poorly determined errors. Some of our clusters only have an upper limit on their luminosities (such as NGC 6352 in Table~\ref{table*:globs}, etc.), and some have predicted background X-ray luminosities that are comparable to the maximum X-ray luminosities, if not higher (such as NGC 3201 in Table~\ref{table*:globs}, etc.). Also, there are several galaxies with large errors in metallicity (NGC 205, 221, and 224), so the most flexible way we found to address these problems in the analysis was to use a Bayesian analysis method, more specifically MCMC as our main sampling fitting method. We use PyMC3\footnote{https://docs.pymc.io} from \citet{Salvatier16} and its implementation of the No U-Turns Sampler (NUTS).  Our analysis uses logarithms of the relevant quantities, which are better handled by MCMC routines than variations by many orders of magnitude among the data. Our code requires symmetric error bars (in log space); we find the 1-sigma error range and use this to compute a midpoint (in log space). When we have only an upper limit on $L_X/M$, we enforce a lower limit of log ($L_X/\Msun$)=25.5.

We use the Gelman-Rubin diagnostic (\^R) to test for convergence. We sample our models for 5000 steps with 1500 burn-in steps and obtain values of \^R within 0.0002 of 1.0000, which indicates that the fit procedure is well-behaved.

We also apply a bootstrapping procedure to test the robustness of our MCMC fitting to our sample selection. Bootstrapping attempts to compensate for stochastic variation in the selection of our sample, by performing the same fit repeatedly with different samples. 
The new samples are constructed by sampling from the original data with replacement until the same number of data points has been reached, but  with some data points potentially duplicated, and others omitted. We perform one hundred thousand iterations of such random subsets.

We also calculate both Pearson and Spearman correlation coefficients for the correlation between each variable and X-ray emissivity. 
The Spearman test is less affected by outliers in the data (as the data is only sorted by rank), while the Pearson test assumes all data have the same errors. Thus, the Spearman test is arguably more appropriate for datasets where some of the data have much larger errors (e.g. see the large errors for E3). The scores for each fit are calculated and tabulated in Table~\ref{table*:coefficients}.

\subsection{Effects of Age, Metallicity, Binary Fraction, and Density}\label{s:effects}

 As discussed above (\S 1, \S 4), there is evidence from previous work, and within our sample, for a correlation of density with X-ray emissivity. Fig.\ref{fig:RoLx} suggests that this effect is clearly visible only at high central densities. We quantify this with our PyMC3 NUTS regression using various subsamples: all objects, all globular clusters, and objects with densities lower or higher than $10^4$ \Msun/pc$^3$. 

For the complete sample, we find 
a very weak correlation between X-ray emissivity per unit mass and stellar density, with a best-fit slope of -0.04$\pm$0.01 and weak Pearson and Spearman coefficients (0.11 and 0.20, p-values 0.49 and 0.24 respectively, Table~\ref{table*:coefficients}). 
However, the globular cluster sample alone shows a very strong correlation, with p-value $<0.01$ of occurring by chance. When dividing the sample into parts with density above and below  $10^4$ \Msun/pc$^3$, we find that neither sample shows strong correlations. We suggest that this is due to density having a strong effect on X-ray emissivity only for the higher-density population; specifically, that only in high-density globular clusters are substantial numbers of X-ray binaries formed dynamically. Fig. \ref{fig:RoLx} suggests that the threshold for this effect lies around $10^4$ \Msun/pc$^3$.

 Next, we test  the remaining parameters (age, metallicity, and binary fraction) individually against X-ray emissivity, for stellar populations with central density less than $10^4$ \Msun\ pc$^{-3}$.
 We are particularly interested in the low-density population, since dynamically-formed globular cluster populations, produced at high density, may have different dependences on these  parameters than populations that are not formed dynamically.

\begin{figure*}
\begin{multicols}{2}
    \includegraphics[width=\columnwidth]{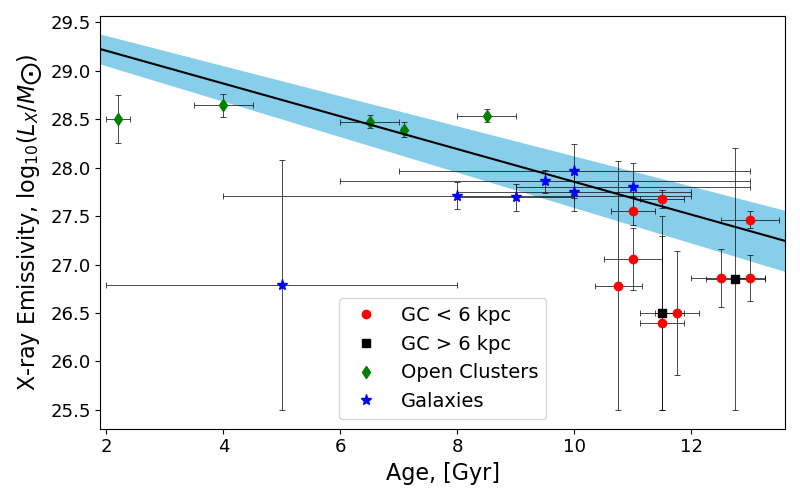}\par 
    \caption{X-ray emissivity vs. age (\citealt{VanDenBerg13A} estimates for globular clusters), using clusters with central luminosity densities less than $10^4~\mathrm{\Msun\  pc^{-3}}$.  
    See Fig.~\ref{fig:MC_BRo} and the text for details.
    }
    \label{fig:MC_ALx_Van}
    \includegraphics[width=\columnwidth]{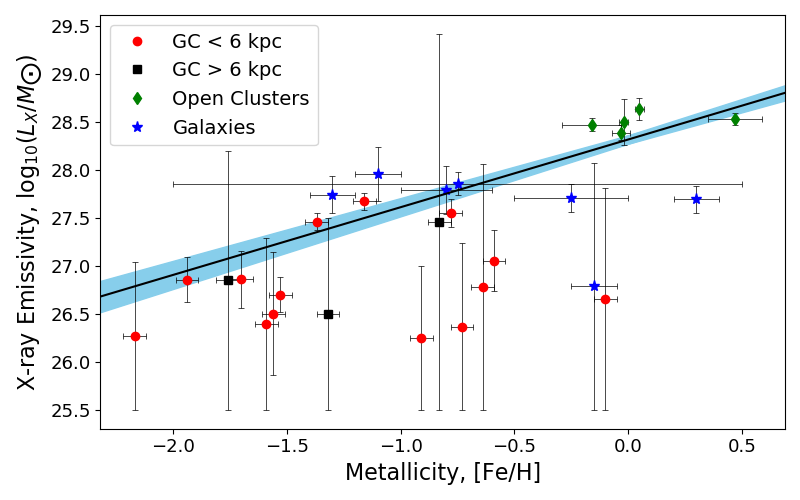}\par 
    \caption{X-ray emissivity plotted vs. metallicity, using clusters with central luminosity densities less than $10^4~\mathrm{\Msun\ pc^{-3}}$. See  Fig.~\ref{fig:MC_BRo} and the text for details.}
    \label{fig:MC_MLx}
    \end{multicols}
\begin{multicols}{2}
    \includegraphics[width=\linewidth]{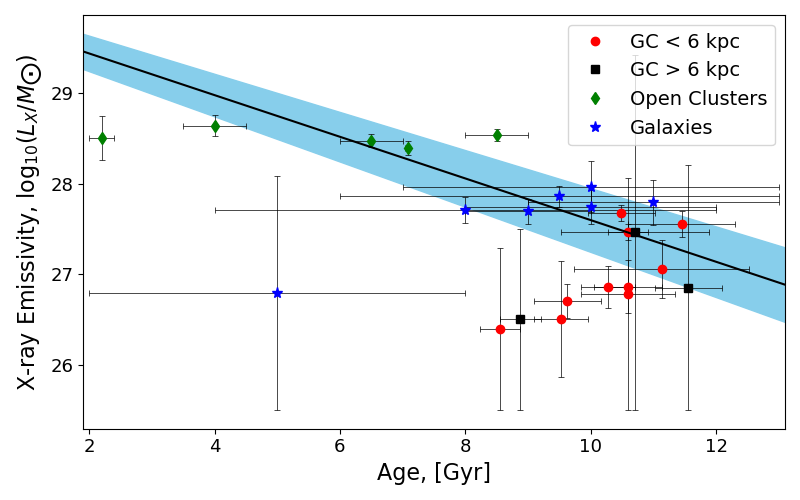}\par
    \caption{X-ray emissivity vs age (\citealt{Marin-Franch09} estimates for globular clusters), using clusters with central luminosity densities less than $10^4~\mathrm{\Msun\  pc^{-3}}$. See  Fig.~\ref{fig:MC_BRo} and the text for details.}
    \label{fig:MC_ALx_Mar}
    \includegraphics[width=\columnwidth]{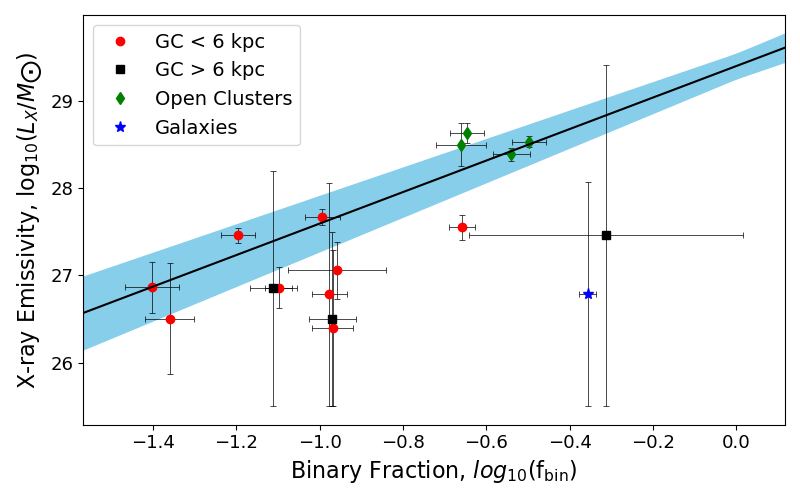}\par
    \caption{X-ray emissivity plotted against binary fraction, using clusters with central luminosity densities less than $10^4~\mathrm{\Msun\  pc^{-3}}$. See Fig.~\ref{fig:MC_BRo} and the text for details.}
    \label{fig:MC_BLx}
\end{multicols}
\end{figure*}

The age of a stellar population is likely to affect its X-ray populations. CV population mass transfer rates are correlated with companion mass, so are likely to decline with age as the masses of available companions decline \citep[e.g.][]{Ivanova06,Stehle97}. At the same time, X-rays from normal stars strongly decline with age as the stars spin down \citep[e.g.][]{Randich97}, but after 1 Gyr the AB X-ray luminosity dominates, and we do not know how this varies with time. (Close ABs spiral together and merge, but wider binary orbits shrink, causing initially wider ABs to produce stronger X-ray emission later).

The measurement of ages of stellar systems is a difficult art, particularly when small differences in ages among old populations are sought. 
Several recent works have measured the (relative, or absolute) ages of globular clusters; these include 
\citet{Marin-Franch09}, \citet{DeAngeli05}, \citet{VanDenBerg13A}, \citet{Hansen13}, and \citet{OMalley17}. 
However, there are substantial differences between the relative ages of globular clusters in these works, suggesting that age measurement is not a settled subject. We use the two studies with the largest overlap with our cluster sample, \citet{Marin-Franch09} and \citet{VanDenBerg13A}, and compare calculations using either sample. 
\citet{Marin-Franch09} give only relative globular cluster ages; we assume an average age of 10.7 Gyrs (which matches the age of 47 Tuc, and the average for outer halo globular clusters in general, from \citet{Salaris02}) to place these on an absolute scale.

Fig.~\ref{fig:MC_ALx_Van} and Fig.~\ref{fig:MC_ALx_Mar} compare X-ray emissivity vs.\ age estimates for our sample (using the two globular cluster age studies). Galaxies often have complex star formation histories, so we plot the average ages of the bulk of the star formation for our galaxies as referenced in Table~\ref{table*:nonglobs}.
For our regression analyses with one parameter, our independent variables are the slope and intercept of a line through the data. 
Our PyMC3 calculations find that the most likely correlation between age and X-ray emissivity is negative.   
The slope of the fit using the \citet{VanDenBerg13A} ages in Fig.~\ref{fig:MC_ALx_Van} is $-0.16\pm0.02$, while the fit using the \citet{Marin-Franch09} ages in Fig.~\ref{fig:MC_ALx_Mar} is $-0.22\pm0.02$. The error range, including slope and intercept errors, is shaded in blue. 
We confirm the negative correlations by calculating the Pearson and Spearman coefficients for the correlation of age vs. $L_X/M$, finding coefficients of -0.64 and -0.68  (each have p-values $<0.01$ of obtaining such a high correlation by chance) using the \citet{VanDenBerg13A} ages, and -0.46 and -0.35 (p-values 0.02 and 0.09, respectively) using the \citet{Marin-Franch09} ages, respectively (Table~\ref{table*:coefficients}).
 The correlations appear to be due principally to  
different populations having different average ages, without apparent correlations between emissivity and age within each population. 
We suspect these apparent correlations are really due to the other fitted parameters (see below).

Considering only the non-GC populations, the  correlation between $L_X/M$ and age is insignificant ($p$=0.39 and 0.23 for the Pearson and Spearman tests respectively),  and appears to be due to  differences between the emissivities and average ages of the galaxy and open cluster populations, which  have wide, overlapping age ranges; see Fig.~\ref{fig:MC_ALx_Van}.
For the GC sample alone, the Marin-Franch and Vandenberg ages give opposite slopes for the $L_X/M$ vs. age relation  (see Fig.~\ref{fig:MC_ALx_Van} and Fig.~\ref{fig:MC_ALx_Mar}); interestingly, the Marin-Franch ages indicate a significant positive correlation ($p$=0.01) between age and $L_X/M$  (Table~\ref{table*:coefficients}), the opposite of what we expected. 
Uncertainties in the age  measurements limit the power of this analysis.

\begin{table}
\centering
\begin{tabularx}{3.3in}{Bsssss}
\hline
Comparison  &  Pearson Coeff.  & Prob. & Spearman Coeff.  & Prob. & Figure \\
\hline
$\mathrm{f_{bin}}$ vs. $\rho_c$ (All) & -0.837 & <0.01 & -0.816 & <0.01 & Fig.~$\ref{fig:MC_BRo}$ \\
\hline
$L_X/M$ vs.:\\
\hline
 $\rho_c$ (All)     & 0.114 & 0.49 & 0.195 & 0.24 & \\
 $\rho_c$ (GCs)     & 0.676 & <0.01 & 0.735 & <0.01 & Fig.~\ref{fig:RoLx} \\
 $\rho_c$ (Non-GCs) & 0.966 & <0.01 & 0.821 & 0.02 & \\
\hline
 Age (V; All)  & -0.644 & <0.01 & -0.677 & <0.01 & \\
 Age (V; GCs)  & 0.016 & 0.96 & 0.000 & 1.000 & Fig.~$\ref{fig:MC_ALx_Van}$ \\
 Age (Non-GCs) & -0.271 & 0.39 & -0.375 & 0.23 & \\
\hline
 Age (MF; All) & -0.462 & 0.02 & -0.346 & 0.09 & \\
 Age (MF; GCs) & 0.670 & 0.01 & 0.735 & <0.01 & Fig.~$\ref{fig:MC_ALx_Mar}$ \\
 \hline
 Fe/H (All)     & 0.620 & <0.01 & 0.562 & <0.01 & \\
 Fe/H (GCs)     & 0.215 & 0.41 & 0.172 & 0.51 & Fig.~$\ref{fig:MC_MLx}$ \\
 Fe/H (Non-GCs) & 0.314 & 0.32 & 0.378 & 0.23 & \\
\hline
 $\mathrm{f_{bin}}$ (All)    & 0.553 & 0.02 & 0.461 & 0.06 & \\
 $\mathrm{f_{bin}}$ (GC)     & 0.464 & 0.13 & 0.259 & 0.42 & Fig.~$\ref{fig:MC_BLx}$ \\
 $\mathrm{f_{bin}}$ (Non-GC) & -0.848 & 0.07 & -0.500 & 0.39 & \\
 \hline
\end{tabularx}
\caption{
Correlation coefficients (for Pearson, Spearman tests) and appropriate p-values for several comparisons. Each comparison includes either (All), just (GCs), or just (Non-GC) data, as specified. Age tests use (MF)=\citet{Marin-Franch09} ages, or (V)=\citet{VanDenBerg13A} ages for GCs. 
Column 6 identifies the relevant figure showing each correlation.}
\label{table*:coefficients}
\end{table}

Metallicity is known to be a factor affecting the frequency of bright LMXBs in extragalactic globular clusters \citep[e.g.][]{Kundu03}, and of their descendants, millisecond pulsars \citep[e.g.][]{Hui11}. A clear effect of metallicity has not been found for faint X-ray binaries so far \citep[e.g.][]{Heinke06b}, but this could be due to a relatively small number of studied sources. \citet{Stehle97} model low-metallicity CV evolution, finding slightly higher mass-transfer rates, which would lead to shorter CV lifetimes, while \citet{Cote18} find larger numbers of accreting WDs in CV-like systems at low metallicity in their models. Thus, there is not yet a clear, accepted prediction for the effects of metallicity on $L_X$/M for CVs.  
We note that the mass-to-light ratio may vary with metallicity (as suggested by stellar population models, e.g. \citealt{McLaughlin05}, but cf. \citealt{Watkins15} and \citealt{Strader11} who find the opposite dependence), which may generate an apparent metallicity effect on X-ray emissivity. 
 \citet{BaumgardtHilker18} do not find a correlation of mass-to-light ratio with metallicity, and we use their mass estimates, which suggests that any metallicity effect is not due to mass-to-light ratio variance.

We plot in Fig.~\ref{fig:MC_MLx} X-ray emissivity vs. the metallicity of the moderate and low-density globular clusters (central density $<10^4~\mathrm{\Msun\  pc^{-3}}$).  A trend towards higher X-ray emissivity among higher metallicity systems is clear, and the trend  may be visible among the globular cluster population as well as the non-GC population. 
Our PyMC3 calculation finds a slope of  0.69$\pm$0.06 for the best fit of metallicity vs. $L_X/M$ for all low-density systems, suggesting a much larger effect than age. We find 
strong evidence of correlation from the Pearson and Spearman tests, with chance probabilities $<$1\% (Table~\ref{table*:coefficients}).
We also computed correlation coefficients for the globular cluster population and non-GC population separately,
but did not detect a significant correlation for either population individually. The non-GC systems have significantly higher X-ray emissivity than globular clusters of the same metallicity, suggesting that another variable plays a larger role.

The fraction of cluster members that are binaries is a clearly plausible parameter to explain the X-ray emissivity variations among different populations. All X-ray emitting populations of old stars are produced in binary systems of some kind; ABs, CVs, LMXBs, and MSPs are binaries, or descend from binaries. A significant caveat is that each type of X-ray emitting system may be produced by binaries of different initial orbital period and component masses, and that the distribution of binaries in a population cannot be completely described by a single number, the binary fraction. In particular, X-ray emitting binaries are generally in relatively short orbits ($<<$1 AU), while the differences in binary fraction among different populations may be strongest among longer-orbit systems, which are vulnerable to disruption at moderate densities \citep{Hut91}. However, systems in close orbits will have previously been longer-period binaries, with periods reduced by angular momentum loss, or (more strongly) by common envelope evolution.

Fig.~\ref{fig:MC_BLx} shows the correlation between binary fraction and X-ray emissivity among our sample. Unfortunately, relatively few of our populations have measured binary fractions, so we have a significantly smaller sample here than in the other comparisons.  
Our PyMC3 calculations give a slope of  1.82$\pm$0.20, the largest fitted slope of any parameter.
Pearson and Spearman tests indicate significant, or marginally significant, correlations (p=0.02 and 0.06 respectively;  Table~\ref{table*:coefficients}).   
We also checked for correlations among the subsamples of all globular clusters and all non-globular clusters, 
but did not find statistically significant (p$<$0.05) results for either independently (these have the smallest sample sizes among our investigated samples).

Finally, we explore  combinations of the promising variables of binary fraction, metallicity, and  density.  
We first performed an analysis including age along with the other parameters, but found that this number of parameters led to a poorly constrained fit.
Given the lack of strong evidence for age alone as a parameter, and the uncertainty in relative ages among different methods, we omit age from our principal analyses.   
We set weights by taking binary fraction to the power $\alpha$, metallicity to the power $\beta$, and density to the power of $\gamma$,

\begin{equation}
    ~~~~~~~~~~~~~~~~\frac{L_X}{M} = \mathrm{(f_{bin})}^\alpha \times (10^{\mathrm{[Fe/H]}})^\beta\times(\mathrm{\rho})^\gamma \times 10^b~,
    \label{eqn:ABM}
\end{equation}

We find it convenient to take logarithms of both sides, so as to perform the fit in linear space allowing easier measurement of the contribution of each variable. This gives

\begin{equation}
\begin{split}
    ~~~~~~~~~~~~~~~~\mathrm{log}_{10}\left(\frac{L_X}{M} \right) = ~& \alpha (\mathrm{log_{10}(f_{bin}})) + \beta (\mathrm{[Fe/H]}) + \\
    ~~~~~~~~~~~~~~~~& \gamma(\mathrm{log}_{10}(\mathrm{\rho})) + b~. 
    \label{eqn:Linear_ABM}
\end{split}
\end{equation}

We set up the priors in the following way:

\begin{ceqn}
\begin{equation}
\begin{split}
    \alpha, \beta, \gamma ~~&\sim~~ \mathcal{N}(0,5) \\
    \mathrm{b} ~~&\sim~~ \mathcal{N}(0,50) \\
    \mathrm{log}_{10}\left(\frac{L_X}{M}\right)^\star ~~&\sim~~ \mathcal{N}\left(\mathrm{log}_{10}\frac{L_X}{M},\sigma_{\mathrm{log}_{10}L_X/M}\right)
    \label{eqn:priors}
\end{split}
\end{equation}
\end{ceqn}

 $\alpha, \beta, \gamma$, and b are the exponents on the variables of binary fraction, metallicity, and density, as defined in Eq.~\ref{eqn:ABM}. We use wide, uninformative priors to emphasize the data's contribution in the likelihood function. We also define a prior, $\mathrm{log}_{10}(L_X/M)^\star$, to describe each individual point in the X-ray emissivity.

The likelihood function is written as

\begin{ceqn}
\begin{equation}
\begin{split}
    p&(y_i ~|~ \mu, \sigma) = \sqrt{\frac{1}{2 \pi \sigma^2}} ~\mathrm{exp}\left( -\frac{(y_i-\mu)^2}{2\sigma^2} \right) \\
    \mathcal{L} = &\prod^{N}_{i=1} p \left( y_i ~|~ \log \left( \frac{L_X}{M} \right) , \sigma_{\log \left( \frac{L_x}{M} \right)} \right) \times {\rm priors}~.
\end{split}
\end{equation}
\end{ceqn}

where $\mu$ equals the expected outcome for the X-ray emissivity, which is defined as the model represented by Equation~\ref{eqn:Linear_ABM}, and $\sigma$ is the error of the X-ray emissivity.

 Sampling this space using the NUTS algorithm, we generate the parameter distributions as listed in Table~\ref{tab:PyMC3_fits_coefficients}  for four different groups -- all data, all low-density objects ($\rho$ < 10$^4$ \Msun/pc$^3$), all high-density objects, and all globular clusters.   
We divide the data by density, due to our hypothesis that high-density environments produce X-ray sources through dynamical processes, with different dependencies than the normal methods of producing these sources at low density. We also check the results for the globular cluster subsample specifically. 

\begin{figure}
    \centering
    \includegraphics[width=\linewidth]{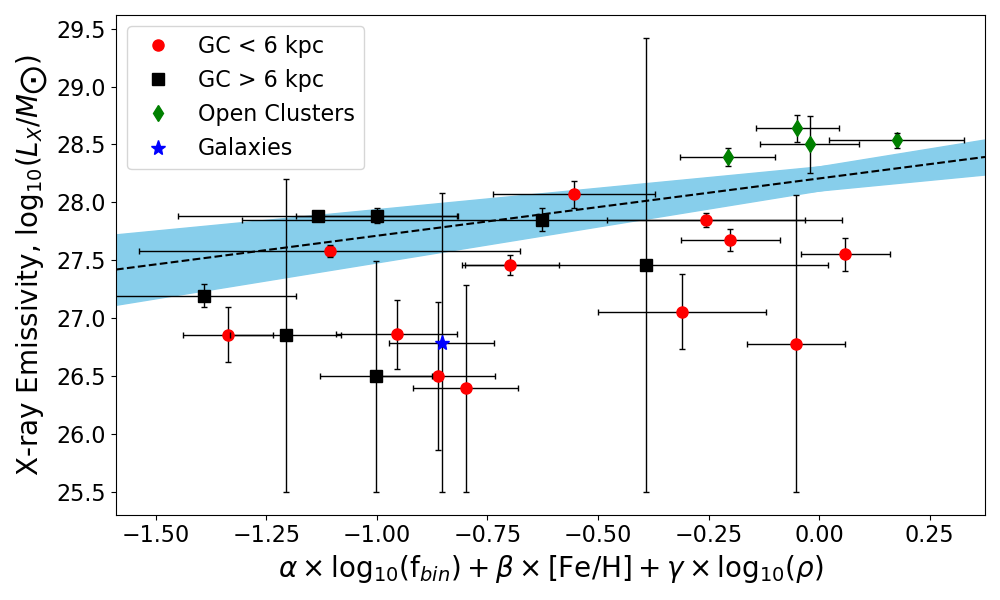}
    \caption{The best-fit combination of binary fraction, metallicity, and density, plotted vs. X-ray emissivity. The exponents used are the best-fitting values of the PyMC3 fit from Table~\ref{tab:PyMC3_fits_coefficients}, "All $\rho$". For the uncertainty description, refer to Fig.~\ref{fig:MC_BRo} and \ref{fig:MC_ALx_Van}.
    }
    \label{fig:PyMC3_LxM}
\end{figure}

Fits to the full dataset ("All $\rho$" in Table~\ref{tab:PyMC3_fits_coefficients}) found two minima in the MCMC chains. However, one of these fits had large negative coefficients on both binary fraction and density. We opted to impose a prior that the binary fraction coefficient be positive, selecting the more physically plausible solution, which also gave a statistically better fit. 
This fit gives strong Pearson and Spearman correlation coefficients (0.54 and 0.47; probabilities <0.01 and 0.02), indicating that the fitted parameters are well-correlated with the X-ray emissivity, although there remains significant scatter (see Fig.~\ref{fig:PyMC3_LxM}).  We find significantly non-zero exponents for binary fraction, metallicity, and density, indicating that all three variables  contribute to the variance in X-ray luminosity. The highest exponent ($1.04^{+0.14}_{-0.15}$) is on binary fraction, suggesting that this is the most important parameter. However, the substantial scatter remaining in the fit suggests that additional as-yet unidentified parameters, or systematic errors in our analysis, affect the data.

 The high-density and low-density fits (Table~\ref{tab:PyMC3_fits_coefficients}) show interesting differences. The best-fit exponents on binary fraction have opposite signs, but neither is  significant. Density is  significant (exponent $0.55^{+0.13}_{-0.14}$) for the high-density sample, but not for the low-density sample, while metallicity is the only significant variable for the low-density sample (exponent $1.01^{+0.21}_{-0.24}$). 
The globular cluster sample does show a significant binary fraction effect, and shows a very strong dependence on density.

We plot the joint distributions of the variables
to observe the correlations between variables.   Fig.~\ref{fig:PyMC3_corner} illustrates that there is a strong anticorrelation between the effects of binary fraction (exponent $\alpha$) and metallicity (exponent $\beta$) for the low-density sample, while the effects of density ($\gamma$) are not strongly correlated with other parameters.

\setlength{\extrarowheight}{1mm}
\begin{table}
    \centering
    \begin{tabular}{c|c|c|c|c}
        \textbf{MCMC}  & All $\rho$ & Low $\rho$ & High $\rho$ & GCs\\
        \hline
        $\alpha$, on $f_{bin}$ & 1.04$^{+0.14}_{-0.15}$  & -0.60$^{+0.54}_{-0.49}$  & 0.32$^{+1.29}_{-1.07}$ & 0.63$^{+0.15}_{-0.18}$ \\
        $\beta$, on [Fe/H]    & 0.60$^{+0.09}_{-0.09}$  & 1.01$^{+0.21}_{-0.24}$  & 0.10$^{+0.16}_{-0.11}$ & 0.15$^{+0.09}_{-0.07}$ \\
        $\gamma$, on $\rho$    & 0.39$^{+0.06}_{-0.07}$  & 0.03$^{+0.09}_{-0.08}$  & 0.55$^{+0.13}_{-0.14}$ & 0.49$^{+0.06}_{-0.07}$ \\
        b, on $L_X/M$         & 28.3$^{+0.1}_{-0.1}$ & 28.0$^{+0.3}_{-0.3}$ & 25.6$^{+2.8}_{-2.0}$ & 26.6$^{+0.2}_{-0.2}$ \\
        \hline
        Pearson         & 0.540                   & 0.754                   & 0.832  & 0.705 \\
        p-value         & <0.01                   & <0.01                   & 0.02 & <0.01   \\
        \hline
        Spearman      & 0.469                   & 0.699                   & 0.571  & 0.770  \\
        p-value        & 0.02                    & <0.01                    & 0.18 & <0.01   \\
    \end{tabular}
    \caption{The best-fitting values and errors (1$\sigma$) for the exponent parameters (and intercept) from our PyMC3 fits, fitting density, binary fraction, and metallicity (see Fig.~\ref{fig:PyMC3_LxM}) The first column of fits include all our samples, the second include all samples with a density less than $10^4$ M$_{\odot}$ pc$^{-3}$, the third is all samples with densities above that, while the last column includes only all the globular cluster samples. 
    }
    \label{tab:PyMC3_fits_coefficients}
\end{table}

\begin{figure}
    \centering
    \includegraphics[width=\linewidth]{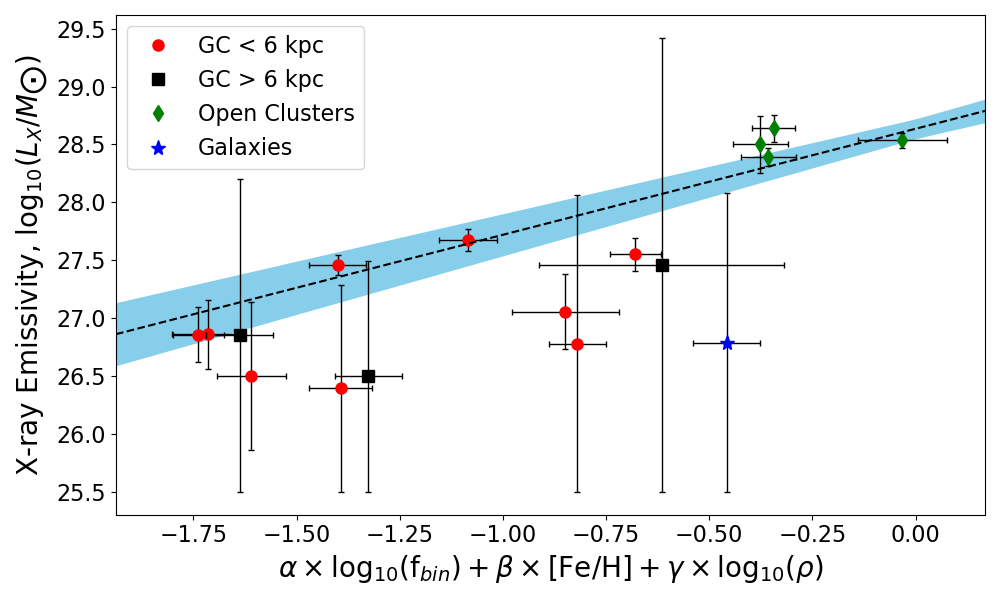}
    \caption{The best-fit combination of binary fraction, metallicity, and density, plotted vs. X-ray emissivity, for low-density populations. The exponents used are the best-fitting values of the PyMC3 fit from Table~\ref{tab:PyMC3_fits_coefficients}, "Low $\rho$". For the uncertainty description, refer to Fig.~\ref{fig:MC_BRo} and \ref{fig:MC_ALx_Van}.}
    \label{fig:PyMC3_low_LxM}
\end{figure}

\begin{table}
    \centering
    \begin{tabular}{c|c|c|c|c}
        \textbf{Bootstrap}  & All $\rho$ & Low $\rho$ & High $\rho$ & GCs \\
        \hline
        $\alpha$, on $f_{\text{bin}}$ & 0.08$\pm$0.48  & 0.0$\pm$1.4 & 0$\pm$13 & 0.31$\pm0.39$ \\
        $\beta$, on [Fe/H]    & 0.33$\pm$0.22  & 0.78$\pm$0.42  & 0.1$\pm$6.8 & 0.11$\pm0.17$ \\
        $\gamma$, on $\rho$    & 0.04$\pm$0.18  & 0.04$\pm$0.26  & 0.3$\pm$9.2 & 0.37$\pm0.17$ \\
        b, on $L_X/M$         & 28.2$\pm$0.4 & 28.3$\pm$0.7 & 26$\pm$45 & 26.6$\pm0.4$ \\
    \end{tabular}
    \caption{The best-fitting values and errors (1$\sigma$) for the exponent parameters (and intercept) from our bootstrap fits, fitting  binary fraction, metallicity, and density. The first column of fits include all our samples, the second include all samples with a density less than $10^4$ M$_{\odot}$ pc$^{-3}$, the third is all samples with densities above that, while the last column includes only all the globular cluster samples.
    }
    \label{tab:Bootstrap_fits_coefficients}
\end{table}

\begin{figure}
    \centering
        \includegraphics[width=\columnwidth]{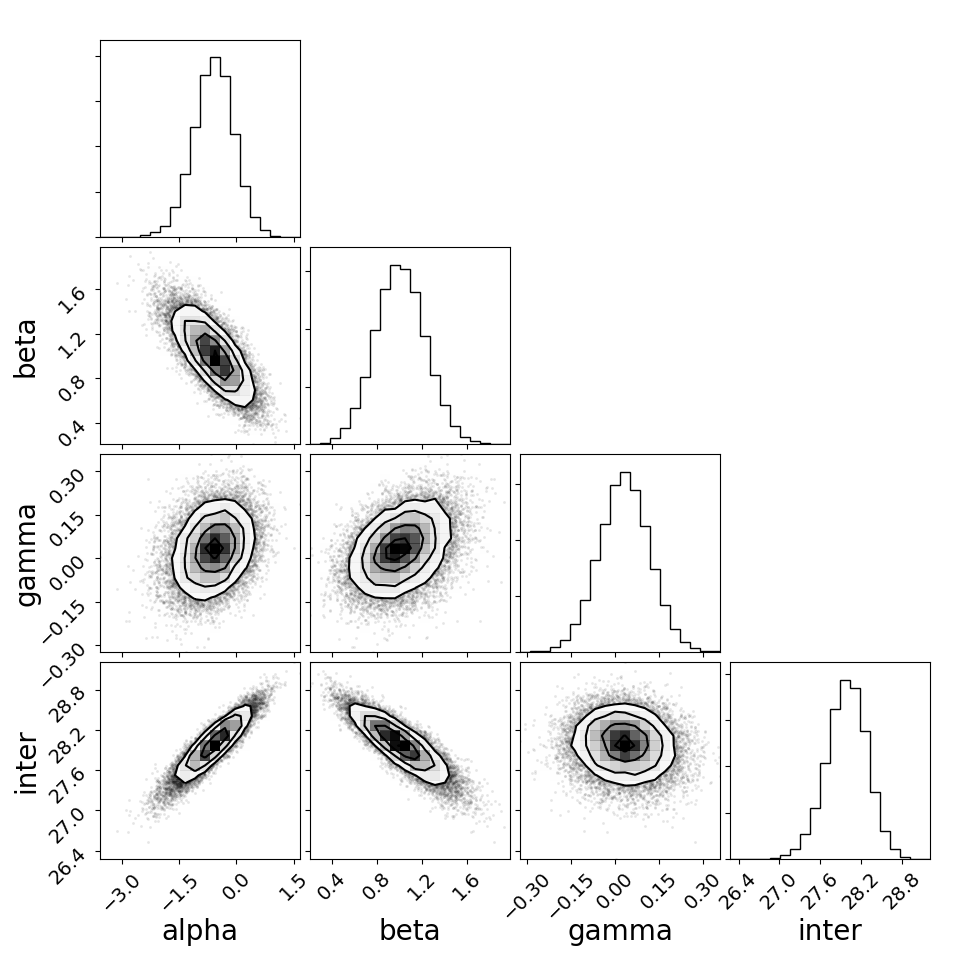}
        \caption{A corner plot \citep{ForemanMackey16}  generated from the ensemble sampling of all populations within the low-density regime. The exponents are with respect to the variables described in Table \ref{tab:PyMC3_fits_coefficients} and \ref{tab:Bootstrap_fits_coefficients}: ``alpha" on f$_{\text{bin}}$, ``beta" on [Fe/H], ``gamma" on $\rho$, and ``inter" on $L_X/M$.}
        \label{fig:PyMC3_corner}
\end{figure}

\begin{figure}
    \centering
    \includegraphics[width=1.05\linewidth]{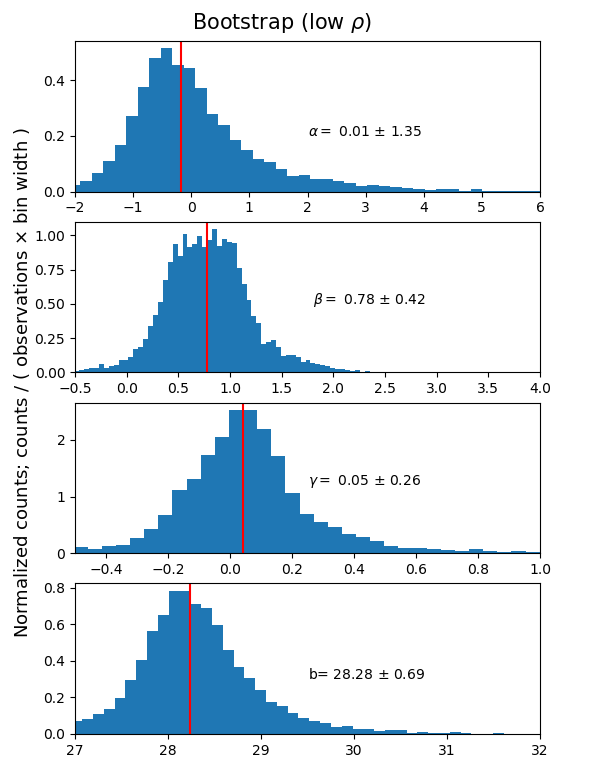}
    \caption{Histograms of the values from our bootstrap test for the exponents $\alpha$, $\beta$, $\gamma$ on the variables binary fraction, metallicity, and density, as well as the intercept (scale factor) for the low $\rho$ limit. The red line indicates the median of the distribution of each variable, while the mean is written in each panel with its 1$\sigma$ uncertainty.}
    \label{fig:bootstrap}
\end{figure}

We also ran bootstrap tests with this paradigm. As described above, our bootstrap tests vary the sample of which objects are included, and randomly locate each datapoint on the x-coordinate with a probability distribution defined by the error distribution of the data. We generate distributions of the best-fit values for  $\alpha$, $\beta$, $\gamma$, and b (the intercept) which we show in Fig.~\ref{fig:bootstrap} for the low-density sample. As the bootstrap tests remove some of the (already quite small) samples, they substantially increase the errors on the parameters. 
 The bootstrap estimates (Table~\ref{tab:Bootstrap_fits_coefficients}) largely agree with our estimates from the MCMC fitting (see Table~\ref{tab:PyMC3_fits_coefficients}). In our bootstrap fits to the full sample ("All $\rho$"), we see that only the metallicity exponent remains nonzero at 1$\sigma$ confidence, while the exponents on binary fraction and age are  consistent with zero. The bootstrap for the low-density sample also retains a significant exponent for metallicity, while the bootstrap on the high-density sample (the smallest sample) does not constrain any parameter. Finally, the globular cluster sample retains a robust dependence on density in the bootstrap fit.

We conclude that our PyMC3 fits find evidence for a dependence of X-ray emissivity on binary fraction, metallicity, and density, with density having a strong effect at higher densities and metallicity a stronger effect at low densities. However, our robust bootstrap fits suggest substantial degeneracies among these parameters,  with only metallicity producing a significant effect for the full and low-density samples, and only density being significant for the globular cluster sample.

\section{Discussion}

We present a somewhat different perspective than  
\citet{Ge15}. \citet{Ge15} carefully compare the X-ray emissivities of four globular clusters with those of several galaxies and open clusters, and argue for the universality of X-ray emissivity in these populations,  while we argue that low-density globular clusters have lower X-ray emissivity than open clusters and Galactic populations.  \citet{Ge15} also identify an increased X-ray emissivity of open clusters compared to galactic field populations. We discuss this important point in more detail below,  though we confess that we do not yet fully understand the source of this difference.  

Key differences between our work and that of \citet{Ge15} include our use of many more globular clusters, and our identification of a  central density above which dynamical formation of X-ray sources substantially enhances the X-ray emissivity. Three of the clusters used by \citet{Ge15} lie above this density, and we therefore attribute their higher X-ray emissivity to dynamical formation of tight accreting binaries involving WDs or NSs (evidenced by the nature of the X-ray emitting binaries, which are dominated in X-ray flux by neutron stars in 47 Tuc and NGC 6266, and bright CVs in NGC 6397, while in e.g. open clusters chromospherically active binaries dominate). We identify the lower X-ray emissivity of $\omega$ Cen as typical of lower-density globular clusters, not the exception. Thus, we see a substantial difference in X-ray emissivity between lower-density globular clusters and all other populations, including open clusters, that needs explanation.  We suggest that binary fraction and/or metallicity may drive this difference.

Recently, \citet{Cheng18} have assembled a very large sample of Chandra observations of globular clusters, and calculated X-ray fluxes for each cluster by  extracting spectra from the entire half-mass radius (using an exterior annulus, or other distant regions, for background). This is simpler  than the point-source method, allowing them to calculate X-ray emissivities for a larger sample of Galactic globular clusters. \citet{Cheng18} also find significantly different results; they find a poorer correlation between X-ray emissivity (measured as $L_X/L_K$) and the stellar interaction rate per unit mass (e.g. their fig. 6), and argue that most GC X-ray sources are primordial binaries, with only a small role for stellar encounter rate.   We agree with Cheng et al. that X-ray emission from many GCs ( the lower-density globular clusters, probably the majority by number) is dominated by primordial binaries, which generally scale with mass.
However,  there are some complexities with some datasets that may influence their conclusions.

For instance, their fig. 6 (right panel) compares $L_X/L_K$ with $\gamma$ (stellar encounter rate per unit mass) among 23 well-studied GCs. The correlation here is not as clear as suggested in other works \citep[e.g.][]{Pooley03}, in large part due to the unusually high values of $L_X/L_K$ adopted for NGC 6397 and NGC 6366.  
 We suggest some systematic errors that may affect 
their estimates of cluster mass (using $K$-band estimates) and of $L_X$. 

\begin{figure}
    \centering
        \includegraphics[width=\linewidth]{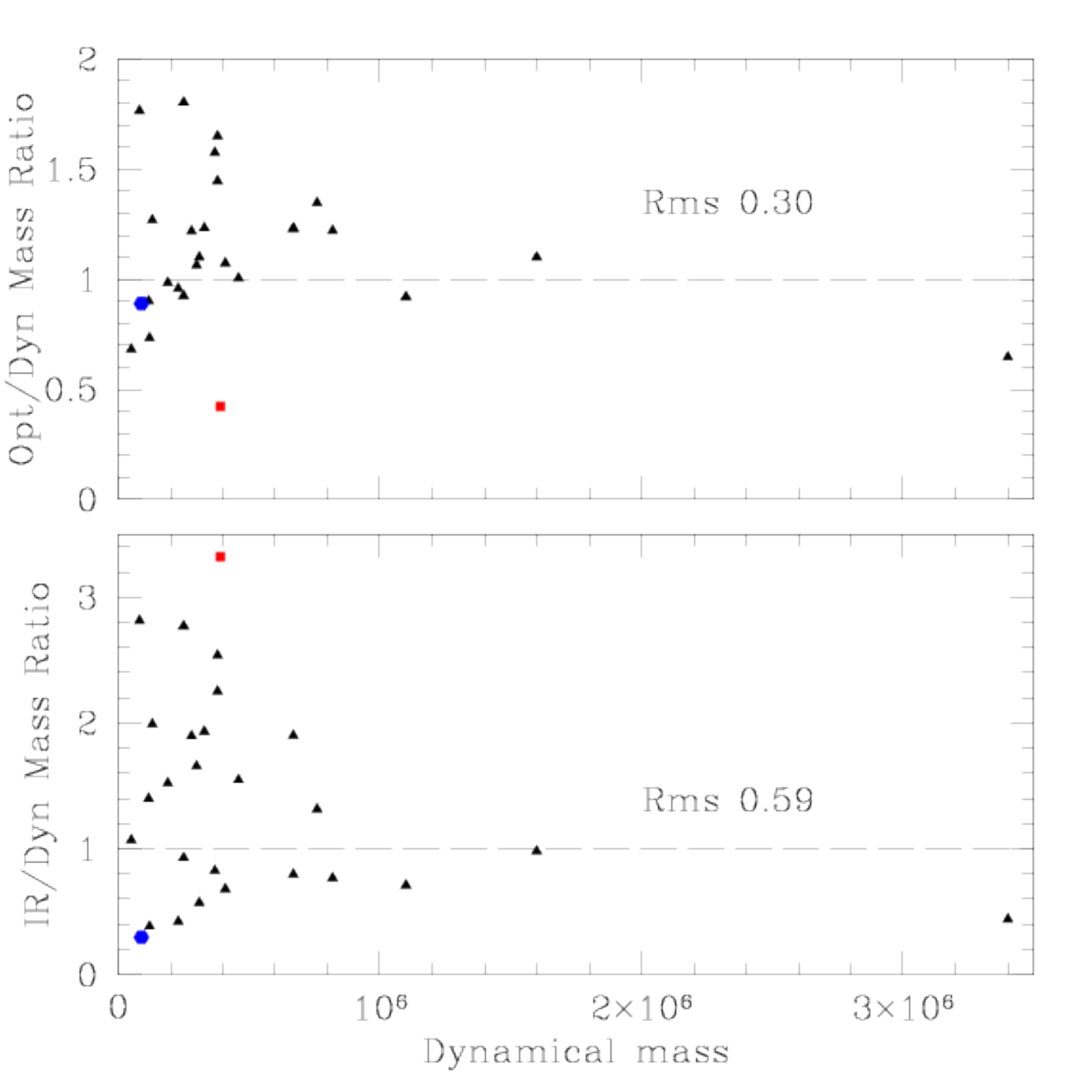}
        \caption{Ratios of dynamical masses of globular clusters \citep{BaumgardtHilker18} to masses calculated from \citet{Harris96} (2010 revision) catalog $M_V$ estimates (top), or by \citet{Cheng18} using $K$ measurements (bottom).  NGC 6397 is indicated with a blue filled hexagon, Terzan 5 by a red square. Rms measurements for both ratios are indicated. The variance is substantially larger when using $K$ measurements.}
        \label{fig:Cheng}
\end{figure}

\citet{Cheng18} estimate cluster masses using $K$-band estimates, while we use 
 mass estimates from \citet{BaumgardtHilker18}.
We compare 
the spread in the ratio of Cheng et al's K-band infrared luminosity estimates to mass, with the spread in the ratio of visible-light luminosity  \citep[][2010 update]{Harris96} to mass, in
Fig.~\ref{fig:Cheng}.  Visible-light estimates appear to be superior, as the rms deviations from dynamical mass estimates are 0.30 and 0.59 for visible and K-band estimates, respectively.

\begin{figure}
    \centering
        \includegraphics[width=\linewidth]{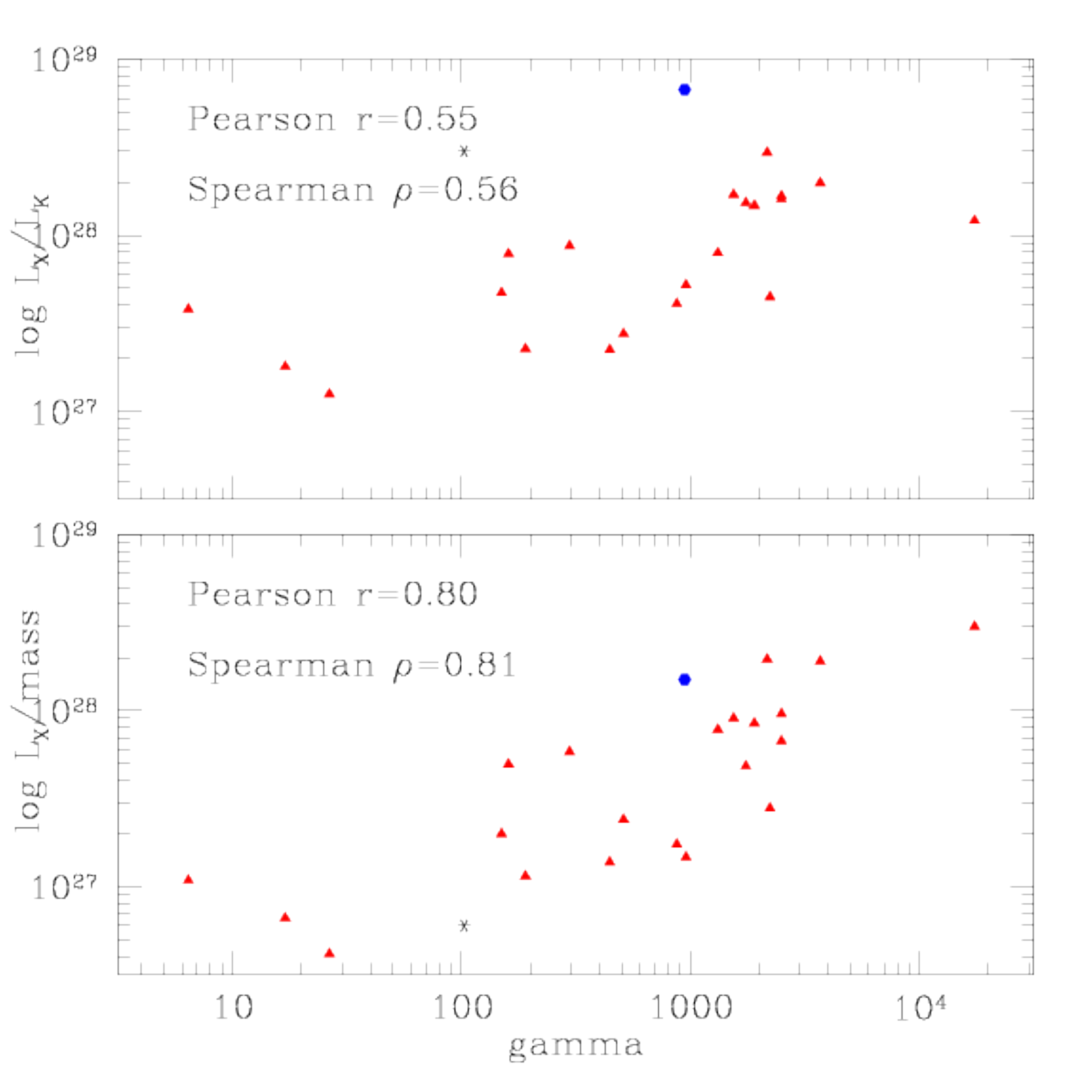}
        \caption{Ratios (y-axis) of $L_X$ to $K$-band luminosity \citep[][top]{Cheng18}, or to masses calculated from visible light
        \citet[][2010 revision, bottom]{Harris96}, compared with the normalized (by mass) stellar interaction rate (x-axis). In the lower panel, we have also adjusted $L_X$ for NGC 6366 (black star), following \S 3.  NGC 6397, particularly affected by a spurious K-band luminosity estimate, is indicated with a blue filled hexagon. Pearson and Spearman correlation coefficients are indicated for both samples. }
        \label{fig:Cheng2}
\end{figure}

\citet{Cheng18} calculate $L_X$ by extracting all X-ray flux within the half-mass radius. For clusters with high X-ray fluxes, this method works well, but for clusters with low X-ray fluxes (those with few X-ray sources, and/or more distant clusters), it is vulnerable to contamination by fore- or background sources.
We suggest that the poorer correlation between normalized encounter rate and normalized $L_X$ in Cheng et al's fig. 6 (left), vs. their fig. 6 (right), 
 may be 
due to  their left figure adding many clusters which are more distant, low-encounter-rate, and/or obscured, and thus adding lower-quality data, obscuring the existing correlation.

For instance, as discussed in \S 3.2.1, a background cluster of galaxies is projected onto the low-encounter-rate cluster NGC 6366, producing $\sim$75\% of the observed X-ray flux within the cluster half-mass radius. We find the X-ray flux from NGC 6366 cluster members to be between 4 and 25\% of the total flux within its half-mass radius.

In Fig.~\ref{fig:Cheng2}, we show two plots similar to Cheng et al.'s fig. 6 (right),  plotting normalized encounter rate vs. normalized $L_X$, for the same 23 clusters. On the top, we reproduce Cheng et al.'s plot, and indicate the Pearson and Spearman correlation coefficients. On the bottom, we normalize $L_X$  using dynamical mass estimates (visible-light estimates give similar values) instead of $K$-band estimates, and rescale the $L_X$ estimate for NGC 6366. The inferred correlation coefficients are significantly higher. This illustrates our argument that careful attention to membership is necessary to get robust measurements of X-ray emissivity, and thus to make these determinations.  On the other hand, Cheng et al's method is necessary to explore unresolved flux, which is quite important for some cluster observations.

\subsection{Metallicity}

 We find significant evidence for a positive effect of metallicity on the X-ray emissivity of low-luminosity X-ray sources across multiple stellar populations, via multiple tests. The explanation for this is not clear. Most explanations that have been offered to explain the larger numbers of bright LMXBs in higher-metallicity globular clusters \citep[e.g.][]{Kim13} involve neutron stars or dynamical interactions, which do not seem plausible for the low-density sample. However, \citet{Ivanova12} identify an increase in the numbers and average masses of red giants in higher-metallicity populations. This effect may help explain increased CV and active binary X-ray emission. 

\subsection{Binary fraction}

\citet{Verbunt01} suggested that the elevated X-ray luminosity per unit mass observed in low-density open clusters could be the result of significant mass-loss \citep[see also][]{vandenBerg13,Ge15,Vats17}.  In the simplest scenario, the total cluster mass is the dominant cluster parameter that decides the number of binaries, with the \textit{primordial} scaling between binary numbers and total cluster mass being roughly linear (e.g. this would arise naturally if the primordial binary fraction is the same in all clusters).  The extra rise in X-ray luminosity per unit mass in low-density open clusters seen in Fig.\ref{fig:MC_BRo} is then due solely to the evaporation of preferentially low-mass \textit{single} stars across the tidal boundary, which inflates the binary fraction.  The rate of stellar evaporation is decided by the rate of two-body relaxation, which scales inversely with the total cluster mass.  In this scenario, the (observable) binary fraction should reflect this process, and thus one might think the calculation should be straightforward. 

However, the binaries that exist in star clusters are limited in orbital period by the density of the cluster, while binaries in the field of galaxies may have much longer orbital periods. The longer orbital period systems found in the field will never interact, and thus do not increase X-ray emissivity. Thus, the details of the binary period distribution, as well as the total binary fraction, will be important in determining X-ray emissivity. The exact details of which binary orbital periods contribute efficiently to increasing X-ray emissivity are a complicated topic, beyond the scope of this work.

If the reduced X-ray emissivity of (low-density) globular clusters can be attributed (largely, or wholly) to their reduced binary fraction, this still leaves open the question of why globular clusters have much lower binary fractions than other populations. Binaries of a given orbital period will tend to be disrupted at high densities, and increasing the density destroys binaries of shorter and shorter periods   
\citep{Heggie75}.
Detailed N-body \citep{Hurley07} and Monte Carlo \citep{Fregeau09} models of globular clusters indicate that the overall globular cluster binary fraction does not decrease with time in realistic present-day globular clusters, at least when only "hard" binaries (those with short orbital periods--which are also those relevant for X-ray emissivity) are considered. The destruction of binaries in globular cluster cores is counteracted by evaporation of single stars from the outside of the cluster. Including "soft" (long-orbit) binaries can lead to binary fractions that decrease with time, but it appears unlikely that these longer-period systems will strongly affect the X-ray emissivity. 

A more likely scenario is that most globular clusters were much denser (up to $10^6$ \Msun\ pc$^{-3}$ on average within the half-mass radius) during the star formation phase.  Such high initial densities, which reduce dramatically as the cluster expands due to gas loss, can substantially reduce cluster binary fractions, especially in those with the largest masses \citep{Sollima08,Parker09, Marks10}, as discussed in particular detail by \citet{Leigh15}. There is substantial evidence that blue straggler stars in most clusters are dominated by binaries where one star has transferred mass to the other \citep{Leigh07,Knigge09,Geller11,Gosnell15}. The numbers of these blue straggler stars generally trace the same decreasing abundance with increasing cluster mass as binary fractions \citep{Piotto04,Leigh13b}, indicating that the abundances of both blue stragglers and X-ray emitting stars in lower-density stellar populations may be driven by the fractions of binaries.

\section{Conclusions}

In this paper, we investigate the relations between the density, binary fraction, and metallicity of old stellar populations, and their X-ray luminosity per unit mass, or X-ray emissivity.  A variety of old populations in galaxies and old clusters show substantially higher $L_X$/M than low- to moderate-density globular clusters, while the higher-density globular clusters show an increase in $L_X$/M due to well-studied dynamical effects \citep[e.g.][]{Pooley03,Bahramian13}.  The substantially higher $L_X$/M of open clusters compared to most globular clusters was first identified by \citet{Verbunt01}, and subsequently extended to a range of environments \citep[e.g.][]{Revnivtsev07,vandenBerg13,Ge15}.   

In order to investigate X-ray emissivity in different stellar populations, we have compiled a wide range of data from the literature.  We identify relatively bright X-ray sources projected onto three low-density environments (the open cluster NGC 6819, the globular cluster NGC 6366, and the Sculptor Dwarf galaxy,) with foreground or background sources, emphasizing the importance of ascertaining membership of X-ray sources among the relevant population.

We show that globular clusters, below a critical density of $\sim10^4$ \Msun pc$^{-3}$, have much lower X-ray emissivities than other populations, while globular clusters above this density have increasing X-ray emissivity with density. At these higher densities, a new population of close binaries is created through dynamical interactions, increasing X-ray emissivity.  
 Comparison of X-ray emissivity with density, age, metallicity, and overall binary fraction suggests that binary fraction and/or metallicity can explain variations in X-ray emissivity in low-density stellar populations, while density has the strongest effect on X-ray emissivity in higher-density stellar populations. 
A straightforward fit to the full sample indicates significant effects from density, metallicity, and binary fraction, but  subjecting the fit to bootstrap tests leaves the effect of binary fraction unclear. 
Further careful study of faint X-ray sources in various stellar populations (in particular, accurate membership of binary fractions in more populations) will elucidate further details of these relations, and test the suggestions here.

\section*{Acknowledgements}
LN thanks M. Mayrand, J. St-Antoine and the OMM staff  for their technical assistance. We acknowledge helpful discussions with R. Arnason, T. Maccarone, \& A. Bahramian, and helpful referee reports. 
COH, GRS, EWR, NI and LN acknowledge financial support from Discovery Grants from the Natural Sciences and Engineering Research Council (NSERC) of Canada, and COH from a Discovery Accelerator Supplement.  LC was supported in part by NASA Chandra grant G06-17040X. NI acknowledges support from the CRC program. NWCL acknowledges the generous support of Fondecyt Iniciacion Grant \#11180005.  AJR is funded by the Australian Research Council through grant number FT170100243.  GRS and CLS were supported in part by NASA Chandra grants GO7-8078X, GO7-8089A, GO8-9053X, and GO9-9085X, and NASA Hubble grants HST-GO-11679.01 and HST-GO-12012.02-A.

The National Radio Astronomy Observatory is a facility of the National Science Foundation (NSF) operated under cooperative agreement by Associated Universities, Inc (AUI). 
Based in part on observations obtained at the Gemini Observatory (and processed using the Gemini IRAF package), which is operated by AUI, under a cooperative agreement with the NSF on behalf of the Gemini partnership: the NSF (United States), the National Research Council (Canada), CONICYT (Chile), Ministerio de Ciencia, Tecnolog\'{i}a e Innovaci\'{o}n Productiva (Argentina), and Minist\'{e}rio da Ci\^{e}ncia, Tecnologia e Inova\c{c}\~{a}o (Brazil). We acknowledge use of the Hubble Archive at STSci, the Chandra Source Catalog, the NRAO VLA archive, the Chandra, Swift, and XMM-Newton data archives, NASA's Astrophysical Data Service, and arXiv.org, maintained by Cornell University Library. 
This work has used data from the European Space Agency (ESA) mission
{\it Gaia} (\url{https://www.cosmos.esa.int/gaia}), processed by the {\it Gaia}
Data Processing and Analysis Consortium (DPAC,
\url{https://www.cosmos.esa.int/web/gaia/dpac/consortium}). Funding for the DPAC
has been provided by national institutions, in particular the institutions
participating in the {\it Gaia} Multilateral Agreement.

Code bibliography:
 XSPEC \citep{Arnaud96}, CIAO \citep{Fruscione06}, XMM SAS (\url{https://www.cosmos.esa.int/web/xmm-newton/sas}), AIPS (\url{http://www.aips.nrao.edu/index.shtml}), Gemini IRAF (\url{https://www.gemini.edu/sciops/data-and-results/processing-software}), PyMC3 \citep{Salvatier16},  corner.py \citep{ForemanMackey16},
SuperMongo (\url{https://www.astro.princeton.edu/~rhl/sm/sm.html}).

\bibliographystyle{mnras}
\bibliography{odd_src_ref_list}

\begin{center}
\begin{table*}
\caption{Properties of non-globular cluster populations}
\begin{tabularx}{\textwidth}{bmmmssYm}
\hline
\hline
Name      & Distance      & $\rho_{c}$   & Mass    & {Age}  & Binary   & {[Fe/H]} & {$L_{X}/M$} \\ 
	      & (kpc)         & (\Msun pc$^{-3}$) & (\Msun) & (Gyrs) & Fraction & & ($10^{27}$ \ergss\Msun$^{-1}$)   \\
\hline
Local     & $<1$          & 0.10 & -         & 5*$\pm$3  & 0.44$\pm$0.02   & -0.15*$\pm$0.1 & $9\pm3$  \\ 
NGC2682/M67 & 0.82        & 32   & 2100$\pm$600       & 4         & 0.227$\pm$0.021 & 0.05$\pm$0.02  & 45$\pm$14  \\ 
NGC 188   & 1.70$\pm$0.07 & 9 & 2300$\pm$460       & 5.5--7.1  & 0.29$\pm$0.03   & -0.03$\pm$0.04 & 25$\pm$6  \\ 
NGC 6791  & 4.1           & 11  & 6000$\pm$1000      & 8--9      & 0.32$\pm$0.03   & 0.47$\pm$0.12  & 35$\pm$8  \\ 
NGC 6819  & 2.34$\pm$0.06 & 54    & 2600               & 2--2.4    & 0.22$\pm$0.03   & -0.02$\pm$0.02 & 37$\pm$20  \\ 
Cr 261    & 2.45$\pm$0.25 & 15    & 6500$\pm$700       & 6--7      & -               & -0.16$\pm$0.13 & 30$\pm$7  \\ 
NGC 185   & 617           & -    & $1.6\times10^8$    & 8--12     & -               & -1.3$\pm$0.1   & 6.1$\pm$2.5 \\ 
NGC 147   & 676           & -    & $1.1\times10^8$    & 7--13     & -               & -1.1$\pm$0.1   & 11.2$\pm$6.4 \\ 
NGC 205   & 824           & -    & $6.8\times10^8$    & 9--13     & -               & -0.8$\pm$0.2   & 7.3$\pm$3.8 \\ 
M31 bulge & 780           & -    & $4\times10^{10}$   & 6--13     & -               & -2--0.5        & 7.5$\pm$2 \\ 
NGC 221/M32 & 794           & 1.0* & $2.1\times10^9$    & 8*(2--10) & -               & -0.3*(-0.5--0) & 5.4$\pm$1.7 \\ 
NGC 3379  & 9800          & -    & $5.7\times10^{10}$ & 8--10     & -               & 0.3            & 5.2$\pm$1.6 \\ 
\hline
\end{tabularx}		
\\[5pt]
Properties of non-globular cluster stellar populations considered. Density indicates central densities for open clusters, density of considered region in galaxies. $L_X$ measured in 0.5-2 keV band.
$L_X$/M for open clusters uses M/2, since only X-ray sources within the half-mass radius are included.
* indicates the average value (for stellar populations showing a range in a parameter). \\
References: 
Local: density, \citet{Holmberg00}, age and [Fe/H], \citet{Nordstrom04}, binary fraction, \citet{Raghavan10}, $L_X$/M, \citet{Sazonov06,Warwick14};
M67:  mass and binary fraction, \citet{Geller15}, age, \citet{Bellini10}, [Fe/H], \citet{Jacobson11}, density, \citet{Hurley05}, $L_X/M$, \citet{vandenBerg04};
NGC 188: mass, \citet{Geller08}, binary fraction, \citet{Geller15}, age \& distance \citet{Meibom09}, [Fe/H], \citet{Jacobson11}, $L_X/M$, \citet{Vats18};
NGC 6791: mass, \citet{Platais11}, age, \citet{Grundahl08}, binary fraction, \citet{Bedin08}, [Fe/H], \citet{Gratton06,Carraro06}, $L_X/M$, \citet{vandenBerg13};
NGC 6819: mass \citet{Kalirai01}, age, \citet{Basu11}, binary fraction \citet{Milliman14}, [Fe/H], \citet{Lee-Brown15},
$L_X/M$, \citet{Gosnell12,Platais13}; 	    
Collinder 261:  age, \citet{Bragaglia06}, [Fe/H], \citet{Friel02}, 
mass \& $L_X/M$, \citet{Vats17};
NGC 185, NGC 147, NGC 205: $L_X/M$, \citet{Ge15}, others \citet{McConnachie12};
M31 bulge: $L_X/M$, \citet{Revnivtsev08}, others \citet{Brown06};
M32:  age, \citet{Monachesi12}, [Fe/H], \citet{Coelho09},  $L_X/M$, \citet{Revnivtsev07};
NGC 3379; age, \citet{Idiart07}, $L_X/M$,\citet{Revnivtsev08}.
\label{table*:nonglobs}
\end{table*}
\end{center}

\begin{table*}
\caption{Properties of globular clusters}
\centering
\begin{tabularx}{\textwidth}{bSSSmmSSSSSmm}
\hline
\hline
Cluster & Expos. & Dist. & log($\rho_{c}$)  & Mass & Binary &  \# Srcs & $L_X$ (half) & $L_{\rm X,max}$        & $L_{\rm X,bg?}$    & $L_{\rm X,min}$ & $2 L_X/M$ & Reference\\
        & (ks)     & (kpc)    & (\Msun/pc$^{3}$) & (\Msun) & \% & (Memb.) &  \multicolumn{4}{c}{($10^{31}~$\ergss)} & ($10^{27}$\ergss/\Msun) & \\
\hline
N104/ 47Tuc      & 281   & 4.5  & 5.15 & $7.6\times10^5$  & 1.8(0.6)  & 300(105) & 295 & 270  & 12  & 246  & 5.82$\pm$1.28       & 1,2,3\\
N3201     & 83  & 4.9  & 2.98 & $1.3\times10^5$  & 10.8(1.2)  & 4(0) & $<$7.5 & 13     & 9   & 0    & *0.84$\pm$0.84 & 4,6 \\
N4372     & 10    & 5.8  & 2.33 & $2.2\times10^5$  & -          & 6(0) & <1.8 & *12    & 30  & 0    & 0.63$\pm$0.63 & 5,6\\
N5139/ $\omega$ Cen & 70    & 5.2  & 3.42 & $3.4\times10^6$  & -          & 81 & 71 & 130   & 25  & 57   & 0.89$\pm$0.39    & 7,8,9\\
N6121/M4  & 26    & 2.2  & 3.91 & $9.1\times10^4$  & 10.2(1.0)  & 21(15) & 26 & 19    & 4   & 18   & 3.64$\pm$0.98        & 10,11\\
N6218/M12 & 22    & 4.8  & 3.50 & $8.1\times10^4$  & 6.4(0.6)   & 6(2) & 14 & 10    & 7   & 10   & 1.76$\pm$0.46        & 12\\
N6254/M10 & 32.6  & 4.4  & 3.81 & $1.9\times10^5$  & 4.4(0.6)   & 11(1) & 6.9 & 13    & 4   & 0.73 & 0.89$\pm$0.81        & 13,6\\
N6304     & 5.3   & 5.9  & 4.76 & $1.5\times10^5$  & -          & 7(6) & 145 & 81    & 5   & 71   & 15.8$\pm$6.3        & 14,15,6\\
N6352     & 19.8  & 5.7  & 3.44 & $6.1\times10^4$  & 10.6(1.0)  & 7(0) & 35 & 19    & 8   & 0    & 5.65$\pm$5.65 & 6 \\
N6366     & 22    & 3.5  & 2.66 & $5.0\times10^4$  & 11.4(3.0)  & 5(2) & $<6$ & 1.9   & 6   & 1.4  & 2.30$\pm$1.50        & 16,6\\
N6397     & 49    & 2.3  & 6.03 & $8.9\times10^4$ & 2.4(0.6) & 79(43) & 66 & 44    & 1.3 & 41   & 15.0$\pm$4.6        & 17,18\\
N6544     & 12    & 2.5  & 5.72 & $1.2\times10^5$  & -          & 7(1) & 8.6 & 6.2  & 1.5 & 3.0  & 1.70$\pm$0.89       & 6,19\\
N6553     & 5.2   & 6.0  & 4.11 & $3.3\times10^5$  & -          & 22(1) & 77 & 47   & 6   & 34   & *4.48$\pm$1.95        & 6,20\\
N6626/M28 & 38    & 5.5  & 5.13 & $3.0\times10^5$  & -          & 30(3) & 288 & 257   & 5   & 247  & 15.7$\pm$3.4        & 21\\
N6656/M22 & 16    & 3.2  & 3.90 & $4.1\times10^5$  & 4.0(0.6)   & 15 & 18 & 29    & 7   & 7.6   & *0.92$\pm$0.57        & 22\\
N6752     & 29    & 4.0  & 5.31 & $2.3\times10^5$  & 0.9(0.6)   & 19(10) & 36 & 47 & 3.0 & 41 & *4.44$\pm$0.94        & 23\\
N6809/M55 & 34    & 5.4  & 2.49 & $1.9\times10^5$  & 8.0(0.6)   & 17(2) & $<$6 & 11.5  & 16  & 4    & *0.93$\pm$0.49        & 24,25\\
N6838/M71 & 52    & 4.0  & 3.10 & $5.3\times10^4$  & 22.0(1.6)  & 29(10) & 13 & 13    & 2.8 & 7    & 7.50$\pm$2.70        & 26,27\\
IC 1276   & S-2.4 & 5.4  & 3.05 & $9.3\times10^4$  & -          &  & $<8$ & - & 10  & 0    & 1.08$\pm$1.08 & 6\\
Pal 6     & 4.6   & 5.9  & 3.73 & $1.4\times10^5$  & -          &  & $<6.8$ & -  & 3   & 0    & 0.48$\pm$0.48 & 6\\
Pal 10    & 11    & 5.9  & 3.78 & $5.5\times10^4$  & -          &  & $<18$ & -  & 2   & 0    & 5.48$\pm$5.48 & 6\\
Terzan 5  & 35    & 5.9  & 6.15 & $3.9\times10^5$  & -          &  & 557 & 900   & 1.5 & 897  & 25.7$\pm$5.1        & 29,30,31,32\\
\hline
\multicolumn{13}{c}{Beyond 6 kpc}\\
\hline
N288      & 55    & 8.8  & 2.05  & $1.2\times10^5$  & 10.8(1.4) &  & 7 & 19    & 22  & 0   & 2.74$\pm$2.74 & 33\\
N2808     & 57    & 9.6  & 4.93  & $8.2\times10^5$  & -         &  & 137 & 70    & 3   & 70  & 2.29$\pm$0.87        & 34,35\\ 
N6093/M80 & 49    & 10.0 & 5.06  & $2.8\times10^5$  & 1.2(0.6)  &  & 101 & 110   & 3   & 104 & 6.90$\pm$1.39        & 36 \\
N6144     & 55    & 8.5  & 2.58  & $5.3\times10^4$ & 7.8(1.0) &  & 42 & 19    & 12  & 0   & 3.98$\pm$3.98 & 28\\
N6341/M92 & 53    & 8.3  & 4.57  & $3.1\times10^5$  & 2.0(0.6)  &  & 30 & 24    & 4.6 & 20  & 1.65$\pm$0.47        & 37\\
N6388     & X-25  & 10.0 & 5.64  & $1.1\times10^6$  & 0.8(0.8)  &  & 459 & 310   & 3   & 304 & 5.59$\pm$1.59        & 38\\
N6440     & 23    & 8.4  & 5.51  & $3.8\times10^5$  & -         &  & 541 & 400   & 0.9 & 398 & 18.7$\pm$4.7        & 39 \\
N6715/M54 & 30    & 26.8 & 4.96  & $1.6\times10^6$  & -         &  & 780 & 250   & 11  & 228 & 6.25$\pm$3.65        & 40\\
N7099/M30 & 49    & 8.0  & 5.28  & $1.3\times10^5$  & 2.4(0.6)  &  & 53 & 56    & 5.3 & 45  & 6.81$\pm$1.54        & 41\\
E3           & 20    & 8.1  & 1.11 & $2.8\times10^3$  & 63(41) 
&  & 37 & 2.8   & 18   & 0    & 53.6$\pm$53.6 & 28,6\\
\hline
\end{tabularx}		
\\[5pt]
Densities and distances from \citet{Harris96}, 2010 update, except for Terzan 5.  Masses from \citet{BaumgardtHilker18}. 
Luminosities are given in the 0.5-2 keV range.
"Exposure" column indicates Chandra exposure times, except that X indicates XMM, S indicates Swift.  Upper limits on $L_{X,max}$ indicate that X-ray flux within the half-mass radius is not yet detected.  Estimates for the expected "$L_X$" due to X-ray background are provided under $L_{X,bg?}$, for comparison. Zeroes for $L_{X,min}$ indicate that no X-ray sources can be confidently identified with the cluster. We quote $2 L_X$/M, since we only measure $L_X$ of sources within the half-mass radius.

References: 
1) \citet{Grindlay01a}, 	
2) \citet{Edmonds03a}, 		
3) \citet{Heinke05b}, 
4) \citet{Webb06}, 
5) \citet{Servillat08a}, 	
6) this work,
7) \citet{Gendre03a},
8) \citet{Haggard09},
9) \citet{Cool13},
10) \citet{Bassa04},
11) \citet{Bassa05},
12) \citet{Lu09}, 
13) \citet{Shishkovsky18},
14) \citet{Guillot09a}, 
15) \citet{Guillot09b},
16) \citet{Bassa08}, 
17) \citet{Bogdanov10},
18) \citet{Cohn10},
19) \citet{Cohen14},
20) \citet{Guillot11}, 
21) \citet{Becker03},
22) \citet{Webb13},
23) \citet{Pooley02a},
24) \citet{Kaluzny05},
25) \citet{Bassa08},
26) \citet{Elsner08},
27) \citet{Huang10},
28) \citet{Lan10},
29) \citet{Valenti07},
30) \citet{Heinke06a},
31) \citet{Lanzoni10},
32) \citet{Prager17},
33) \citet{Kong06},
34) \citet{Servillat08a},
35) \citet{Servillat08b}, 
36) \citet{Heinke03c}, 
37) \citet{Lu11},
38) \citet{Nucita08},
39) \citet{Pooley02b},
40) \citet{Ramsay06}, 
41) \citet{Lugger07}.
\label{table*:globs}
\end{table*}

\appendix
\section{Details on individual globular clusters}

{\bf NGC 3201:}
\citet{Webb06} report an XMM observation of NGC 3201, with four sources located within the half-mass radius, though not concentrated in the core.  An archival Chandra observation of NGC 3201 (ObsID 11031) finds 22 sources projected within the half-mass radius of the cluster with total $L_X$(0.5-2 keV) (assuming the cluster $N_H$ and a power-law of photon index 2) of $1.3\times10^{32}$ erg/s, though all could be background sources. \citet{Cheng18} measure the total X-ray flux from NGC 3201 to be $<7.5\times10^{31}$ erg/s (correcting to the 0.5-2 keV band using their choice of a $\Gamma=2$ power-law), consistent with our analysis.

{\bf NGC 4372:}
An XMM observation of NGC 4372 reported by \citet{Servillat08a} reported no X-ray sources within the half-mass radius. A recent archival Chandra observation (PI Chomiuk) reveals six faint sources within NGC 4372's half-mass radius. Assuming a power-law of photon index 1.8, with the cluster $N_H$, gives an estimate of $L_X=1.2\times10^{32}$ erg/s for their luminosity, though it is unclear whether any of these X-ray sources are associated with the cluster.  Using the "double-subtraction" method to analyze the total emission from the half-mass radius, we measure the  X-ray flux from NGC 4372 to be $<1.8\times10^{31}$ erg/s, consistent with our point-source analysis.

{\bf $\omega$ Cen/NGC 5139:}
For $\omega$ Cen, we use a half-mass radius of 5' (following the 2010 revision of the Harris catalog), and accept the optical identifications of cluster members listed in \citet{Cool13} within that radius, to determine the minimum $L_X$ from the cluster ($L_{X,min}$(0.5-2 keV)=$5.7\times10^{32}$ erg/s).  The \citet{Cheng18} estimate of the total 0.5-2 keV $L_X$ is $7.1\times10^{32}$ erg/s, which is  basically consistent.

{\bf M12/NGC 6218:}
For M12, we accept the identification of the X-ray source CX1  as a secure cluster member by \citet{Lu09}, supported by \citet{Gottgens19}.  We use the {\it HST} proper motion catalog of \citet{Nardiello18} to check the membership status of the other sources. We find that the star CX2b (a sub-subgiant) is not a cluster member, while CX2c (a candidate CV) is a cluster member, and verify its blue colours in the ultraviolet.  We confirm CX2 as a CV and cluster member, but are not able to ascertain membership for sources CX3-6. We obtain $L_{X,min}=9\times10^{31}$ erg/s, while the \citealt{Cheng18} total $L_X$ estimate is $1.4\times10^{32}$ erg/s.

{\bf M10/NGC 6254:}
NGC 6254/M10 has one secure cluster member X-ray source reported so far, the radio source reported by \citet{Shishkovsky18}, with $L_X$(0.5-2)$=7.3\times10^{30}$ erg/s. We identify a further 10 sources in M10, spread across the half-mass radius, 
 giving a maximum point source  $L_{X,max}$(0.5-2)$=1.3\times10^{32}$ erg/s. We also perform a double-subtraction spectral analysis of the Chandra data within the half-mass radius, finding  $L_X$(0.5-2)=$6.9\pm2.5\times10^{31}$ erg/s.

{\bf NGC 6304:}
 We use the 97 ks Chandra observation of NGC 6304 \citep{Guillot13}, and find 17 X-ray sources within the half-mass radius between 0.5 and 2 keV, concentrated towards the cluster core. 
The brightest of these is the quiescent LMXB reported by \citet{Guillot09a,Guillot09b},  which has a 0.5-2 keV $L_X$ of $6.1\times10^{32}$ erg/s (from a direct spectral fit using an absorbed NSATMOS hydrogen atmosphere neutron star model).  Including the other detected point sources gives $L_X=8.1\times10^{32}$ erg/s. However, \citet{Cheng18} measure $1.45^{+0.08}_{-0.05}\times10^{33}$ erg/s (corrected to 0.5-2 keV) using all Chandra data within the half-mass radius.
Our own measurement of the total $L_X$ within the half-mass radius for this cluster gives $1.2^{+0.4}_{-0.3}\times10^{33}$ erg/s, between the other two estimates.

 \citet{Guillot09a} suggested that two X-ray sources lying outside the cluster half-mass radius, XMMU 171411-293159 and XMMU 171421-292917, are cluster members and quiescent LMXBs, and showed that they are clearly  associated with the stars 2MASS 17141152-2931594 and 2MASS 17142095-2929163. 
However, these stars now have Gaia parallax distances ($800\pm100$ pc; 732$\pm20$ pc) and proper motions (pmRA=-3.79$\pm0.09$ mas/yr, pmDec=-5.93$\pm0.06$ mas/yr; and pmRA=0.81$\pm0.07$ mas/yr, pmDec=-2.15$\pm0.04$ mas/yr) inconsistent with the distance (5.9 kpc) and \citet{GaiaGC} proper motion (pmRA=-3.95$\pm0.01$ mas/yr, pmDec=-1.125$\pm0.007$ mas/yr) of NGC 6304, so are clearly not members of NGC 6304.

{\bf NGC 6352:}
 We analyze an archival 19.8 ks Chandra observation of NGC 6352 (ObsID 13674), finding ten likely X-ray sources in the cluster, the membership of which are uncertain (they are not clustered in the core).   We find a total possible point source luminosity of  $L_X=1.8\times10^{32}$ erg/s, while \citet{Cheng18} measure a total of  $L_X=3.5\pm0.7\times10^{32}$ erg/s.

{\bf NGC 6366:}
The unusual extended X-ray source CX1b in NGC 6366 \citep{Bassa08} is discussed in detail in \S 3.2 below, where it is shown to be a background cluster of galaxies. 
 In \S 3.2, we present evidence that the giant star associated with CX1a is the correct optical counterpart and is associated with the cluster. 

The half-mass radius of 2.92' of NGC 6366 also includes CX2, CX3, CX4, CX5, and CX14 (note that \citealt{Bassa08} used a smaller half-mass radius, leaving CX14 outside).  Of these, we can verify only CX5 as a cluster member; the optical counterpart has a proper motion, measured with both HST \citep[using the ACS and WFC3 surveys,][]{Nardiello18} and Gaia, consistent with the cluster proper motion, indicating cluster membership. CX5's counterpart is a  sub-subgiant in the CMD; as the fraction of sub-subgiants that are X-ray sources is quite large \citep{Geller17}, we consider it very likely that this star is the true counterpart to CX5.
The other sources we take to be possible members, giving us a point source $L_X=1.4\times10^{31}-3.1\times10^{31}$ erg/s.
\citet{Cheng18} measure $L_X$(0.5-2)$=1.5\pm0.5\times10^{32}$ erg/s. 
We double check this measurement using the double subtraction method, and measure $L_X$(0.5-2)$=1.2^{+0.4}_{-0.6}\times10^{32}$ erg/s. 
However, this includes a background cluster of galaxies; subtracting its flux ($L_X=6\times10^{31}$ if it were in NGC 6366) leaves us with only an upper limit of $L_X$(0.5-2)$<6\times10^{31}$ erg/s, consistent with the point source measurement.

 {\bf NGC 6388:}
 We use the XMM flux measurement of NGC 6388 from \citet{Nucita08}, which is consistent with the \Chandra\ estimate \citep{Maxwell12}, as the XMM flux includes all the X-ray emission from the core of this rich cluster.

{\bf NGC 6544:}
Significant background flaring occurred during the 16.5 ks \Chandra\ observation of NGC 6544 (ObsID 5435); removing this leaves 12 ks.  Five clear sources can be seen within the half-mass radius, plus diffuse emission concentrated within a couple arcsec of the cluster centre location of \citet{Cohen14}.  Since the total X-ray flux substantially exceeds the expected cosmic background, we estimate the X-ray luminosity of the cluster by subtracting the latter. The \citet[][2010 edition]{Harris96} catalog reports NGC 6544 as the cluster with the highest central density, $10^{6.06}$ \Lsun\ pc$^{-3}$. However, \citet{Cohen14} report a new photometric and structural study of NGC 6544 that gives a  larger core and lower central density, which we calculate to be  $10^{5.45}$ \Lsun\ pc$^{-3}$.
 \citet{Cheng18} give a total luminosity, $L_X$(0.5-2)$=9\pm2\times10^{31}$ erg/s, about twice our estimated point source luminosity $L_X=4.7\times10^{31}$ erg/s.

{\bf NGC 6553:}
\citet{Guillot11} report XMM and Chandra observations of NGC 6553, identifying 21 X-ray sources within the half-mass radius. They identify one secure cluster member (a qLMXB, their source 3), and show that one  source (\#5) is not a member. 
We calculate the 0.5-2 keV minimum unabsorbed flux from their spectral fit to source \#3. For other objects within the half-mass radius, we are not able to reproduce the fluxes listed in their Table 2 from their count rates and reported spectral models, so we instead calculate fluxes assuming a power-law of photon index 2, from the pn count rates provided by Guillot et al., and checked using an independent analysis of the longer Chandra observation, finding a total point source $L_X=4.7\times10^{32}$ erg/s. \citet{Cheng18} measure a flux from the half-mass radius of $7.7\times10^{32}$ erg/s.   

{\bf M28/NGC 6626:}
 We used the longest (142 ks) Chandra exposure of M28 \citep{Bogdanov11}, and detected 66 sources in the 0.5-2 keV range. These gave a maximum point source $L_X=2.57\times10^{33}$ erg/s, in good agreement with the total flux estimate from \citet{Cheng18} of $L_X$(0.5-2)$=2.88\times10^{33}$ erg/s.

{\bf M22/NGC 6656:}
\citet{Webb13} report 15 X-ray sources detected by Chandra within M22's half-mass radius.   We regard only the optically identified CV1 \citep{Anderson03,Bond05,Webb13} and the millisecond pulsar PSR J1836-2354A \citep{Lynch11,Webb13} as verified cluster sources,  giving a minimum  $L_X$(0.5-2)$=7.6\times10^{31}$ erg/s, and a maximum of $2.9\times10^{32}$ erg/s.  \citet{Cheng18} give an intermediate estimate of the flux within the half-mass radius of $1.8\times10^{32}$ erg/s.

{\bf M55/NGC 6809:}
M55 has one CV (CX1) that has shown a recorded outburst \citep{Kaluzny05}, and one source (CX7) with an optical counterpart located in the sub-subgiant region, where spurious matches are very rare \citep{Bassa08}.  Bassa et al's optical counterpart to CX7 is present in Gaia DR2, and its proper motion is consistent with NGC 6809's proper motion \citep{GaiaGC}. 
We take CX1 and CX7 to be secure cluster members, giving a minimum $L_X=4\times10^{31}$ erg/s and a maximum of $1.2\times10^{32}$ erg/s. \citet{Cheng18} gives a limit on the total flux from the half-mass radius of $6\times10^{31}$ erg/s.  

{\bf M71/NGC 6838:}
 \citet{Huang10} presented a detailed search for optical counterparts in M71. 
We regard the following as secure cluster members: 
s08 (radio-detected millisecond pulsar),
s04, s06, s07, s18, s20, s27a (giant,  subgiant, or sub-subgiant stars with proper motions reported by \citealt{Nardiello18} and Gaia DR2 that agree with the \citealt{GaiaGC} cluster proper motion), 
s19a (sub-subgiant, proper motion verified by HST, \citealt{Nardiello18}),
and s29 (CV, identified by strong UV excess vs. redder $V$-$I$ colours).
We exclude the foreground sources s02, s12, s16, s22, s24, and s26; we confirm that the HST proper motions of the first 4 are discordant with the cluster motion. 
The X-ray point source luminosity associated with the cluster is then in the range of $L_X=5.9\times10^{31}$--$1.4\times10^{32}$ erg/s, consistent with the total luminosity estimate of $1.3\times10^{32}$ erg/s of \citet{Cheng18}.

{\bf E3 and NGC 6144:}
No X-ray sources projected on the globular clusters  E3 or NGC 6144 \citep{Lan10} can be confidently identified as members of their clusters.  E3 CX03 is identified by Lan et al. as a background galaxy. We identify a G=15 Gaia \citep{GaiaDR2} source with a parallax distance of 553$\pm$15 pc (thus a clear foreground source) with E3 CX01, leaving only E3 CX02 ($L_X=2.8\times10^{31}$ erg/s) as a candidate cluster member.
 Extracting the total flux from E3's half-mass radius gives a total $L_X$(0.5-2)$=3.7\pm1.2\times10^{32}$ erg/s, substantially larger than the resolved source measurement. 

For NGC 6144, the suggested counterparts to CX1 and CX2  \citep{Lan10}  show blue colors in UV HST filters (supporting a CV nature), but \citet{Nardiello18} cannot estimate their membership probability, so we are not certain they are not AGN. \citet{Cheng18} estimate a total half-mass flux of $4.2\times10^{32}$ erg/s.

{\bf Palomar 6, Palomar 10, IC 1276:}
No X-ray sources are detected within the half-mass radii of  
Palomar 10  (ObsID 8945), Palomar 6 (ObsIDs 9986 and 10010), or IC 1276 (observed only by Swift).  We set upper limits for these using the total possible excess X-ray counts within the half-mass radius of each cluster.   \citet{Cheng18} give an upper limit of $L_X<1.8\times10^{32}$ erg/s for Palomar 10. We estimate an upper limit of $L_X<6.8\times10^{31}$ erg/s for Palomar 6 from \Chandra\  ObsID 10010, and an upper limit of $L_X<8\times10^{31}$ erg/s for IC 1276 from a 2.4 ksec Swift/XRT observation.

\bsp	
\label{lastpage}
\end{document}